\pdfoutput=1
\documentclass[11pt]{article}

\usepackage[english]{babel}
\usepackage{xcolor}
\usepackage{graphicx}

\usepackage[colorlinks,allcolors=blue,linktocpage=true]{hyperref}

\usepackage{cite}
\usepackage{subfigure}
\usepackage{amsmath}

\usepackage{placeins}
\usepackage{booktabs}

\usepackage[top=15mm,bottom=12mm,left=30mm,right=30mm,foot=12mm,includefoot]{geo
metry}

\allowdisplaybreaks[3]

\newcommand{\sect}[1]{section~#1}
\newcommand{\app}[1]{appendix~#1}

\newcommand{\fig}[1]{figure~#1}
\newcommand{\figs}[1]{figures~#1}

\newcommand{\tab}[1]{table~#1}

\newcommand{\eqn}[1]{equation~#1}
\newcommand{\eqs}[1]{equations~#1}

\newcommand{\gev}{\operatorname{GeV}}

\newcommand{\fm}{\operatorname{fm}}

\newcommand{\ms}{\mskip 1.5mu}
\newcommand{\bs}{\mskip -1.5mu}

\newcommand{\msbar}{$\overline{\text{MS}}$ }

\newcommand{\conv}[1]{\underset{x_{#1}}{\otimes}}

\newcommand{\pr}[2]{{}^{#1}\bs #2}      % with small backspace
\newcommand{\prb}[2]{{}^{#1}\! #2}      % with big backspace
\newcommand{\prn}[2]{{}^{#1} #2}        % without backspace

\newcommand{\Rp}{R^{\ms \prime}}

\newcommand{\Rbar}{\overline{R}}
\newcommand{\Rpbar}{\overline{R}{}^{\ms \prime}}

\newcommand{\Pro}[2]{P_{\rule{0pt}{1.6ex}#1}^{#2}}

\graphicspath{{graphs/}, {plots/}}

\begin{document}

\begin{flushright}
DESY 21-148 \\
UUITP-46/21 \\
\href{https://arxiv.org/abs/2109.14304}{arXiv:2109.14304 [hep-ph]}
\end{flushright}

\begin{center}
\vspace{4\baselineskip}
\textbf{\Large Double parton distributions out of bounds in colour space} \\
\vspace{3\baselineskip}
M.~Diehl$^{\ms 1}$, J.~R.~Gaunt$^{\ms 2}$, P.~Pichini$^{\ms 1, 3}$ and P.~Pl{\"o}{\ss}l$^{\ms 1}$
\end{center}

\vspace{2\baselineskip}

${}^{1}$ Deutsches Elektronen-Synchrotron DESY, Notkestr.~85, 22607 Hamburg, Germany \\
${}^{2}$ Department of Physics and Astronomy, University of Manchester, Manchester, M13 9PL, \\ \phantom{${}^{2}$} United Kingdom \\
${}^{3}$ Department of Physics and Astronomy, Uppsala University, Box 516,
75120 Uppsala, Sweden \\
\vspace{3\baselineskip}

\parbox{0.9\textwidth}{
}

We investigate the positivity of double parton distributions with a non-trivial dependence on the parton colour.  It turns out that positivity is not preserved by leading-order evolution from lower to higher scales, in contrast to the case in which parton colour is summed over.  We also study the positivity properties of the distributions at small distance between the two partons, where they can be computed in terms of perturbative splitting kernels and ordinary parton densities.

\vfill

\newpage

\tableofcontents

\begin{center}
\rule{0.6\textwidth}{0.3pt}
\end{center}

%%%%%%%%%%%%%%%%%%%%%%%%%%%%%%%%%%%%%%%%%%%%%%%%%%%%%%%%%%%%%%%%%%%%%%%%%%

\section{Introduction}

Parton distribution functions and related quantities are crucial ingredients for computing hadronic cross sections at high energies, and they are the main quantities that describe the structure of hadrons at the level of quarks and gluons.  It is hence important to know and understand their general properties.  One of these properties is positivity.  For ordinary parton distributions (PDFs) this is just the statement $f_a(x) \ge 0$ if the parton $a$ and the hadron are unpolarised.  In the polarised case, this generalises to a set of inequalities, namely the well-known Soffer bounds on polarised distributions \cite{Soffer:1994ww}.  Corresponding bounds have been formulated for transverse-momentum dependent distributions (TMDs) in \cite{Bacchetta:1999kz}, for impact parameter distributions in \cite{Diehl:2005jf}, and for double parton distributions (DPDs) in \cite{Diehl:2013mla}.  The latter appear in the description of double parton scattering and contain a wealth of information about correlations between two partons in a hadron; for a recent review we refer to the monograph \cite{Bartalini:2018qje}.  DPDs have a non-trivial dependence not only on the  polarisation of the partons, but also on their colour, and corresponding positivity bounds have been derived in \cite{Kasemets:2014yna}.

Positivity bounds can be of considerable practical value.  They may be used as constraints in fits of PDFs, and in the context of spin physics, ans\"atze that saturate certain bounds are often used to estimate the maximal allowed size of spin asymmetries.  A corresponding strategy for DPDs in spin or colour space appears all the more attractive because our current knowledge of these distributions is very incomplete.

Whether positivity bounds on parton distributions actually hold turns out to be a non-trivial question.  At leading order (LO) accuracy, the positivity of PDFs can quite directly be deduced from the positivity of cross sections, whilst the situation is more involved at next-to-leading order (NLO) in the strong coupling \cite{Altarelli:1998gn,Candido:2020yat}.  When derived along such lines, positivity holds for renormalisation scales $\mu$ that are high enough for the approximations of leading-twist dominance and of the perturbative expansion to be valid.  To formulate a corresponding approach for DPDs would be complicated due to the large number of involved degrees of freedom (two pairs of partons in each colliding hadron and two hard-scattering subprocesses), and we shall not pursue such an avenue here.

Intuitively, the positivity bounds on parton distributions are a consequence of their parton-model interpretation as number densities or linear combinations of number densities.  At a more formal level, one may use light-cone quantisation and write appropriate linear combinations of distributions as squared operator matrix elements that are summed over unobserved degrees of freedom. Equivalently, one can represent the distributions in terms of light-cone wave functions.  This has for instance been done for ordinary PDFs in \cite{Brodsky:1989pv} and for impact parameter distributions in \cite{Diehl:2002he}.  A corresponding representation holds for DPDs (see \cite{Gaunt:2012ths} for the momentum space version, which can readily be adapted to the case of definite transverse parton position).  A limitation of this approach is that it does not account for the renormalisation of ultraviolet divergences in the matrix elements (at least not in customary schemes such as $\overline{\text{MS}}$) nor for subtleties related with Wilson lines or the definition of light-cone gauge at infinite light-cone distances.  It has long been realised that ultraviolet subtractions can in principle invalidate the positivity of distributions.  A detailed discussion and examples can be found in the recent paper \cite{Collins:2021vke}.

An important result is that LO DGLAP evolution to higher scales conserves the positivity of PDFs, both in the unpolarised sector \cite{Durand:1986te,Collins:1988wj} and in the polarised one \cite{Barone:1997fh, Bourrely:1997bx}.  This means that if PDFs satisfy the positivity bounds at a certain scale $\mu$, these bounds remain valid at higher scales when the PDFs are evolved at leading order.  Discussions for NLO evolution can be found in \cite{Vogelsang:1997ak, Martin:1997rz}.  Conversely, experience shows that PDFs eventually turn negative when evolved down to very low scales.  Examples for this can for instance be found in \cite{Diehl:2019fsz}.  In \cite{Diehl:2013mla} it was shown that positivity of spin dependent but colour summed DPDs is conserved by LO DGLAP evolution to higher scales.  The first goal of the present paper is to investigate whether the same holds for the bounds derived in \cite{Kasemets:2014yna} for DPDs with non-trivial colour dependence.  In addition to DGLAP evolution, we will also consider Collins-Soper evolution in a rapidity variable, which appears when the parton colours are not summed over.  Following \cite{Kasemets:2014yna} we will limit ourselves to unpolarised partons throughout this work.

For small transverse distances $y$ between the two partons, DPDs can be computed in terms of ordinary parton densities and  kernels for the perturbative splitting of one parton into the two observed partons and (at higher orders) additional unobserved ones.  Using the results of the recent two-loop calculation in \cite{Diehl:2021wpp}, we can investigate to which extent positivity of DPDs in colour space is realised in the small $y$ limit.  This is the second goal of our work.

This paper is organised as follows.  In \sect{\ref{sec:bases}}, we specify the two parametrisations for the colour structure of DPDs used in this work, specify the property of positivity, and discuss the perturbative splitting mechanism for DPDs at LO accuracy.  In \sect{\ref{sec:cs-evol}}, we analyse whether Collins-Soper evolution from smaller to larger scales conserves positivity of DPDs, and in \sect{\ref{sec:dglap-evol}} we perform a corresponding analysis for LO DGLAP evolution.  The perturbative splitting mechanism at NLO accuracy is investigated in \sect{\ref{sec:splitting}}.  Our results are summarised in \sect{\ref{sec:sum}}, and some technical formulae are collected in an appendix.

\section{Colour structure of DPDs}
\label{sec:bases}

In this section, we discuss the general colour structure of quark and gluon DPDs and state the hypothesis of positivity in colour space.  Throughout this work, we consider distributions for unpolarised partons.

\begin{figure}[t]
\begin{center}
\subfigure[$F_{q \bar{q}}^{i i' \ms j j'}$]{\includegraphics[width=0.44\textwidth]{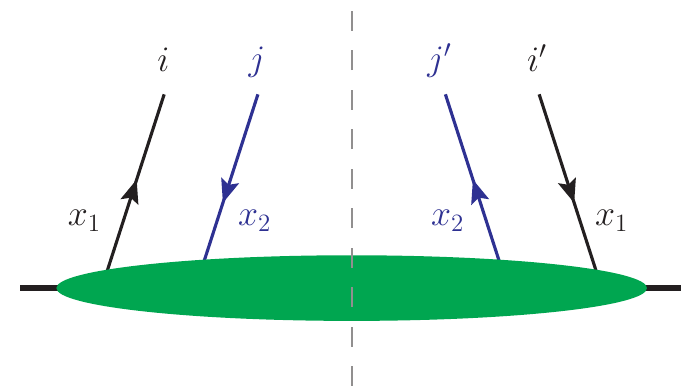}}
\hspace{2.6em}
\subfigure[$F_{q g}^{i i' \ms a a '}$]{\includegraphics[width=0.44\textwidth]{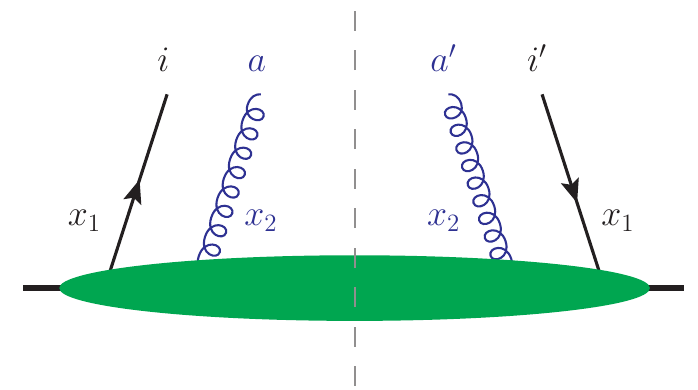}}
\end{center}
\caption{\label{fig:distrib} Assignment of colour labels for a quark-antiquark distribution (a) and a quark-gluon distribution (b).  The dashed vertical line indicates the final state cut of the scattering process in which the distributions appear.}
\end{figure}

As illustrated in \fig{\ref{fig:distrib}}, the colour structure of a DPD can be described in terms of four colour indices, one for each parton field in its definition.  We generically write $F_{a_1 a_2}^{r_1^{} r_1'\, r_2^{} r_2'}$, where $a_1$ and $a_2$ denote the two parton flavours.  The colour indices are in the fundamental or adjoint representation as appropriate, with $r_1$ and $r_2$ referring to the partons in the amplitude of the scattering process and $r_1'$ and $r_2'$ to the partons in the conjugate amplitude.  As described in \cite{Buffing:2017mqm}, the definition of a DPD involves the hadronic matrix element of two twist-two operators, as well as a soft factor given as the matrix element of Wilson line operators in the vacuum.  Both matrix elements contain ultraviolet divergences that need to be renormalised.  One can take different renormalisation scales $\mu_1$ and $\mu_2$ for the two partons, and the dependence of the DPD on these scales is given by DGLAP equations, which will be discussed in \sect{\ref{sec:dglap-evol}}.  The soft factor removes rapidity divergences in the hadronic matrix element, in a similar way as in the definition of transverse-momentum dependent distributions \cite{Collins:2011zzd}.  This leads to a dependence of the DPD on a rapidity scale $\zeta_p$, which is described by a Collins-Soper equation as discussed in \sect{\ref{sec:cs-evol}}.

%%%%%%%%%%%%%%%%%%%%%%%%%%%%%%%%%%%%%%%%

\paragraph{The \texorpdfstring{$s$}{s} and \texorpdfstring{$t$}{t} channel colour bases.}

The four colour indices of a DPD must be coupled to an overall colour singlet.  This can be achieved by first coupling the colour of two parton pairs to an irreducible representation and then coupling these two representations to an overall singlet.  Depending on the choice of parton pairs, we consider two bases for the colour coupling.

In the $s$ channel basis, we pair the partons in the amplitude and in the conjugate amplitude.  The projection on irreducible representations can then be written as
\begin{align}
\label{s-channel-F}
{F}^{R \Rp}_{a_1 a_2} &=
  \frac{1}{m(R)}\;
  \Pro{\Rbar\, \Rpbar}{r_1^{} r_2^{}\, r_1' r_2'}\,
   F_{a_1 a_2 \phantom{R}}^{r_1^{} r_1'\, r_2^{} r_2'}
\qquad \text{for } (a_1 a_2) \neq (q \bar{q}), (\bar{q} q)
\end{align}
with colour indices $r_1^{}, r_2^{}, r_1', r_2'$ in the fundamental or adjoint representation as appropriate.
The multiplicity $m(R)$ of the representation $R$ is $m(1) = 1$, $m(8) = m(A) = m(S) = 8$, $m(\overline{3}) = m(3) = 3$, etc.  The normalisation of ${F}^{R \Rp}_{a_1 a_2}$ in \eqref{s-channel-F} follows the choice made in \cite{Kasemets:2014yna} and corresponds to eq.~(6) in \cite{Mekhfi:1985dv}.
We denote the conjugate of a representation $R$ by $\Rbar$, where it is understood that some representations like the singlet and the octet are their own conjugate.

The matrix $P_{\ms\Rbar\, \Rpbar}$ in \eqref{s-channel-F} projects the colour of the parton pair $a_1 a_2$ on the representation $R$ in the amplitude and on the representation $\Rp$ in the conjugate amplitude.  Its explicit form is given in the appendix.
Quark-antiquark distributions $F_{q \bar{q}}^{R \Rp}$ are defined as in \eqref{s-channel-F}, but with transposed indices $r_2'\ms r_1'$ in $P_{\ms\Rbar\, \Rpbar}$, which ensures that covariant indices are always contracted with contravariant ones.  Likewise, the definition of $F_{\bar{q} q}^{R \Rp}$ has transposed indices $r_2^{}\ms r_1^{}$.

In the $t$ channel basis, we pair the partons with momentum fractions $x_1$ and $x_2$.  Following \cite{Diehl:2011yj, Buffing:2017mqm}, we write
\begin{align}
\label{t-channel-F}
\pr{R_1 R_2}{F}_{a_1 a_2} &=
  \mathcal{N}_{a_1} \mathcal{N}_{a_2}\,
  \varepsilon(R_1)\, \varepsilon(R_2)\,
  \frac{1}{\sqrt{m(R_1)}}\;
  \Pro{\Rbar_1 \Rbar_2}{r_1^{} r_1'\, r_2^{} r_2'}\,
   F_{a_1 a_2 \phantom{R}}^{r_1^{} r_1'\, r_2^{} r_2'}
\qquad \text{for } a_1, a_2 \neq \bar{q}
\end{align}
with prefactors $\mathcal{N}_{q} = \sqrt{3}$, $\mathcal{N}_{g} = \sqrt{8}$, $\varepsilon(A) = i$ and $\varepsilon(R) = 1$ for $R \neq A$.  The matrix  $P_{\ms\Rbar_1 \Rbar_2}$  in \eqref{t-channel-F} projects the colour indices $r_i^{} r_i'$ of parton $a_i$ on the representation $R_i$ for $i=1,2$.
For distributions with antiquarks, one needs to transpose the corresponding colour indices in $P_{\ms\Rbar_1 \Rbar_2}$.  The correct ordering is hence $r_1'\ms r_1^{}$ if $a_1^{} = \bar{q}$ and $r_2'\ms r_2^{}$ if $a_2^{} = \bar{q}$.

Combining the definition \eqref{s-channel-F} with the completeness relation \eqref{complete-rel} and the explicit form of the singlet projector $P_{11}$, one readily finds that
\begin{align}
   \label{singlet-sum}
\pr{11}{F}_{a_1 a_2} &= \sum_{R} m(R) \, F^{R \Rbar}_{a_1 a_2}
\end{align}
for all parton combinations, where the sum runs over all relevant colour representations $R$.

Throughout this work, we fix $N=3$ for the number of colours.  The colour factors used in later results thus have the values $C_F = 4/3$, $C_A = 3$, and $T_F = 1/2$.
The accessible colour representations for different parton combinations are given in \tab{\ref{tab:reps}}.  We note that a quark and a gluon can couple to $\overline{6}$ (see e.g.\ \tab{24} in \cite{Slansky:1981yr}) rather than to $6$ as stated in \eqn{(8c)} of \cite{Mekhfi:1985dv}.  As shown in \cite{Buffing:2017mqm}, the $t$ channel distributions $\pr{R_1 R_2}{F}$ are real valued, except for the decuplet sector in the pure gluon case, in which case one has $(\pr{10\ms \overline{10}}{F}_{g g} )^* = \pr{\overline{10}\ms 10}{F}_{g g}$.  In the $s$ channel basis, this translates into all distributions $F^{R \Rp}$ being real, except for the mixed octet combinations, where one finds $( F_{g g}^{A S} )^* =  F_{g g}^{S A}$.

\begin{table}
\begin{center}
$$
\renewcommand{\arraystretch}{1.15}
\begin{array}{c c c}
                & s~\text{channel} & t~\text{channel} \\ \hline
a_1 a_2         & R \Rp & R_1 R_2 \\ \hline
q \bar{q}       & 11, 88  & 11, 88 \\
q q             & \overline{3}\ms 3, 6\ms \overline{6} & 11, 88  \\
\bar{q} \bar{q} & 3\ms \overline{3}, \overline{6}\ms 6 & 11, 88 \\
g q             & 3\ms \overline{3}, \overline{6}\ms 6,
                  15\ms \overline{15} & 11, S8, A8 \\
g \bar{q}       & \overline{3}\ms 3, 6\ms \overline{6},
                  \overline{15}\ms 15 & 11, S8, A8 \\
g g             & \multicolumn{2}{c}{11, SS, AA, S\bs A, AS,
                  10\ms \overline{10}, \overline{10}\ms 10, 27\ms 27} \\
\hline
\end{array}
$$
\caption{\label{tab:reps} Combinations of colour representations in the $s$ channel distributions $F^{R \Rp}_{a_1 a_2}$ and the $t$ channel distributions $\pr{R_1 R_2}{F}_{a_1 a_2}$.  Two adjoint indices can couple to a symmetric ($S$) or an antisymmetric ($A$) octet.  For two-gluon distributions, the colour combinations are identical in the $s$ and $t$ channels.  Interchanging $a_1 \leftrightarrow a_2$ implies interchanging $R_1 \leftrightarrow R_2$ while keeping $R \Rp$ unchanged.}
\end{center}
\end{table}

%%%%%%%%%

\paragraph{Density interpretation and positivity.}
The parton model interpretation of DPDs can be obtained in the same way as for single parton distributions by expressing the field operators in terms of creation and annihilation operators in light-cone quantisation, neglecting all complications from Wilson lines and from renormalisation.  Details can for instance be found in \cite{Collins:2011zzd}.

The $t$ channel distributions are normalised such that $\pr{11}{F}_{a_1 a_2}(x_1,x_2,y)$ is the probability density for finding partons $a_1$ and $a_2$ with momentum fractions $x_1$ and $x_2$ at a transverse distance $y$ from each other, with the density measure being $dx_1\, dx_2\, d^2 y$.  The colours and polarisations of both partons are summed over in $\pr{11}{F}_{a_1 a_2}$.  Correspondingly, the $s$ channel distribution $F_{a_1 a_2}^{R \Rbar}$ is the probability density for finding the parton pair in one of the $m(R)$ states of the colour representation $R$.  This provides an intuitive interpretation of the relation \eqref{singlet-sum}.

The positivity property for DPDs in full colour space is then the statement that
\begin{align}
F^{R \Rbar}_{a_1 a_2} &\ge 0
& \text{for all } R ,
\end{align}
which of course implies the weaker condition $\pr{11}{F}_{a_1 a_2} \ge 0$.  Note that we define ``positivity'' as including the value zero.

Note that in the pure gluon channel, the distributions in the $s$ channel basis include the cases $F^{A S}_{g g}$ and $F^{S\bs A}_{g g}$, which correspond not to densities but to interference terms in colour space (and which may be complex valued as noted above).  Given the large number of accessible colour channels in that case, we will not consider two-gluon DPDs in the remainder of this work, concentrating on the pure quark-antiquark sector and on mixed quark-gluon or antiquark-gluon distributions.

%%%%%%%%%

\paragraph{Basis transformations.}
Whilst the $s$ channel basis is natural for considering positivity, the evolution of DPDs in the renormalisation and rapidity scales is much simpler in the $t$ channel basis.  We hence need the explicit transformations between the two representations.  In the pure quark sector, the transformations between $s$ and $t$ channel bases read
\begin{align}
   \label{quark-trfs}
\begin{pmatrix}
   F_{q q}^{\overline{3}\ms 3} \\[0.3em]
   F_{q q}^{6\ms \overline{6}}
\end{pmatrix}
&= \mathbf{M}_{q q} \,
   \begin{pmatrix}
      \pr{11}{F}_{q q}^{} \\[0.4em]
      \pr{88}{F}_{q q}^{}
   \end{pmatrix} \,,
&
\begin{pmatrix}
   F_{q \bar{q}}^{11} \\[0.3em]
   F_{q \bar{q}}^{88}
\end{pmatrix}
&= \mathbf{M}_{q \bar{q}} \,
   \begin{pmatrix}
      \pr{11}{F}_{q \bar{q}}^{} \\[0.4em]
      \pr{88}{F}_{q \bar{q}}^{}
   \end{pmatrix} \,,
\end{align}
with
\begin{align}
\mathbf{M}_{q q} &= \frac{1}{9} \,
   \begin{pmatrix}
      1 & - \sqrt{2}\, \\[0.2em]
      1 & \frac{1}{\sqrt{2}}
   \end{pmatrix} \,,
&
\mathbf{M}_{q \bar{q}} &= \frac{1}{9} \,
   \begin{pmatrix}
      1 & \sqrt{8} \\[0.2em]
      1 & - \frac{1}{\sqrt{8}}
   \end{pmatrix} \,,
&
\nonumber \\[0.3em]
\mathbf{M}_{q q}^{-1} &= 3 \,
   \begin{pmatrix} 1 & 2 \\[0.2em] - \sqrt{2} & \sqrt{2} \end{pmatrix} \,,
&
\mathbf{M}_{q \bar{q}}^{-1} &=
   \begin{pmatrix} 1 & 8 \\[0.2em] \sqrt{8} & -\sqrt{8} \end{pmatrix} \,.
\end{align}
The transformation matrix for two antiquarks is $\mathbf{M}_{\bar{q} \bar{q}} = \mathbf{M}_{q q}$.
For $g q$ and $g \bar{q}$ distributions, one has
\begin{align}
   \label{quark-gluon-trfs}
\begin{pmatrix} F_{g q}^{3\ms \overline{3}} \\[0.3em]
   F_{g q}^{\overline{6}\ms 6} \\[0.3em]
   F_{g q}^{15\ms \overline{15}}
\end{pmatrix}
&= \mathbf{M}_{g q} \,
   \begin{pmatrix}
      \pr{11}{F}_{g q}^{} \\[0.4em]
      \pr{S8}{F}_{g q}^{} \\[0.4em]
      \pr{A8}{F}_{g q}^{}
   \end{pmatrix} \,,
&
\begin{pmatrix}
   F_{g \bar{q}}^{\overline{3}\ms 3} \\[0.3em]
   F_{g \bar{q}}^{6\ms \overline{6}} \\[0.3em]
   F_{g \bar{q}}^{\overline{15}\ms 15}
\end{pmatrix}
&= \mathbf{M}_{g \bar{q}} \,
   \begin{pmatrix}
      \pr{11}{F}_{\bar{q} g}^{} \\[0.4em]
      \pr{S8}{F}_{g \bar{q}}^{} \\[0.4em]
      \pr{A8}{F}_{g \bar{q}}^{}
   \end{pmatrix}
\end{align}
with
\begin{align}
\mathbf{M}_{g q} &= \frac{1}{24} \,
   \begin{pmatrix}
      1 & \sqrt{\frac{5}{2}} & - \frac{3}{\sqrt{2}} \\[0.5em]
      1 & - \sqrt{\frac{5}{2}} & - \frac{1}{\sqrt{2}} \\[0.5em]
      1 & \sqrt{\frac{1}{10}} & \frac{1}{\sqrt{2}}
   \end{pmatrix} \,,
&
\mathbf{M}_{g \bar{q}} &= \frac{1}{24} \,
   \begin{pmatrix}
      1 & \sqrt{\frac{5}{2}} & \frac{3}{\sqrt{2}} \\[0.5em]
      1 & - \sqrt{\frac{5}{2}} & \frac{1}{\sqrt{2}} \\[0.5em]
      1 & \sqrt{\frac{1}{10}} & - \frac{1}{\sqrt{2}}
   \end{pmatrix} \,,
\nonumber \\[0.4em]
\mathbf{M}_{g q}^{-1} &= 3 \,
   \begin{pmatrix} 1 & 2 & 5 \\[0.4em]
      \sqrt{\frac{5}{2}} & - \sqrt{10} & \sqrt{\frac{5}{2}} \\[0.4em]
      - \frac{3}{\sqrt{2}} & - \sqrt{2} & \frac{5}{\sqrt{2}}
   \end{pmatrix} \,,
&
\mathbf{M}_{g \bar{q}}^{-1} &= 3 \,
   \begin{pmatrix}
      1 & 2 & 5 \\[0.4em]
      \sqrt{\frac{5}{2}} & - \sqrt{10} & \sqrt{\frac{5}{2}} \\[0.4em]
      \frac{3}{\sqrt{2}} & \sqrt{2} & - \frac{5}{\sqrt{2}}
   \end{pmatrix} \,.
\end{align}
All transformation matrices fulfil the symmetry property $\mathbf{M}_{a_1 a_2} = \mathbf{M}_{a_2 \ms a_1}$.  The expressions of $\mathbf{M}_{q q}$, $\mathbf{M}_{q \bar{q}}$, and $\mathbf{M}_{g q}$ were already given in \cite{Kasemets:2014yna}.

%%%%%%%%%

\paragraph{Perturbative splitting at leading order.}

\begin{figure}
   \begin{center}
   \subfigure[\label{fig:qqbar-LO}]{\includegraphics[width=0.24\textwidth]{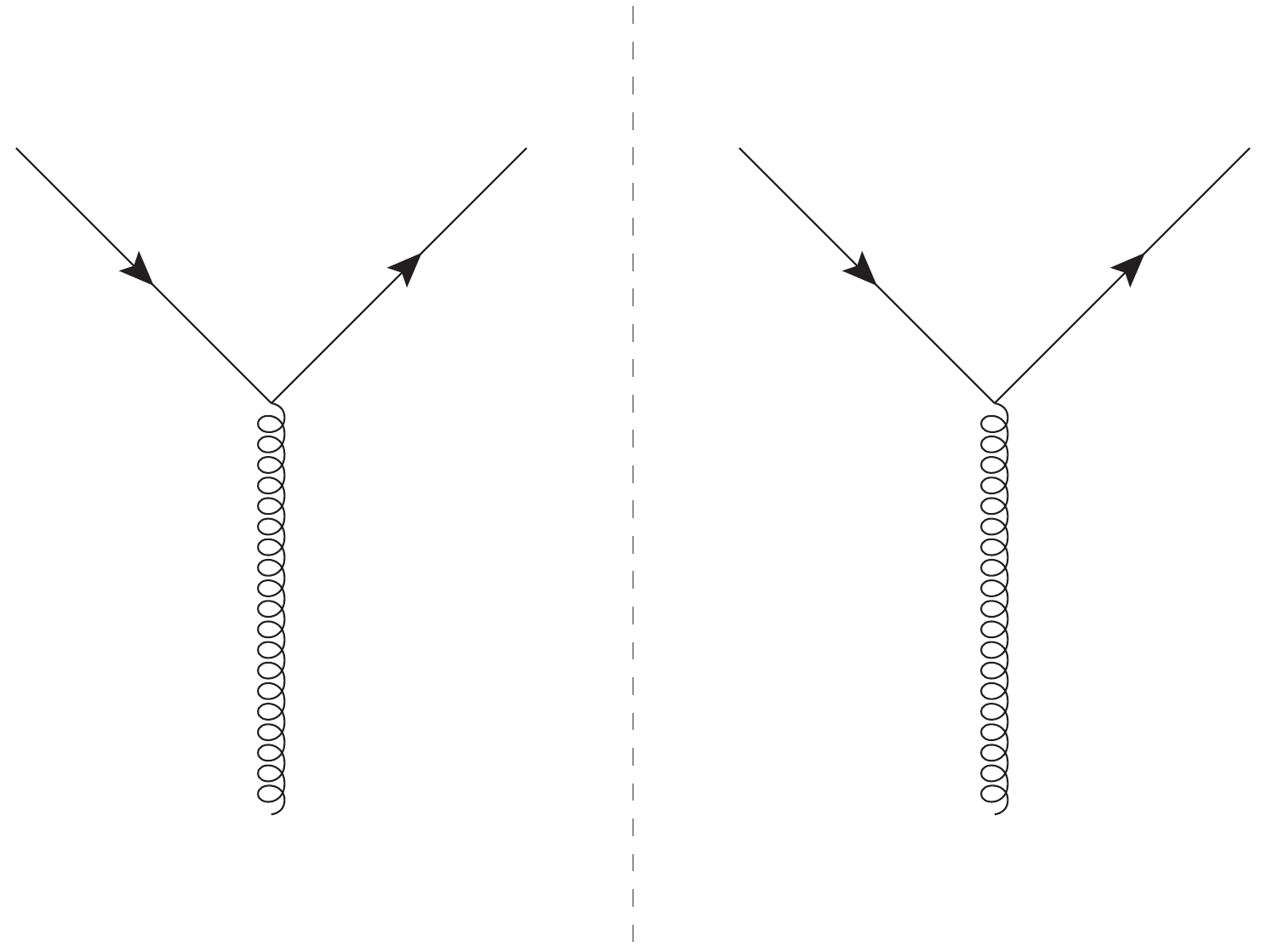}}
   \hspace{3em}
   \subfigure[\label{fig:qqbar-NLO-re}]{\includegraphics[width=0.24\textwidth]{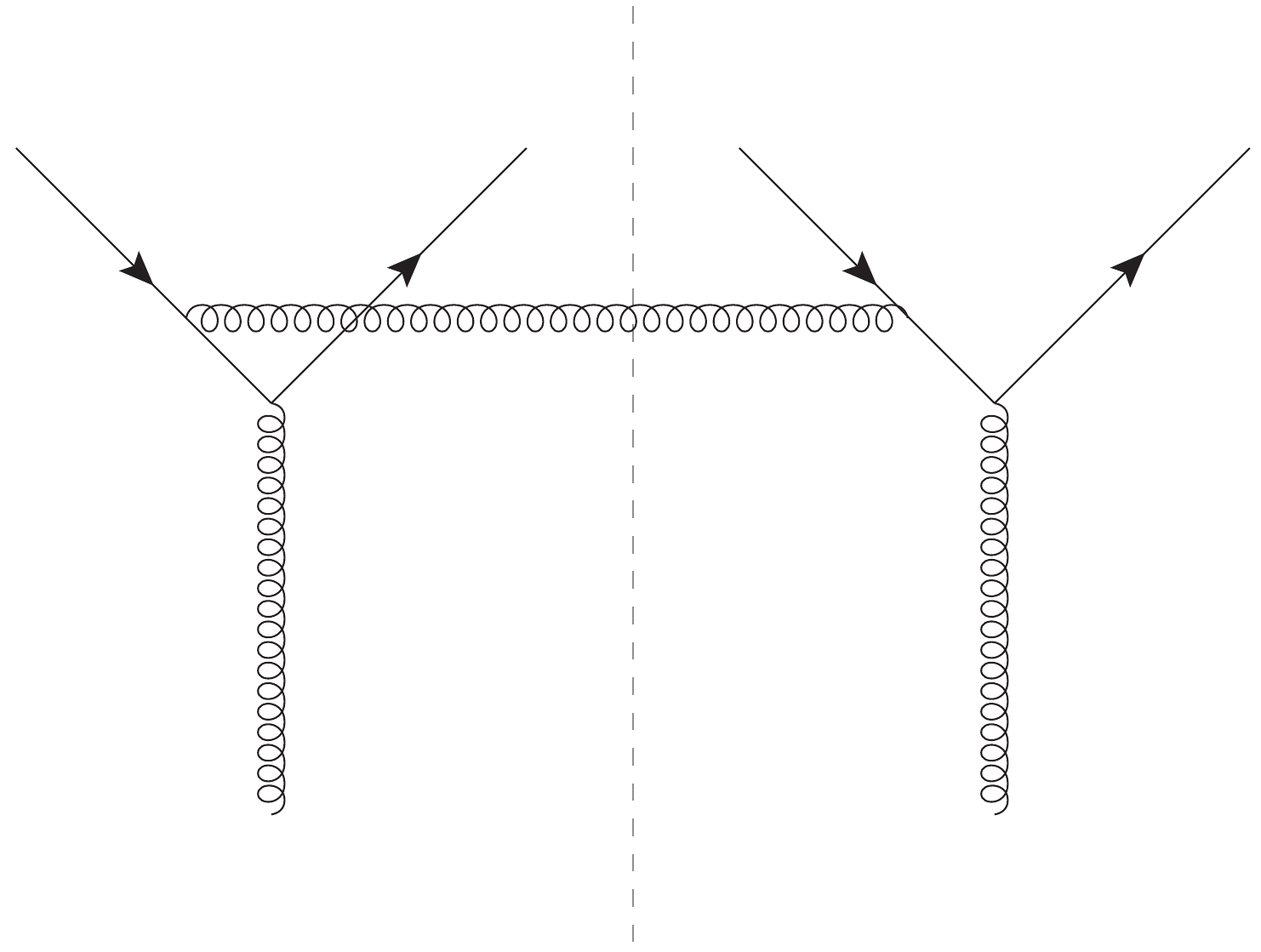}}
   \hspace{3em}
   \subfigure[\label{fig:qqbar-NLO-vi}]{\includegraphics[width=0.24\textwidth]{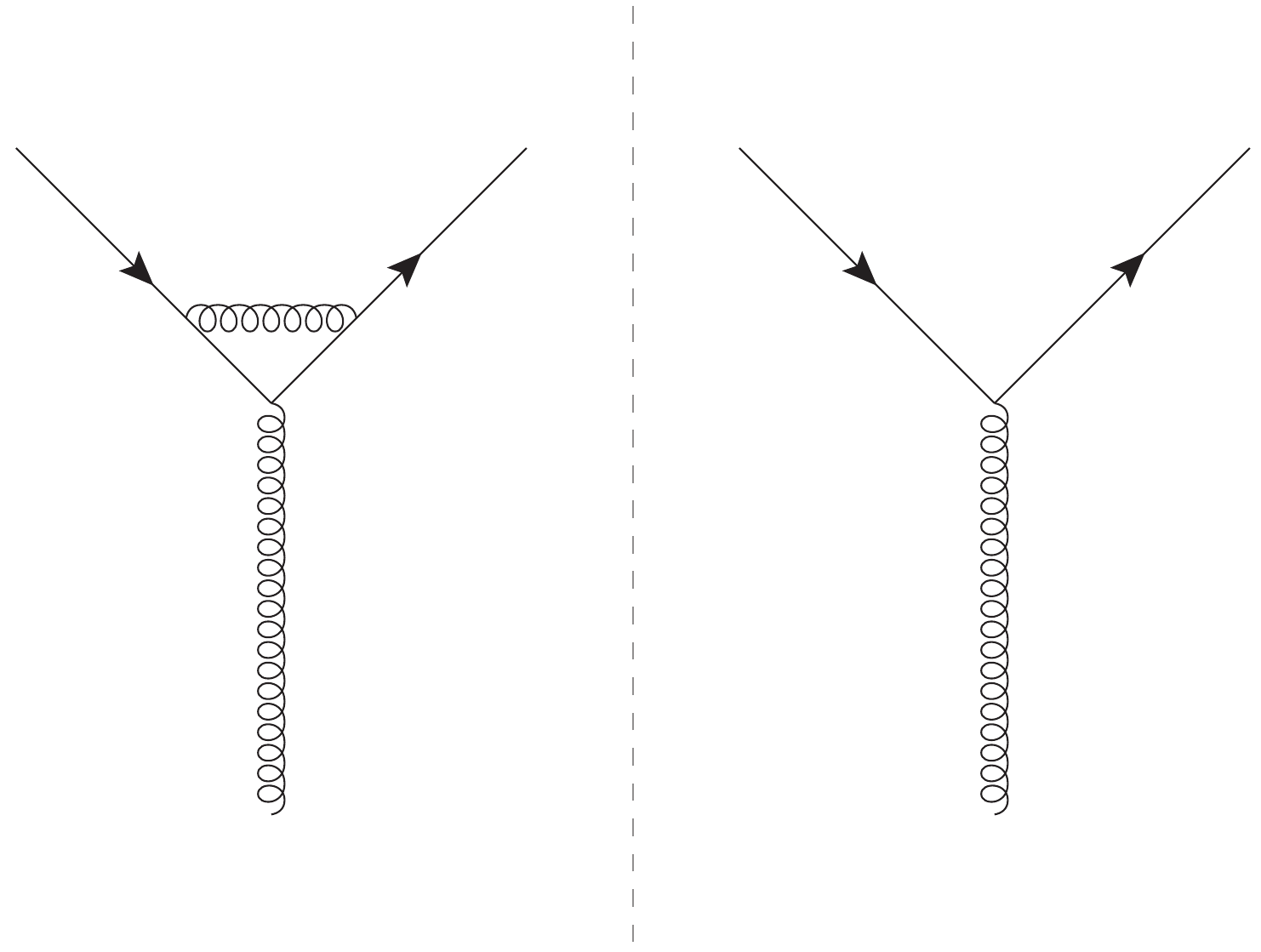}}
   \caption{\label{fig:split-g}  Graph for the perturbative splitting $g \to q \bar{q}$ in a DPD.  Panel (a) shows the LO graph, and panels (b) and (c) are examples of real and virtual NLO graphs.  The dashed vertical line indicates the final state cut of the scattering process in which the distributions appear.}
   \end{center}
\end{figure}

If the transverse distance $y$ between the two partons is small, DPDs can be computed in terms of a perturbative splitting process and ordinary PDFs.  Example graphs for the perturbative splitting are shown in \fig{\ref{fig:split-g}}.  This mechanism is interesting in our context because it generates a non-trivial colour dependence.

Let us take a closer look at the splitting process at one-loop order, postponing the discussion of two-loop accuracy to \sect{\ref{sec:splitting}}.  The splitting formula at leading order reads \cite{Diehl:2011yj, Buffing:2017mqm}
\begin{align}
   \label{split-LO}
\pr{R_1 R_2}{F}_{a_1 a_2}(x_1, x_2, y, \mu, \mu)
&= \frac{a_s(\mu)}{\pi\ms y^2}\;
   \pr{R_1 R_2}{V}_{a_1 a_2, a_0}^{(1)}(u) \; \frac{f_{a_0}(x, \mu)}{x} \,,
\end{align}
where $f_{a_0}(x, \mu)$ is the PDF for parton $a_0$ and we defined
\begin{align}
a_s(\mu) &= \frac{\alpha_s(\mu)}{2\pi}
\end{align}
and
\begin{align}
   \label{xu-defs}
x &= x_1 + x_2 \,,
&
u &= \frac{x_1}{x_1 + x_2} \,.
\end{align}
Note that \eqref{split-LO} is an approximation for small $y$ and receives  corrections suppressed by a power of $a_s$ or $y^2 \Lambda^2$, where $\Lambda$ is a hadronic scale.

The splitting kernels $\pr{11}{V}_{a_1 a_2, a_0}^{(1)}(z)$ for the colour singlet channel are equal to the usual LO DGLAP splitting functions without the distributional parts (plus prescription and delta function) at $z=1$.  One therefore has $\pr{11}{V}_{a_1 a_2, a_0}^{(1)}(z) > 0$.  This implies that $\pr{11}{F}_{a_1 a_2} > 0$ at the scale $\mu$ where the LO splitting formula \eqref{split-LO} is evaluated, provided of course that the PDFs are positive.  To avoid large $a_s^2$ corrections, one should take $\mu \sim 1/y$.

The LO splitting kernels for other colour channels are proportional to $\pr{11}{V}_{a_1 a_2, a_0}^{(1)}(z)$, and it is easy to transform them to the $s$ channel colour basis.  The result is
\begin{align}
   \label{split-LO-colour}
F_{a_1 a_2}^{R \Rbar} &=
\begin{cases}
\pr{11}{F}_{a_1 a_2} \big/ m(R_0) & \text{for } R = R_0 \,, \\
0                                 & \text{for } R \neq R_0 \,,
\end{cases}
\end{align}
where $R_0$ is the colour representation of the initial parton $a_0$ of the splitting process (with $R_0 = A$ for $g\to g g$).  The parton pair $a_1 a_2$ must be in the representation $R_0$ because the LO splitting graphs are disconnected between the amplitude and conjugate amplitude (see \fig{\ref{fig:qqbar-LO}}).  The result \eqref{split-LO-colour} then follows from the relation \eqref{singlet-sum}.  With $\pr{11}{F}_{a_1 a_2} > 0$ one hence finds positivity of DPDs in colour space when taking the LO splitting approximation.  We will see in \sect{\ref{sec:splitting}} whether this still holds at NLO.

\section{Collins-Soper evolution}
\label{sec:cs-evol}

In this section, we investigate how Collins-Soper evolution affects the positivity of DPDs.

Collins-Soper evolution of DPDs does not mix different colour representations in the $t$ channel basis, where one has
\begin{align}
\frac{\partial}{\partial \log\zeta_p}\,
   \prn{R_1 R_2}{F}_{a_1 a_2}(x_1,x_2,y,\mu_1,\mu_2,\zeta_p)
&= \frac{\pr{R_1}{J}(y,\mu_1,\mu_2)}{2} \;
   \prn{R_1 R_2}{F}_{a_1 a_2}(x_1,x_2,y,\mu_1,\mu_2,\zeta_p) \,.
\end{align}
Here we displayed all arguments of the functions.  The Collins-Soper kernel $\pr{R_1}{J}$ depends only on the multiplicity of $R_1$ (which is equal to the multiplicity of $R_2$) but not on the parton types.  Note that colour singlet distributions in the $t$ channel are $\zeta_p$ independent, i.e. $\pr{1}{J} = 0$.   For all parton combinations except $g g$, the only non-trivial kernel needed is hence the one for the colour octet.  Remarkably, this kernel satisfies the exact relation \cite{Vladimirov:2016qkd}
\begin{align}
\pr{8}{J}(y,\mu,\mu) &= K_g(y,\mu) \,,
\end{align}
where $K_g(y,\mu)$ is the Collins-Soper kernel for the evolution of single-gluon TMDs.

Let us discuss the sign of the Collins-Soper kernel, which will be important in the following.  The renormalisation group equation for the Collins-Soper kernel is solved by
\begin{align}
\pr{8}{J}(y,\mu_1,\mu_2)
&= \pr{8}{J}(y,\mu_0,\mu_0)
   - \int_{\mu_0}^{\mu_1} \frac{d\mu}{\mu}\; \pr{8}{\gamma}_J(\mu)
   - \int_{\mu_0}^{\mu_2} \frac{d\mu}{\mu}\; \pr{8}{\gamma}_J(\mu)
\end{align}
with a positive anomalous dimension that is proportional to the cusp anomalous dimension for adjoint Wilson lines:
\begin{align}
   \label{gamma-J8}
\pr{8}{\gamma}_J(\mu)
&= a_s(\mu)\; \pr{8}{\gamma}_J^{(0)} + \mathcal{O}(a_s^2) \,,
&
\pr{8}{\gamma}_J^{(0)} &= 2 C_A \,.
\end{align}
At given $y$ one can hence always achieve a negative $\pr{8}{J}$ by taking the scales $\mu_1$ and $\mu_2$ sufficiently high.

To make a more specific statement, we first consider small distances $y$, where one can compute the kernel in perturbation theory and obtains
\begin{align}
   \label{J8}
\pr{8}{J}(y,\mu,\mu) &= - a_s(\mu)\;
   \pr{8}{\gamma}_J^{(0)}\, \log \frac{\mu^2 \ms y^2}{b_0^2}
   + \mathcal{O}(a_s^2) \,,
\end{align}
where $b_0 = 2 e^{-\gamma} \approx 1.12$ and $\gamma$ is the Euler-Mascheroni constant.  Bearing in mind that there are higher-order terms in \eqref{J8}, we see that the transition from positive to negative $\pr{8}{J}$ happens at $\mu$ around $b_0 /y$, as long as $y$ remains in the perturbative regime.

Not much is known about $\pr{8}{J}$ or $K_g$ for $y$ in the nonperturbative domain.  The situation is different for the Collins-Soper kernel $K_q$ for quark TMDs.  Several phenomenological extractions find that $K_q(y,\mu)$ is negative for large $y$, see e.g.\ \cite{Collins:2014jpa} (\fig{6}), \cite{Bacchetta:2019sam}, and \cite{Scimemi:2019cmh} (\fig{23}).  Furthermore a number of lattice determinations, covering a distance range between about $0.1 \fm$ and $0.8 \fm$, find that $K_q(y,\mu) < 0$ at $\mu=2 \gev$, see \fig{7} in \cite{Shanahan:2020zxr}, \fig{5} in \cite{Zhang:2020dbb}, and \fig{8} in \cite{Schlemmer:2021aij}.

We find it plausible to assume a qualitatively similar behaviour of $K_g(y,\mu)$ and $K_q(y,\mu)$ as functions of $y$ (at perturbatively small $y$, one actually has $K_g / K_q \approx C_A / C_F$ up to corrections of order $\alpha_s^4$, see footnote~10 in \cite{Buffing:2017mqm}).  Under this assumption, we conclude that $\pr{8}{J}(y,\mu_1,\mu_2) < 0$ for scales $\mu_1$ and $\mu_2$ sufficiently larger than $\max( b_0/y, 2 \gev )$.

%%%%%%%%%%%%%%%%%%%%%%%%%%%%%%%%

\paragraph{Collins-Soper evolution in the $s$ channel.}
In the $s$ channel, different colour representations mix under Collins-Soper evolution.  Starting with the $q q$ channel and using the basis transform given in the previous section, we get
\begin{align}
   \label{CS-qq}
\frac{\partial}{\partial \log\zeta_p}\,
\begin{pmatrix}
   F_{q q}^{\overline{3}\ms 3} \\[0.2em]
   F_{q q}^{6\ms \overline{6}}
\end{pmatrix}
&= \mathbf{M}_{q q} \; \frac{\partial}{\partial \log\zeta_p}\,
   \begin{pmatrix}
      \pr{11}{F}_{q q} \\[0.2em]
      \pr{88}{F}_{q q}
   \end{pmatrix}
 = \mathbf{M}_{q q} \; \frac{1}{2}
   \begin{pmatrix} 0 & 0 \\[0.2em] 0 & \pr{8}{J} \end{pmatrix} \,
   \begin{pmatrix}
      \pr{11}{F}_{q q} \\[0.2em]
      \pr{88}{F}_{q q}
   \end{pmatrix}
\nonumber \\
&= \frac{\pr{8}{J}}{2} \; \hat{\mathbf{J}}_{q q} \,
   \begin{pmatrix}
      F_{q q}^{\overline{3}\ms 3} \\[0.2em]
      F_{q q}^{6\ms \overline{6}}
   \end{pmatrix}
\end{align}
with
\begin{align}
   \label{J-qq}
\hat{\mathbf{J}}_{q q}
 = \mathbf{M}_{q q} \,
   \begin{pmatrix} 0 & 0 \\[0.1em] 0 & 1 \end{pmatrix} \,
   \mathbf{M}_{q q}^{-1}
&= \frac{1}{3} \,
   \begin{pmatrix} 2 & -2 \\[0.1em] -1 & 1 \end{pmatrix} \,.
\end{align}
Restoring all arguments, we find that the evolution equation is solved by
\begin{align}
   \label{CS-qq-solved}
\begin{pmatrix}
   F_{q q}^{\overline{3}\ms 3} \\[0.2em]
   F_{q q}^{6\ms \overline{6}}
\end{pmatrix}(x_1,x_2,y,\mu_1,\mu_2,\zeta_p)
&= \mathbf{U}_{q q}(\alpha) \,
   \begin{pmatrix}
      F_{q q}^{\overline{3}\ms 3} \\[0.2em]
      F_{q q}^{6\ms \overline{6}}
   \end{pmatrix} (x_1,x_2,y,\mu_1,\mu_2,\zeta_0)
\end{align}
with the matrix exponential
\begin{align}
\mathbf{U}_{q q}(\alpha)
 = \exp\bigl( - \alpha \, \hat{\mathbf{J}}_{q q} \bigr)
&= \frac{1}{3} \,
   \begin{pmatrix}
      1 + 2 e^{-\alpha} & 2 (1 - e^{-\alpha}) \\[0.1em]
      1 - e^{-\alpha} & 2 + e^{-\alpha}
   \end{pmatrix} \,,
\end{align}
where we abbreviate
\begin{align}
   \label{alpha-def}
\alpha &= - \frac{\pr{8}{J}(y,\mu_1,\mu_2)}{2} \, \log\frac{\zeta_p}{\zeta_0} \,.
\end{align}
The Collins-Soper equation for the other parton combinations we consider has the same form as \eqref{CS-qq} with appropriate changes in the colour labels and matrices.  In the quark-antiquark case, one has
\begin{align}
   \label{J-qqbar}
\hat{\mathbf{J}}_{q \bar{q}}
 = \mathbf{M}_{q \bar{q}} \,
   \begin{pmatrix} 0 & 0 \\[0.1em] 0 & 1 \end{pmatrix} \,
   \mathbf{M}_{q \bar{q}}^{-1}
&= \frac{1}{9} \,
   \begin{pmatrix} 8 & -8 \\[0.1em]
      -1 & 1
   \end{pmatrix} \,,
\nonumber \\[0.3em]
\mathbf{U}_{q \bar{q}}(\alpha)
 = \exp\bigl( - \alpha \, \hat{\mathbf{J}}_{q \bar{q}} \bigr)
&= \frac{1}{9} \,
   \begin{pmatrix}
      1 + 8 e^{-\alpha} & 8 (1 - e^{-\alpha}) \\[0.1em]
      1 - e^{-\alpha} & 8 + e^{-\alpha}
   \end{pmatrix} \,,
\end{align}
and for $g q$ distributions we find
\begin{align}
\hat{\mathbf{J}}_{g q}
 = \mathbf{M}_{g q} \,
   \begin{pmatrix}
      0 & 0 & 0 \\[0.1em] 0 & 1 & 0 \\[0.1em] 0 & 0 & 1
   \end{pmatrix} \,
   \mathbf{M}_{g q}^{-1}
&= \frac{1}{8} \,
   \begin{pmatrix}
      7 & -2 & -5 \\[0.2em]
      -1 & 6 & -5 \\[0.2em]
      -1 & -2 & 3
   \end{pmatrix} \,,
\nonumber \\[0.4em]
\mathbf{U}_{g q}(\alpha)
 = \exp\bigl( - \alpha \, \hat{\mathbf{J}}_{g q} \bigr)
&= \frac{1}{8} \,
   \begin{pmatrix}
      1 + 7 e^{-\alpha} & 2 (1 - e^{-\alpha}) & 5 (1 - e^{-\alpha}) \\[0.1em]
      1 - e^{-\alpha} & 2 + 6 e^{-\alpha} & 5 (1 - e^{-\alpha}) \\[0.1em]
      1 - e^{-\alpha} & 2 (1 - e^{-\alpha}) & 5 + 3 e^{-\alpha} \,.
   \end{pmatrix} \,,
\end{align}
where $\alpha$ is always given by \eqref{alpha-def}.
We furthermore get $\hat{\mathbf{J}}_{\bar{q} \bar{q}} = \hat{\mathbf{J}}_{q q}$ and $\hat{\mathbf{J}}_{g \bar{q}} = \hat{\mathbf{J}}_{g q}$ and corresponding equalities for the evolution matrices $\mathbf{U}_{a_1 a_2}$.  The evolution equations for distributions $F_{q_1 q_2}$, $F_{q_1 \bar{q}_2}$, etc.\ with unequal flavours involve the same matrices as their counterparts for equal flavours.

We see that for all parton combinations $a_1 a_2$ except $g g$ (which we do not consider) all elements of the evolution matrix $U_{a_1 a_2}(\alpha)$ are positive for $\alpha > 0$.  This is the case for forward evolution ($\zeta_p > \zeta_0$), provided that $\pr{8}{J} < 0$, which is the case when $\mu_1$ and $\mu_2$ are sufficiently large.  Under this condition, Collins-Soper evolution to higher scales thus preserves positivity.

With the notation
\begin{align}
\bigl( \hat{\mathbf{J}}_{a_1 a_2}\ms F_{a_1 a_2} \bigr){}^{R \Rbar}
&= \sum_{\Rp}
   \bigl( \hat{\mathbf{J}}_{a_1 a_2} \bigr){}^{R \Rbar,\, \Rp \bs \Rpbar} \,
   F_{a_1 a_2}^{\Rp \bs \Rpbar}
\end{align}
the Collins-Soper equation in the $s$ channel basis reads
\begin{align}
\frac{\partial}{\partial \log\zeta_p}\ms F_{a_1 a_2}^{R \Rbar}
&= \frac{\pr{8}{J}}{2} \,
   \bigl( \hat{\mathbf{J}}_{a_1 a_2}\ms F_{a_1 a_2}
   \bigr){}^{R \Rbar}_{\phantom{a_1}}
\end{align}
for all parton combinations considered here.  Using that $( \hat{\mathbf{J}}_{a_1 a_2} )^2 = \hat{\mathbf{J}}_{a_1 a_2}$, we can write its solution in the form
\begin{align}
   \label{CS-solution}
F_{a_1 a_2}^{R \Rbar}(\zeta_p)
&= F_{a_1 a_2}^{R \Rbar}(\zeta_0) + ( e^{-\alpha} - 1 ) \,
   \bigl( \hat{\mathbf{J}}_{a_1 a_2}\ms F_{a_1 a_2}(\zeta_0)
   \bigr){}^{R \Rbar} \,.
\end{align}
Writing the relation \eqref{singlet-sum} at rapidity scale $\zeta_p$ and using that $\pr{11}{F}_{a_1 a_2}$ is independent of $\zeta_p$, we get
\begin{align}
   \label{CS-sum}
\sum_{R} m(R)\, F_{a_1 a_2}^{R \Rbar}(\zeta_p)
&= \pr{11}{F}_{a_1 a_2}(\zeta_0) \,.
\end{align}
We can now discuss the behaviour of Collins-Soper evolution for large negative $\alpha$, which is relevant when evolving backward with $\pr{8}{J} < 0$, and when evolving forward at scales $\mu_1$ and $\mu_2$ that are so low that $\pr{8}{J} > 0$.  In the regime where the factor $e^{-\alpha}$ in \eqref{CS-solution} is much larger than $1$, the condition \eqref{CS-sum} implies that the evolved distributions $F_{a_1 a_2}^{R \Rbar}(\zeta_p)$ must be positive in some colour channels and negative in others.  An exception to this statement is the case where the initial conditions satisfy $(\hat{\mathbf{J}}_{a_1 a_2}\ms F_{a_1 a_2}(\zeta_0))^{R \Rbar} = 0$ for all $R$, so that all distributions are independent of~$\zeta_p$.  In the $t$ channel basis, this is tantamount to all distributions other than $\pr{11}{F}_{a_1 a_2}$ being zero.  Apart from this special case, evolution to large negative $\alpha$ always leads to a violation of positivity.

\section{DGLAP evolution}
\label{sec:dglap-evol}

In this section, we investigate how leading-order DGLAP evolution affects the positivity of DPDs.  We limit ourselves to the evolution of two-quark and quark-antiquark distributions, which are the simplest cases as far as mixing and the number of colour channels are concerned.  We first discuss evolution in $\mu_1$ at fixed $\mu_2$ and $\zeta_p$ and then evolution in all three scales simultaneously.

%%%%%%%%%%%%%%%%%%%%%%%%%%%%%%%%%

\subsection{Evolution in the scale of one parton}

Let us consider evolution in the renormalisation scale of one parton, which we take to be the first one without loss of generality.  For the parton combinations of interest, the LO evolution equations in the $t$ channel basis read
\begin{align}
   \label{dglap-t}
& \frac{\partial}{\partial \log\mu_1^2}\,
  \pr{R_1 R_2}{F}_{q a}(x_1,x_2,y,\mu_1,\mu_2,\zeta_p)
 = \pr{R_1 R_1}{P}_{q q}(z,\mu_1,x_1^2 \zeta_p) \conv{1}
   \pr{R_1 R_2}{F}_{q a}(z,x_2,y,\mu_1,\mu_2,\zeta_p)
\nonumber \\
&\qquad
   + \sum_{R'} \pr{R_1 \Rp}{P}_{q g}(z,\mu_1,x_1^2 \zeta_p) \conv{1}
     \pr{\Rp R_2}{F}_{g a}(z,x_2,y,\mu_1,\mu_2,\zeta_p) \,,
\nonumber \\
& \frac{\partial}{\partial \log\mu_1^2}\,
   \pr{R_1 R_2}{F}_{\bar{q} a}(x_1,x_2,y,\mu_1,\mu_2,\zeta_p)
 = \pr{R_1 R_1}{P}_{q q}(z,\mu_1,x_1^2 \zeta_p) \conv{1}
   \pr{R_1 R_2}{F}_{\bar{q} a}(z,x_2,y,\mu_1,\mu_2,\zeta_p)
\nonumber \\
&\qquad
   + \sum_{R'} \varepsilon^2(R')\, \pr{R_1 \Rp}{P}_{q g}(z,\mu_1,x_1^2 \zeta_p)
     \conv{1} \pr{\Rp R_2}{F}_{g a}(z,x_2,y,\mu_1,\mu_2,\zeta_p) \,,
\end{align}
where the second parton $a$ is a quark or an antiquark, and where $R' = 1$ if $R_1 = 1$ and $R'= A,S$ if $R_1 = 8$.  In the second equation we have made use of the charge conjugation relations between $\bar{q}g$ and $q g$ splitting kernels; this gives a sign factor $\varepsilon^2(A) = -1$ for the antisymmetric octet.

The evolution equations involve the Mellin convolution
\begin{align}
   \label{mellin-conv}
P(z, \ldots) \conv{1} F(z, x_2, \ldots)
&= \int_{x_1 / (1 - x_2)}^1 \frac{d z}{z} \, P(z, \ldots) \,
   F\biggl( \frac{x_1}{z}, x_2, \ldots \biggr) \,,
\end{align}
whose lower integration boundary reflects the support region of DPDs in the momentum fractions: $F(z_1, z_2, \ldots)$ is zero for $z_1 + z_2 > 1$.   The evolution kernels can be written as
\begin{align}
   \label{LO-kernels}
\prb{R R'}{P}_{q b}(z,\mu,\zeta)
&= a_s(\mu) \, c_{R R'}\, \widetilde{P}_{q b}(z)
   + \Biggl[ \frac{\gamma_q(\mu)}{2}
      + \frac{\pr{R}{\gamma}_{J}(\mu)}{4}\, \log \frac{\mu^2}{\zeta}
     \Biggr] \,
     \delta_{R \Rp}\, \delta_{q b}\, \delta(1-z) + \mathcal{O}(a_s^2) \,,
\end{align}
where $b = q,g$, $\gamma_q(\mu) = 3 C_F\ms a_s(\mu) + \mathcal{O}(a_s^2)$, $\pr{1}{\gamma}_J(\mu) = 0$, and $\pr{8}{\gamma}_J(\mu)$ is given in \eqref{gamma-J8}.
The $z$ dependent part of \eqref{LO-kernels} involves the familiar splitting functions
\begin{align}
\widetilde{P}_{q q}(z) &= C_F\, \frac{1 + z^2}{(1 - z)_+} \,,
&
\widetilde{P}_{q g}(z) &= T_F\, \bigl[ z^2 + (1-z)^2 \bigr]
\end{align}
and the colour factors
\begin{align}
c_{1 1} &= 1 \,,
&
c_{8 8 } &= - 1/8\,,
&
c_{8 S} &= \sqrt{5} /4 \,,
&
c_{8 A} &= 3/4 \,.
\end{align}
Whilst our analysis is limited to the LO approximation of the splitting kernels, our results do not depend on whether one uses the LO or the NLO approximation for the anomalous dimension $\pr{8}{\gamma}_J(\mu)$, which is associated with Sudakov double logarithms.  Taking such anomalous dimensions at NLO corresponds to next-to-leading logarithmic (NLL) approximation (see e.g.~\tab{1} in \cite{Berger:2010xi} for single hard scattering and \sect{6.6} in \cite{Buffing:2017mqm} for DPS).  If one takes $\pr{8}{\gamma}_J(\mu)$ at two-loop accuracy, one may also want to use the two-loop rather than the one-loop running of $\alpha_s(\mu)$.  Our arguments in the present work do not depend on that choice.

We will shortly need to know the sign of the Mellin convolutions \eqref{mellin-conv}.  If $F(z_1, z_2, \ldots) \ge 0$ for all $z_1, z_2$, then we obviously have $\widetilde{P}_{q g}(z, \ldots) \otimes F(z, x_2, \ldots) \ge 0$.  On the other hand, the plus-prescription for $\widetilde{P}_{q q}$ involves a negative term proportional to $F(x_1, x_2, \ldots)$:
\begin{align}
   \label{plus-prescription}
\widetilde{P}_{q q}(z) \conv{1} F(z, x_2, \ldots)
= C_F \lim_{\epsilon\to 0^+} \Biggl[\ms
&
   \int_{x_1 / (1-x_2)}^{1 - \epsilon} \frac{d z}{z} \,
   \frac{1 + z^2}{1-z} \, F\biggl( \frac{x_1}{z}, x_2, \ldots \biggr)
\nonumber \\
& - F(x_1, x_2, \ldots)
      \int _0^{1 - \epsilon} d z\; \frac{2}{1-z}
   \ms\Biggr] \,.
\end{align}
It can therefore have any sign, even if $F(z_1, z_2, \ldots) \ge 0$ for all $z_1, z_2$.

\begin{figure}
\begin{center}
   \subfigure[$x_2 = x_1$]{\includegraphics[height=0.315\textwidth,
      trim=0 0 0 18,clip=true]{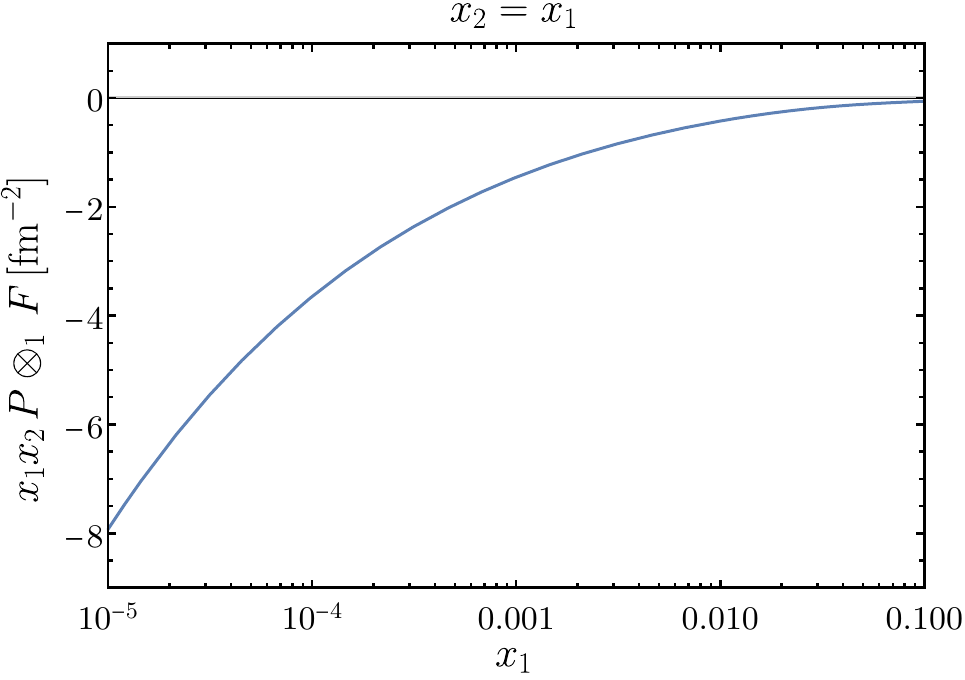}
   \hspace{0.8em}
   \includegraphics[height=0.32\textwidth,
      trim=24 0 0 18,clip=true]{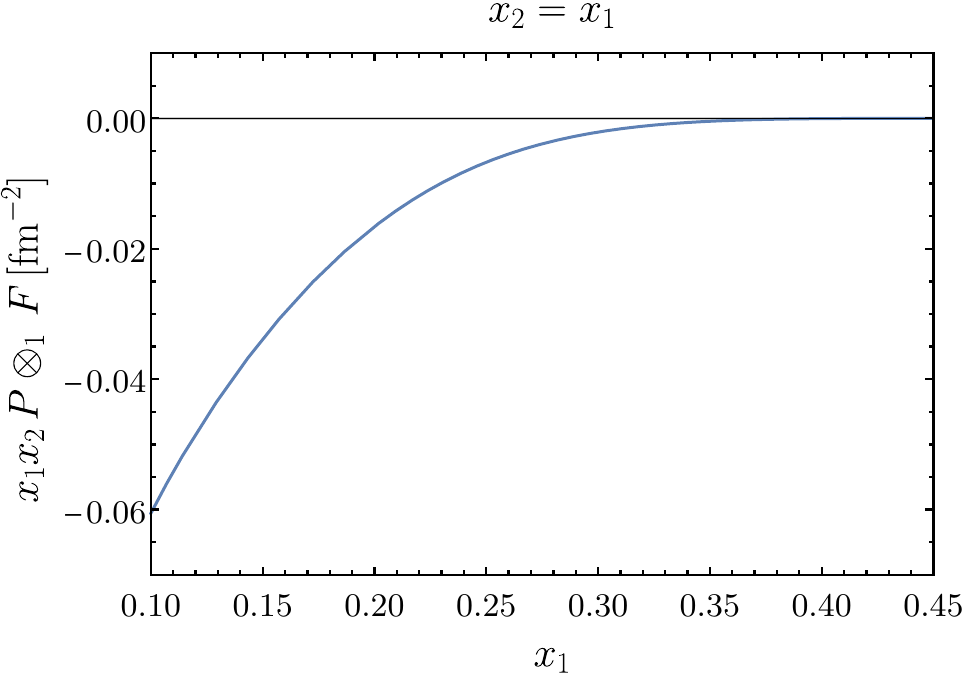}}
\\[1.5em]
   \subfigure[$x_2 = 0.1$]{\includegraphics[height=0.32\textwidth,
      trim=0 0 0 21,clip=true]{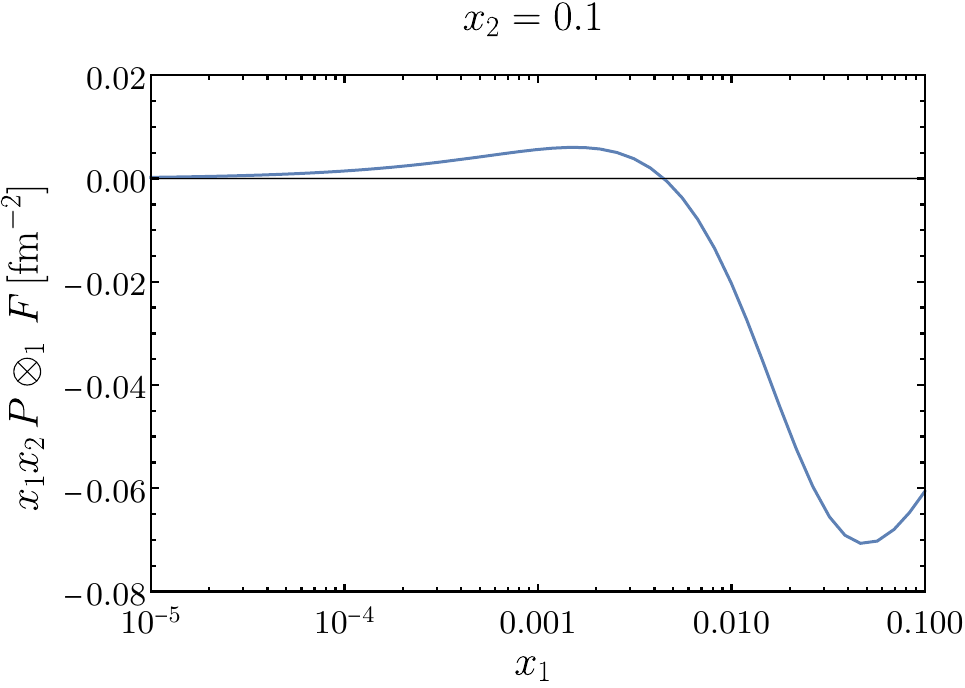}
   \hspace{0.8em}
   \includegraphics[height=0.32\textwidth,
      trim=24 0 0 20,clip=true]{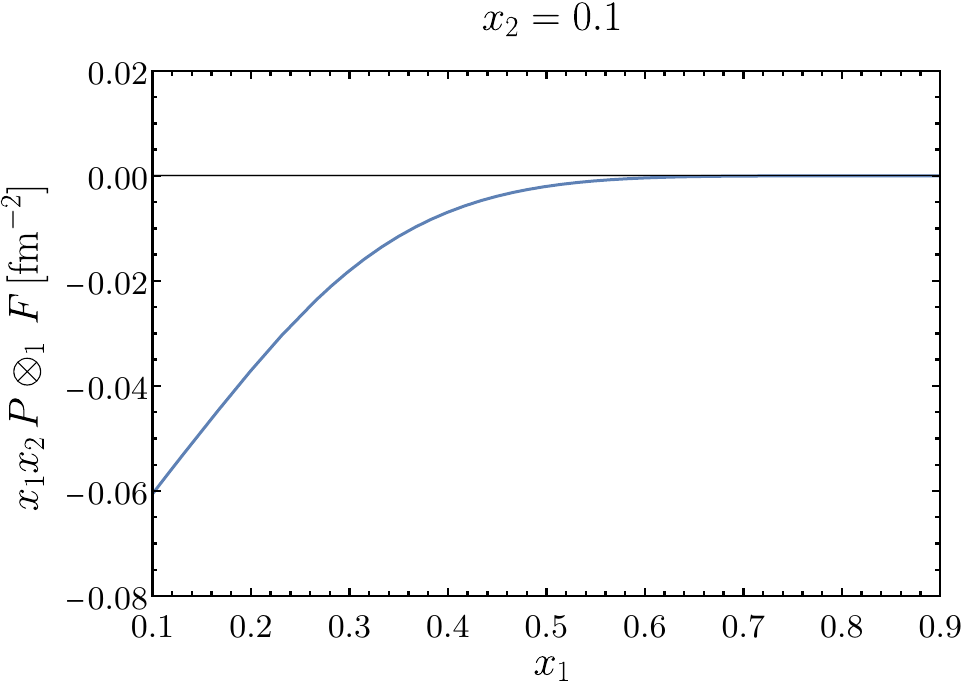}}
\\[1.5em]
   \subfigure[$x_2 = 10^{-3}$]{\includegraphics[height=0.315\textwidth,
      trim=0 0 0 24,clip=true]{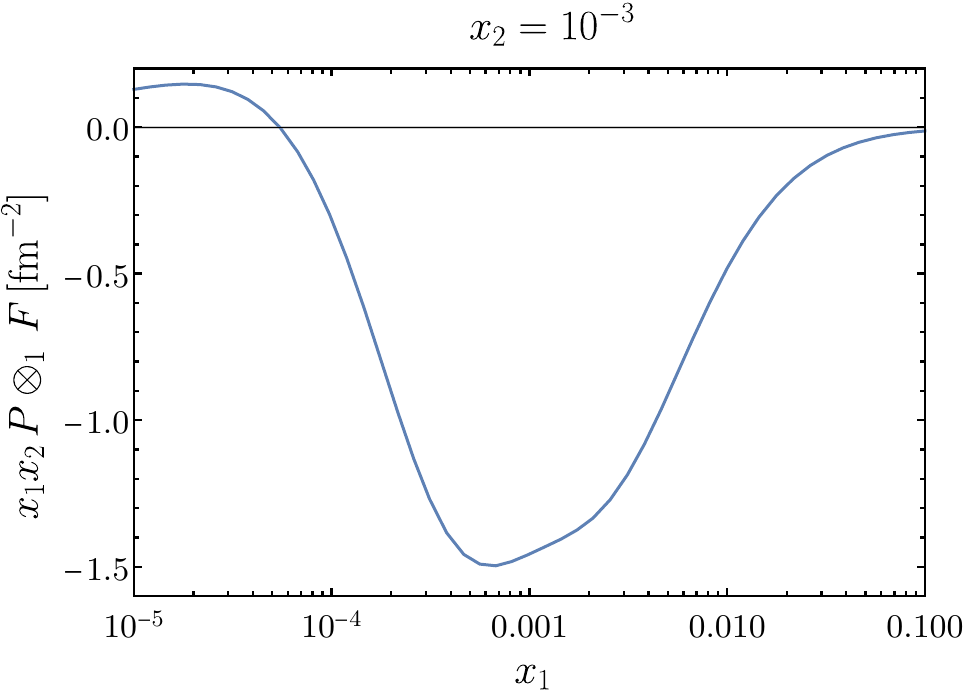}
   \hspace{0.8em}
   \includegraphics[height=0.32\textwidth,
      trim=24 0 0 24,clip=true]{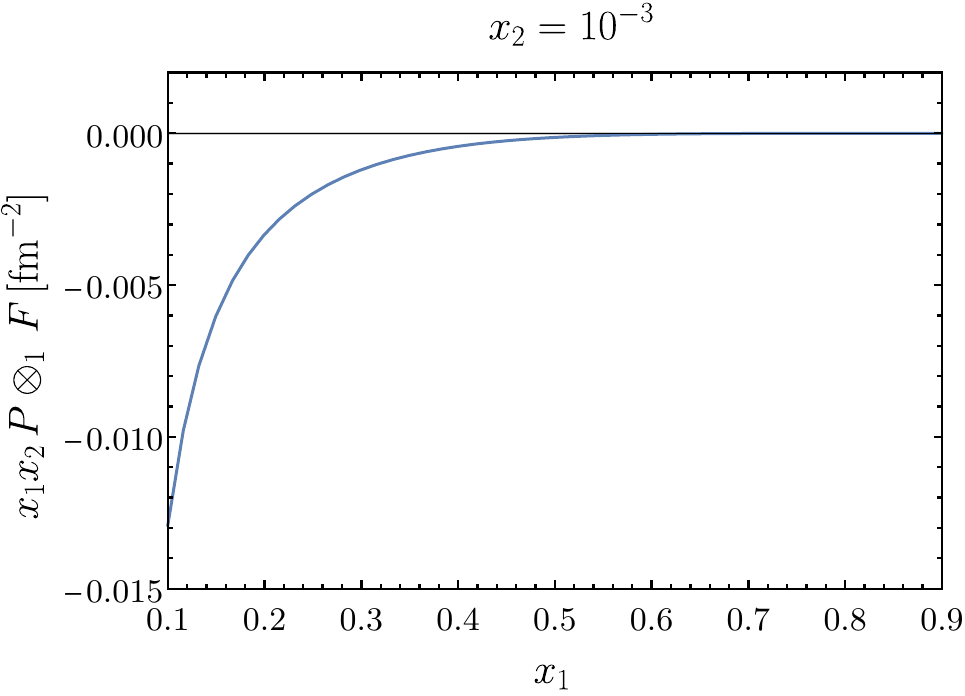}}
\caption{\label{fig:Pqq-Fqqbar} The convolution $x_1 x_2\, P(z,\mu) \,\otimes_{x_1}\! \pr{11}{F}_{q \bar{q}}(z, x_2, y, \mu, \mu)$ with $P(z,\mu) = a_s(\mu)\ms \widetilde{P}_{q q}(z)$ and $\pr{11}{F}_{q \bar{q}}$ computed from the LO splitting formula \protect\eqref{split-LO} at $\mu = b_0/y = 10 \gev$.  A weighting factor $x_1 x_2$ is included to keep details visible at higher momentum fractions.}
\end{center}
\end{figure}

To illustrate this, let us consider the DPDs computed with the LO splitting formula \eqref{split-LO}.  We take the PDFs of the \texttt{CT14lo} PDF set \protect\cite{Dulat:2015mca}, using the LHAPDF interface \protect\cite{Buckley:2014ana} via \texttt{ManeParse} \cite{Clark:2016jgm}.  We evaluate the DPD at $\mu = b_0/y = 10 \gev$ and verify that the PDFs are positive at that scale.  For the strong coupling, we use the value $\alpha_s(10 \gev) = 0.178$ provided by the PDF set.  As shown in \fig{\ref{fig:Pqq-Fqqbar}}, the convolution of $\widetilde{P}_{q q}$ with $\pr{11}{F}_{q \bar{q}}$ so obtained is indeed negative in a large region of the momentum fractions.

The evolution equations in the $s$ channel basis are derived along the same lines as the Collins-Soper equation in \eqref{CS-qq}.  If the first parton is a quark, we get
\begin{align}
   \label{dglap-qq}
\frac{\partial}{\partial \log\mu_1^2}\,
\begin{pmatrix}
   F_{q q}^{\overline{3}\ms 3} \\[0.3em]
   F_{q q}^{6\ms \overline{6}}
\end{pmatrix}
&= a_s(\mu_1) \, \widetilde{P}_{q q} \conv{1} \widehat{\mathbf{P}}_{q q, q}
   \begin{pmatrix}
      F_{q q}^{\overline{3}\ms 3} \\[0.3em]
      F_{q q}^{6\ms \overline{6}}
   \end{pmatrix}
   + a_s(\mu_1) \, \widetilde{P}_{q g} \conv{1} \widehat{\mathbf{P}}_{q g, q}
   \begin{pmatrix}
      F_{g q}^{3\ms \overline{3}} \\[0.3em]
      F_{g q}^{\overline{6}\ms 6} \\[0.3em]
      F_{g q}^{15\ms \overline{15}}
   \end{pmatrix}
\nonumber \\[0.4em]
& \quad
   + \frac{\gamma_q(\mu_1)}{2} \,
   \begin{pmatrix}
      F_{q q}^{\overline{3}\ms 3} \\[0.3em]
      F_{q q}^{6\ms \overline{6}}
   \end{pmatrix}
   + \frac{\pr{8}{\gamma}_J(\mu_1)}{4} \, \log \frac{\mu_1^2}{x_1^2 \zeta_p} \;
   \hat{\mathbf{J}}_{q q}
   \begin{pmatrix} F_{q q}^{\overline{3}\ms 3} \\[0.3em]
      F_{q q}^{6\ms \overline{6}}
   \end{pmatrix} \,,
\\[0.6em]
   \label{dglap-qqbar}
\frac{\partial}{\partial \log\mu_1^2}\,
\begin{pmatrix} F_{q \bar{q}}^{11} \\[0.2em]
   F_{q \bar{q}}^{88}
\end{pmatrix}
&= a_s(\mu_1) \, \widetilde{P}_{q q} \conv{1}
      \widehat{\mathbf{P}}_{q q, \bar{q}}
   \begin{pmatrix} F_{q \bar{q}}^{11} \\[0.2em]
      F_{q \bar{q}}^{88}
   \end{pmatrix}
   + a_s(\mu_1) \, \widetilde{P}_{q g} \conv{1}
      \widehat{\mathbf{P}}_{q g, \bar{q}}
   \begin{pmatrix} F_{g \bar{q}}^{\overline{3}\ms 3} \\[0.3em]
      F_{g \bar{q}}^{6\ms \overline{6}} \\[0.3em]
      F_{g \bar{q}}^{\overline{15}\ms 15}
   \end{pmatrix}
\nonumber \\[0.4em]
& \quad
   + \frac{\gamma_q(\mu_1)}{2} \,
   \begin{pmatrix} F_{q \bar{q}}^{11} \\[0.2em]
      F_{q \bar{q}}^{88}
   \end{pmatrix}
   + \frac{\pr{8}{\gamma}_J(\mu_1)}{4} \, \log \frac{\mu_1^2}{x_1^2 \zeta_p} \;
   \hat{\mathbf{J}}_{q \bar{q}}
   \begin{pmatrix} F_{q \bar{q}}^{11} \\[0.2em]
      F_{q \bar{q}}^{88}
   \end{pmatrix} \,,
\end{align}
where the function arguments of $\widetilde{P}$ and $F$ are as those of $P$ and $F$ in \eqref{dglap-t}.  The colour mixing matrices read
\begin{align}
   \label{P-mat-qq}
\widehat{\mathbf{P}}_{q q, q}
&= \mathbf{M}_{q q} \, \begin{pmatrix} 1 & 0 \\ 0 & -1/8 \end{pmatrix}
   \mathbf{M}_{q q}^{-1}
 = \frac{1}{8} \,
   \begin{pmatrix} 2 & 6 \\ 3 & 5 \end{pmatrix} \,,
\nonumber \\[0.4em]
\widehat{\mathbf{P}}_{q q, \bar{q}}
&= \mathbf{M}_{q \bar{q}} \, \begin{pmatrix} 1 & 0 \\ 0 & -1/8 \end{pmatrix}
   \mathbf{M}_{q \bar{q}}^{-1}
 = \frac{1}{8} \,
   \begin{pmatrix} 0 & 8 \\ 1 & 7 \end{pmatrix} \,,
\end{align}
and
\begin{align}
   \label{P-mat-qg}
\widehat{\mathbf{P}}_{q g, q}
&= \mathbf{M}_{q q} \,
   \begin{pmatrix} 1 & 0 & 0 \\ 0 & \sqrt{5}/4 & 3/4 \end{pmatrix}
   \mathbf{M}_{g q}^{-1}
 = \frac{1}{6} \,
   \begin{pmatrix}
      4 & 12 &  0 \\
      1 &  0 & 15
   \end{pmatrix} \,,
\nonumber \\[0.4em]
\widehat{\mathbf{P}}_{q g, \bar{q}}
&= \mathbf{M}_{q \bar{q}} \,
   \begin{pmatrix} 1 & 0 & 0 \\ 0 & \sqrt{5}/4 & 3/4 \end{pmatrix}
   \mathbf{M}_{g \bar{q}}^{-1}
 = \frac{1}{24} \,
   \begin{pmatrix}
      64 &  0 &  0 \\
       1 & 18 & 45
   \end{pmatrix} \,,
\end{align}
whilst $\hat{\mathbf{J}}_{q q}$ and $\hat{\mathbf{J}}_{q \bar{q}}$ are given in \eqref{J-qq} and \eqref{J-qqbar}, respectively.
With
\begin{align}
\widehat{\mathbf{P}}_{\bar{q} g, \bar{q}}
&= \mathbf{M}_{\bar{q} \bar{q}} \,
   \begin{pmatrix} 1 & 0 & 0 \\ 0 & \sqrt{5}/4 & -3/4 \end{pmatrix}
   \mathbf{M}_{g \bar{q}}^{-1}
 = \widehat{\mathbf{P}}_{q g, q} \,,
\nonumber \\
\widehat{\mathbf{P}}_{\bar{q} g, q}
&= \mathbf{M}_{\bar{q} q} \,
   \begin{pmatrix} 1 & 0 & 0 \\ 0 & \sqrt{5}/4 & -3/4 \end{pmatrix}
   \mathbf{M}_{g q}^{-1}
 = \widehat{\mathbf{P}}_{q g, \bar{q}} \,,
\end{align}
we find that the evolution equations for $F_{\bar{q}\bar{q}}$ and $F_{\bar q q}$ are respectively obtained from \eqref{dglap-qq} and \eqref{dglap-qqbar} by interchanging $q \leftrightarrow \bar{q}$ in the DPDs and swapping their representation labels, i.e.\ $F_{q q}^{\overline{3}\ms 3} \to F_{\bar{q}\bar{q}}^{3\ms \overline{3}}$, $F_{g q}^{3\ms \overline{3}} \to F_{g \bar{q}}^{\overline{3}\ms 3}$, etc.  The splitting kernels, anomalous dimensions, and colour mixing matrices remain the same.

The evolution equations have the same form for distributions with unequal flavours, i.e.\ one may replace $F_{q q} \to F_{q_1 q_2}$ and $F_{g q} \to F_{g q_2}$ in \eqref{dglap-qq}, or $F_{q \bar{q}} \to F_{q_1 \bar{q}_2}$ and $F_{g \bar{q}} \to F_{g \bar{q}_2}$ in \eqref{dglap-qqbar}, whilst keeping the splitting kernels and colour mixing matrices unchanged.

Before analysing the effect of the evolution equations \eqref{dglap-qq} and \eqref{dglap-qqbar} on positivity, let us recall the situation for colour singlet distributions in the $t$ channel basis.  The LO evolution equation for $\pr{11}{F}_{q a}$ reads
\begin{align}
   \label{colour-singlet-dglap}
\frac{\partial}{\partial \log\mu_1^2}\, \pr{11}{F}_{q a}(x_1, x_2, \ldots)
&= a_s(\mu_1) \, \widetilde{P}_{q q} \conv{1} \pr{11}{F}_{q a}
   + a_s(\mu_1) \, \widetilde{P}_{q g} \conv{1} \pr{11}{F}_{g a}
   + \frac{\gamma_q(\mu_1)}{2} \, \pr{11}{F}_{q a} \,,
\end{align}
where $a$ is a quark or an antiquark.  If $\pr{11}{F}_{q a}$ and $\pr{11}{F}_{g a}$ are non-negative for all momentum fractions, the terms involving $\widetilde{P}_{q g}$ or $\gamma_q$ in \eqref{colour-singlet-dglap} are non-negative as well and hence conserve positivity.  As we have shown in \fig{\ref{fig:Pqq-Fqqbar}}, the first term in \eqref{colour-singlet-dglap} can be negative.  However, the part that is responsible for a negative sign is proportional to $\pr{11}{F}_{q a}(x_1, x_2, \ldots)$ itself, as is easily seen in \eqref{plus-prescription}.  This negative contribution hence decreases in magnitude when $\pr{11}{F}_{q a}(x_1, x_2, \ldots)$ approaches zero from above, and closer inspection shows that it cannot lead to a violation of positivity.  This is shown in more detail in \app{B} of \cite{Diehl:2013mla}.  Overall, positivity is hence conserved for LO evolution of $\pr{11}{F}_{q a}$ to higher scales, and the same can be shown for all other parton combinations.

We now analyse the sign of the different terms on the r.h.s.\ of \eqref{dglap-qq} and \eqref{dglap-qqbar} under the assumption that the $s$ channel distributions on the r.h.s.\ are non-negative for all momentum fractions.  For brevity, we write $D_1 F = \partial F / \partial \log \mu_1^2$.
\begin{enumerate}
\item The scale variation $D_1 F_{q a}^{R \Rbar}$ receives contributions $a_R\ms \widetilde{P}_{q q} \otimes F_{q a}^{R \Rbar}$ and $b_R\ms \widetilde{P}_{q q} \otimes F_{q a}^{\Rp \bs \Rpbar}$ with coefficients $a_R \ge 0$, $b_R > 0$, where $\Rp \neq R$.  A negative contribution from the term with $a_R$ cannot lead to a violation of positivity, as just discussed.  However, the contribution from the term with $b_R$ can remain large and negative even if $F_{q a}^{R \Rbar}$ approaches zero.  This term can therefore lead to a zero crossing of the distribution as one evolves to higher scales.
\item The matrices $\widehat{\mathbf{P}}_{q g, a}$ have no negative elements, so that the terms with $\widetilde{P}_{q g}$ are all non-negative.
\item The terms with $\gamma_q$ are all non-negative.
\item The contributions with $\pr{8}{\gamma}_J$ to $D_1 F_{q q}^{\overline{3} 3}$ and $D_1 F_{q q}^{6 \overline{6}}$ have opposite sign, as well as those to $D_1 F_{q\bar{q}}^{11}$ and $D_1 F_{q\bar{q}}^{88}$.  This follows from the form of $\hat{\mathbf{J}}_{q q}$ and $\hat{\mathbf{J}}_{q \bar{q}}$ in \eqref{J-qq} and \eqref{J-qqbar}.  Whether these contributions can violate positivity depends on the sign of $\log \mu_1^2 /(x_1^2\ms \zeta_p)$.
\end{enumerate}
It follows that LO DGLAP evolution of DPDs is not guaranteed to conserve their positivity in colour space, in contrast to the colour singlet distributions $\pr{11}{F}_{a_1 a_2}$.  This is one of our main results.

\begin{figure}
\begin{center}
   \subfigure[$x_2 = x_1$]{\includegraphics[height=0.276\textwidth,
      trim=0 0 100 20,clip=true]{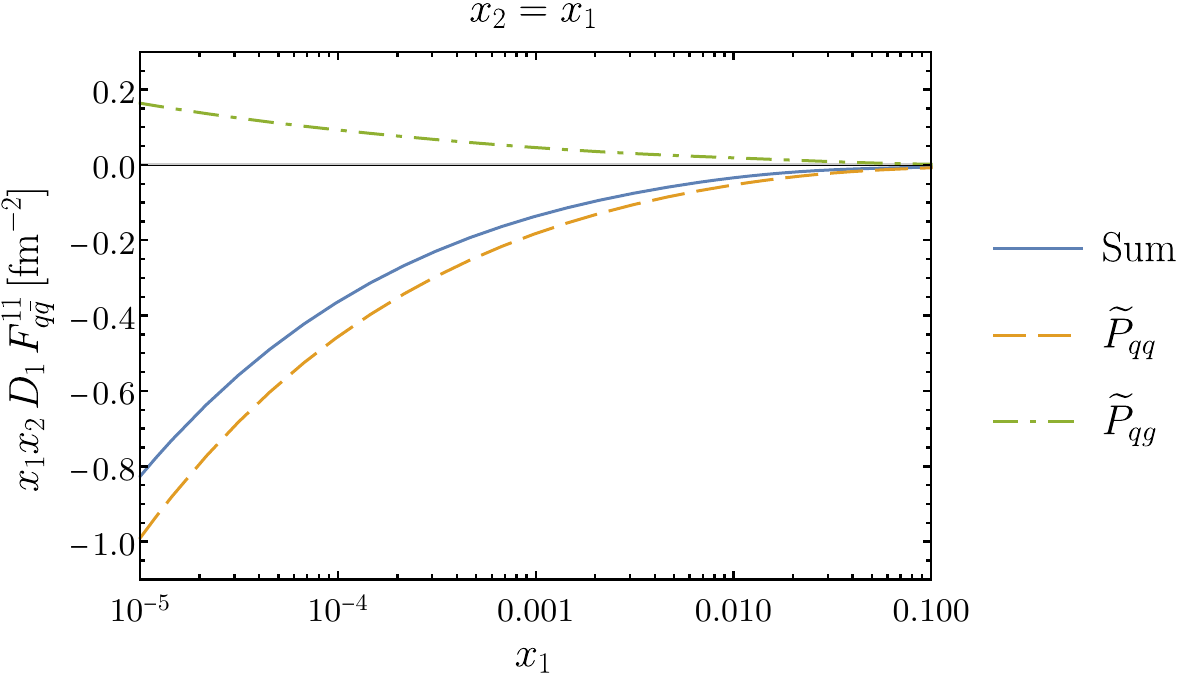}
   \hspace{1.2em}
   \includegraphics[height=0.28\textwidth,
      trim=30 0 0 24,clip=true]{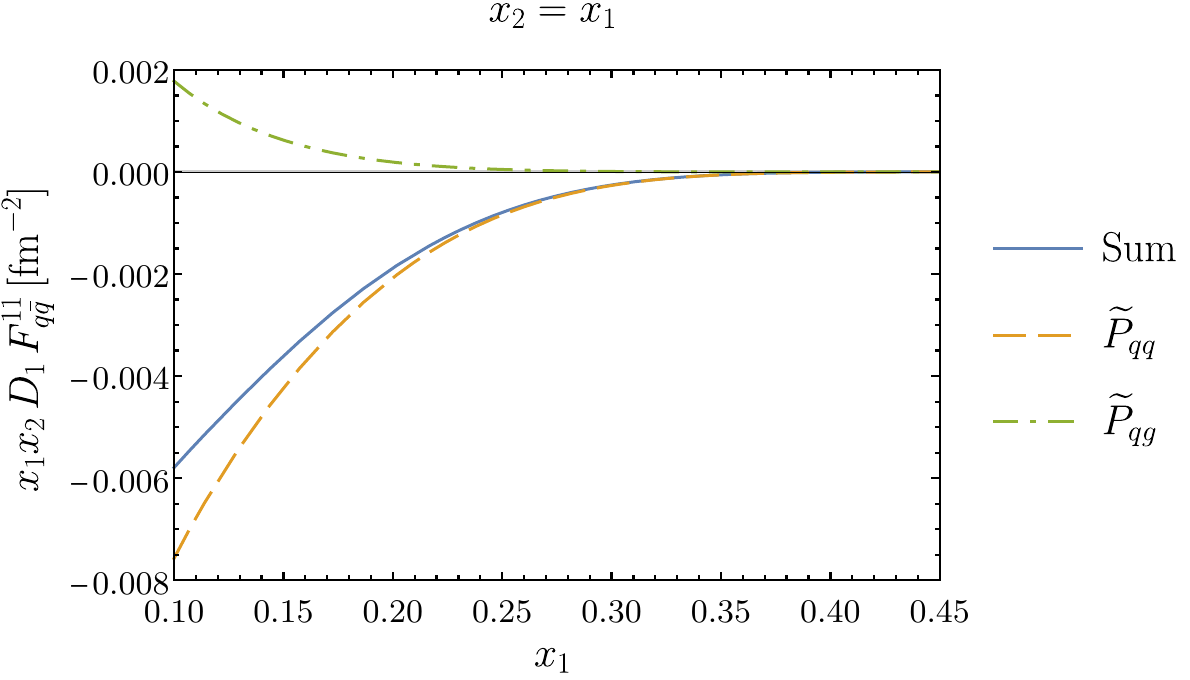}}
\\[1.5em]
   \subfigure[$x_2 = 0.1$]{\includegraphics[height=0.28\textwidth,
      trim=0 0 100 20,clip=true]{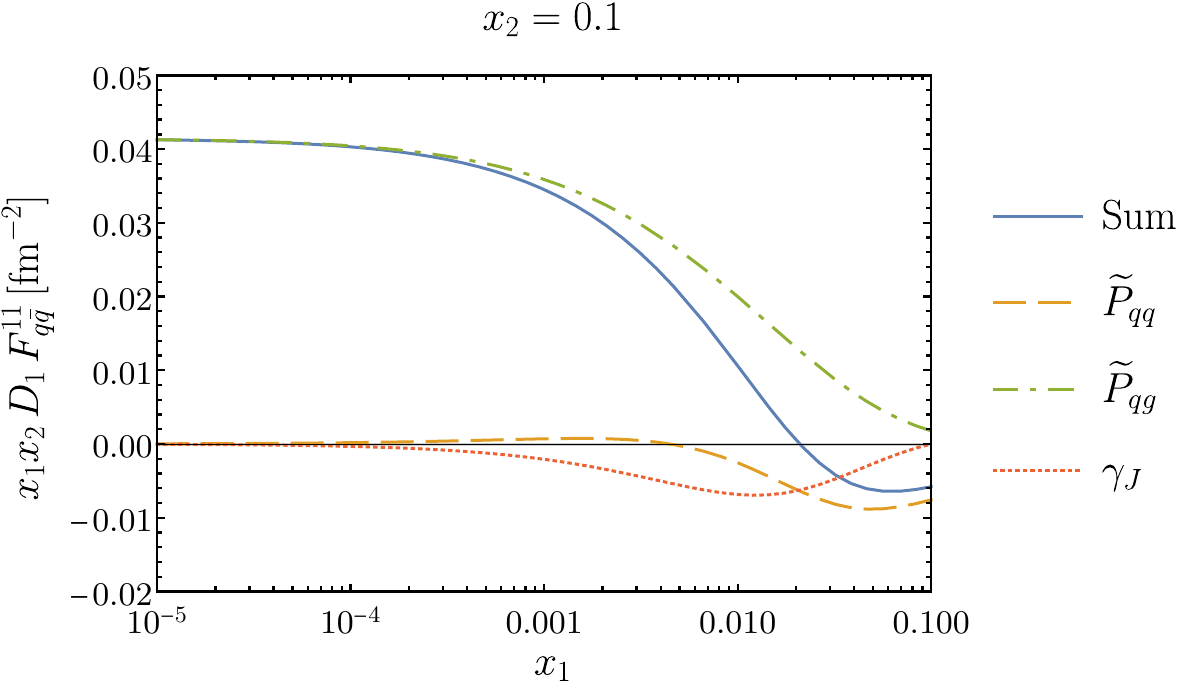}
   \hspace{1.2em}
   \includegraphics[height=0.28\textwidth,
      trim=30 0 0 20,clip=true]{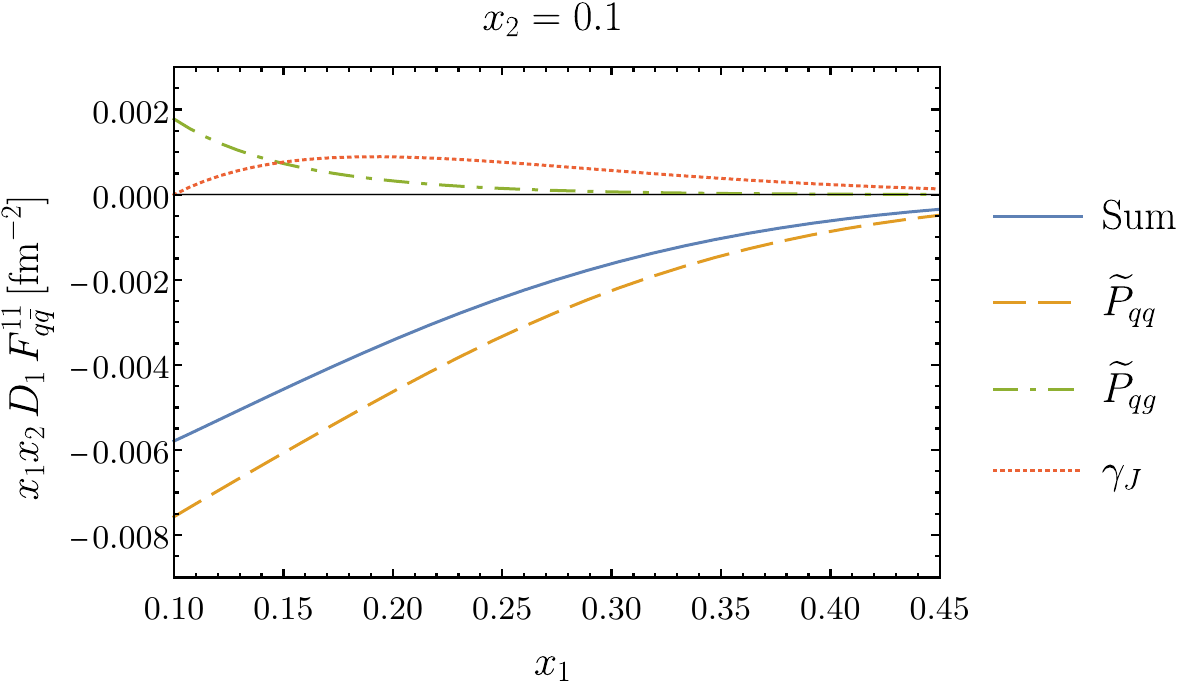}}
\\[1.5em]
   \subfigure[$x_2 = 10^{-3}$]{\includegraphics[height=0.276\textwidth,
      trim=0 0 100 24,clip=true]{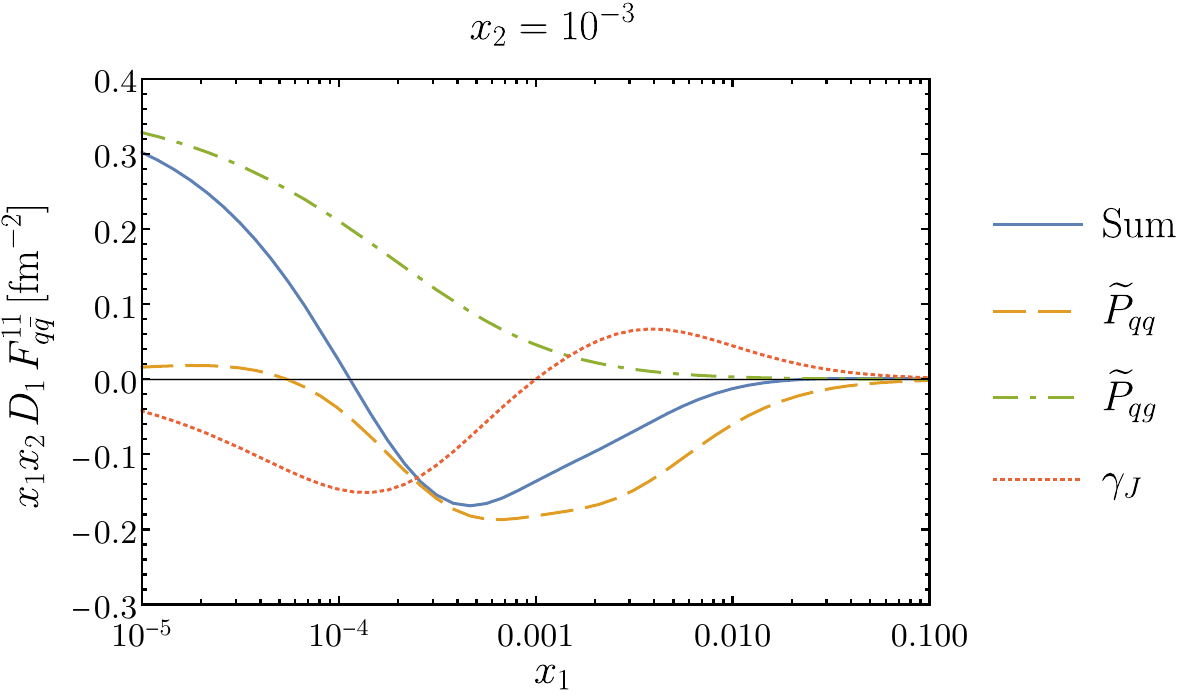}
   \hspace{1.2em}
   \includegraphics[height=0.28\textwidth,
      trim=30 0 0 24,clip=true]{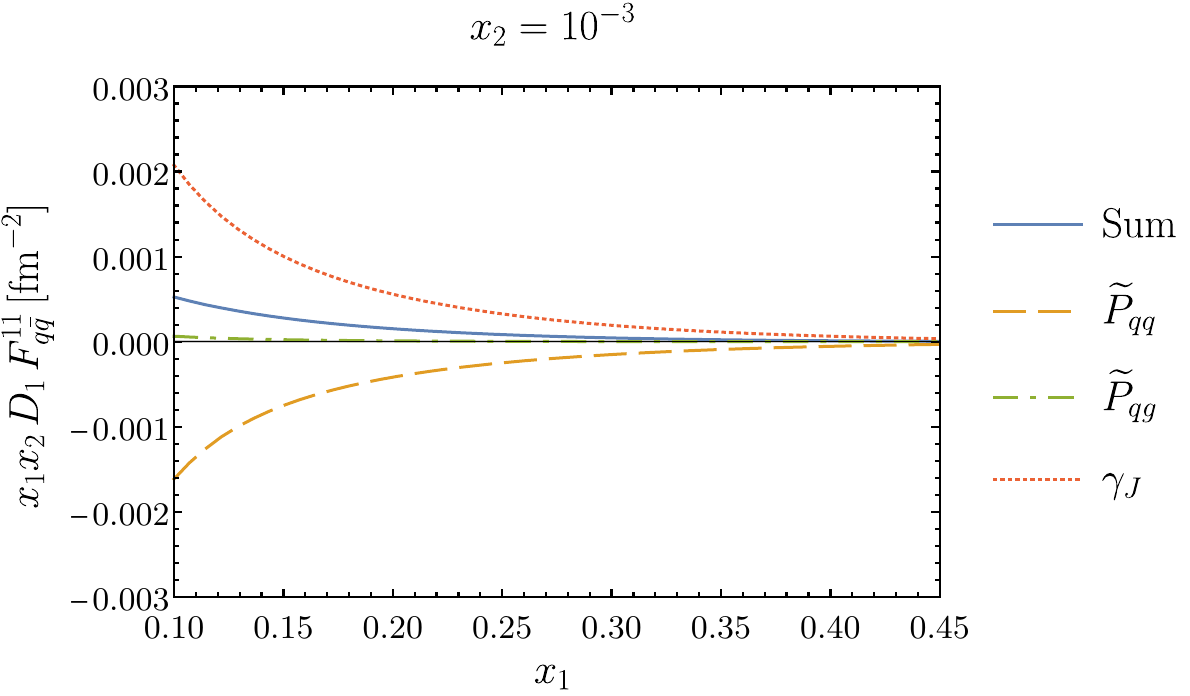}}
\caption{\label{fig:Fqqbar-evol} The right-hand side of \protect\eqref{Fqqbar-LO-evol} with $q \bar{q} = u \bar{u}$, evaluated at $\mu = b_0/y = 10 \gev$ and $\zeta_p = \mu^2 / (x_1 x_2)$ with LO splitting DPDs.  The individual terms going with $\widetilde{P}_{q q}$, $\widetilde{P}_{q g}$, and $\gamma_J$ are shown as well.
The PDFs used in the splitting formula are specified in the text below \protect\eqref{plus-prescription}.  A weighting factor $x_1 x_2$ is included to keep details visible at higher momentum fractions.}
\end{center}
\end{figure}

As a numerical illustration, let us take the initial conditions provided by the perturbative splitting mechanism at LO.  According to \eqref{split-LO-colour}, we have $F_{q \bar{q}}^{1 1} = F_{g \bar{q}}^{6 \overline{6}} = F_{g \bar{q}}^{\overline{15}\ms 15} = 0$ at the scale $\mu$ where the splitting formula is evaluated.  Inserting this in the evolution equation \eqref{dglap-qqbar} and taking the LO approximation \eqref{gamma-J8} of $\pr{8}{\gamma}_J$, we obtain
\begin{align}
   \label{Fqqbar-LO-evol}
\frac{\partial}{\partial \log\mu_1^2} \, F_{q \bar{q}}^{1 1}(\mu_1, \mu_2)
= a_s(\mu) \Biggl[\ms
&
\widetilde{P}_{q q} \conv{1} F_{q \bar{q}}^{88}(\mu,\mu)
+ \frac{8}{3}\,
   \widetilde{P}_{q g} \conv{1} F_{g \bar{q}}^{\overline{3}\ms 3}(\mu,\mu)
\nonumber \\
&
- \frac{2}{9} \; \pr{8}{\gamma}_J^{(0)} \ms
  \log \frac{\mu^2}{x_1^2 \zeta_p} \,
  F_{q \bar{q}}^{88}(\mu,\mu) \ms\Biggr]
\end{align}
at the point $\mu_1 = \mu_2 = \mu$.  We evaluate the r.h.s.\ numerically with the same settings as in \fig{\ref{fig:Pqq-Fqqbar}} and the choice
\begin{align}
   \label{zeta-choice}
\zeta_p = \mu^2 / (x_1 x_2) \,,
\end{align}
which is natural for the perturbative splitting mechanism \cite{Diehl:2021wpp}.  With this choice, the term with $\zeta_p$ is zero at $x_1 = x_2$.  At that point, a negative value of \eqref{Fqqbar-LO-evol} must hence be due to the term with $\widetilde{P}_{q q}$.  We see in \fig{\ref{fig:Fqqbar-evol}} that there are regions in $x_1$ and $x_2$ for which $D_1 F_{q \bar{q}}^{1 1} < 0$ at the scale $\mu$.  With $F_{q \bar{q}}^{1 1} = 0$ at that scale, positivity is thus explicitly violated by evolution to a higher scale for the first parton.

%%%%%%%%%%%%%%%%%%%%%%%%%%%%%%%%%

\subsection{Simultaneous evolution in all scales}

When computing double parton scattering cross sections, the choice of scales $\mu_1$, $\mu_2$ and $\zeta_p$ is driven by the kinematics of the process.  In particular, taking $\mu_1 \neq \mu_2$ is natural for processes with two hard scales of very different size.  On the other hand, the simplest setting for the physical interpretation of a DPD is with all relevant scales set equal.  We therefore investigate the evolution of DPDs with a common renormalisation scale $\mu = \mu_1 = \mu_2$ for both partons and the rapidity scale given by $\zeta_p = \mu^2 / (x_1 x_2)$ as in \eqref{zeta-choice}.

Let us briefly comment on the factor $x_1 x_2$ in \eqref{zeta-choice}.  As explained in \cite{Buffing:2017mqm, Diehl:2021wpp}, the definition of $\zeta_p$ involves the rapidity regulator and the plus-momentum of the target proton.  By contrast, $x_1 x_2\ms \zeta_p$ refers to the regulator and the plus-momenta of the two extracted partons, and in this sense is more closely related to the scales $\mu_1$ and $\mu_2$ that refer to the renormalisation of the operators associated with the two partons.  This motivates our choice \eqref{zeta-choice}, along with the fact that $x_1 x_2\ms \zeta_p$ is the combination of variables appearing at higher orders in the perturbative splitting formula for DPDs, see \eqref{split-master} and \eqref{log-defs}.

Combining the DGLAP equations for the first and second parton with the Collins-Soper equation \eqref{CS-qq} yields
\begin{align}
   \label{dglap-qq-common}
\frac{\partial}{\partial \log\mu^2}\,
\begin{pmatrix}
   F_{q q}^{\overline{3}\ms 3} \\[0.3em]
   F_{q q}^{6\ms \overline{6}}
\end{pmatrix}
&= a_s(\mu) \, \widetilde{P}_{q q} \conv{1} \widehat{\mathbf{P}}_{q q, q}
   \begin{pmatrix}
      F_{q q}^{\overline{3}\ms 3} \\[0.3em]
      F_{q q}^{6\ms \overline{6}}
   \end{pmatrix}
   + a_s(\mu) \, \widetilde{P}_{q q} \conv{2} \widehat{\mathbf{P}}_{q q, q}
   \begin{pmatrix}
      F_{q q}^{\overline{3}\ms 3} \\[0.3em]
      F_{q q}^{6\ms \overline{6}}
   \end{pmatrix}
\nonumber \\[0.4em]
& \quad
   + a_s(\mu) \, \widetilde{P}_{q g} \conv{1} \widehat{\mathbf{P}}_{q g, q}
   \begin{pmatrix}
      F_{g q}^{3\ms \overline{3}} \\[0.3em]
      F_{g q}^{\overline{6}\ms 6} \\[0.3em]
      F_{g q}^{15\ms \overline{15}}
   \end{pmatrix}
   + a_s(\mu) \, \widetilde{P}_{q g} \conv{2} \widehat{\mathbf{P}}_{q g, q}
   \begin{pmatrix}
      F_{q g}^{3\ms \overline{3}} \\[0.3em]
      F_{q g}^{\overline{6}\ms 6} \\[0.3em]
      F_{q g}^{15\ms \overline{15}}
   \end{pmatrix}
\nonumber \\[0.4em]
& \quad
   + \gamma_q(\mu) \,
   \begin{pmatrix}
      F_{q q}^{\overline{3}\ms 3} \\[0.3em]
      F_{q q}^{6\ms \overline{6}}
   \end{pmatrix}
   + \frac{\pr{8}{J}(y,\mu,\mu)}{2} \; \hat{\mathbf{J}}_{q q} \,
   \begin{pmatrix} F_{q q}^{\overline{3}\ms 3} \\[0.3em]
      F_{q q}^{6\ms \overline{6}}
   \end{pmatrix} \,.
\end{align}
With $\zeta_p$ chosen as in \eqref{zeta-choice}, the term with $\pr{8}{\gamma}_J$ in \eqref{dglap-qq} has cancelled against the corresponding term from the evolution equation in $\mu_2$.

The sign of the r.h.s.\ of \eqref{dglap-qq-common} can be analysed along the same lines as in the previous subsection:
\begin{enumerate}
\item The terms with $\widetilde{P}_{q q}$ can lead to a violation of positivity because of the negative part in the plus-distribution and of the positive off-diagonal entries in the matrix $\widehat{\mathbf{P}}_{q q, q}$.
\item The terms with $\widetilde{P}_{q g}$ conserve positivity.
\item The terms with $\gamma_q$ conserve positivity.
\item If $\pr{8}{J} < 0$ then the corresponding terms do not violate positivity, because negative contributions are proportional to the DPD being evolved, whilst the DPD in the other colour channel gives a positive contribution.  This is consistent with our findings for Collins-Soper evolution in \sect{\ref{sec:cs-evol}}.
\end{enumerate}
The leading order expression \eqref{J8} of $\pr{8}{J}(y,\mu,\mu)$ contains an explicit logarithm $\log (\mu^2 y^2)$.  In the leading double logarithmic approximation, one therefore has to keep only the last term in the evolution equation \eqref{dglap-qq-common}, which then reduces to the LO Collins-Soper equation with $\mu^2 = x_1 x_2\ms \zeta_p$.

The evolution equation for $F_{q \bar{q}}$ has the same form as \eqref{dglap-qq-common} and involves the colour mixing matrices $\widehat{\mathbf{P}}_{q q, \bar{q}}$, $\widehat{\mathbf{P}}_{q g, \bar{q}}$, and $\hat{\mathbf{J}}_{q \bar{q}}$.  Its discussion proceeds in full analogy.

In summary, we find that evolution to higher scales can violate positivity  of DPDs in colour space, both when one evolves in the renormalisation scale of one parton and when one evolves in all scales simultaneously.

\section{DPDs from parton splitting at two-loop accuracy}
\label{sec:splitting}

In \sect{\ref{sec:bases}} we saw that the perturbative splitting mechanism gives DPDs that satisfy positivity if the splitting is computed at LO, i.e.\ at one-loop accuracy.  This is not surprising, since the LO splitting formula in the $s$ channel basis can be written as a squared matrix element (in the mixed representation of definite plus-momentum and transverse position for the observed partons).  Starting from two loops, the splitting formula has explicit logarithms of the renormalisation scale $\mu$ and of the rapidity parameter $\zeta_p$, which respectively result from subtractions for ultraviolet and rapidity divergences.  As a consequence of these subtractions, positivity is no longer guaranteed.  It is then natural to ask whether the resulting DPDs at small $y$ violate positivity in colour space, and if so, by how much.  We address this question in the present section, using the results of the recent two-loop calculation in \cite{Diehl:2021wpp}.

The generalisation of the LO splitting formula \eqref{split-LO} to higher orders has the form
\begin{align}
\label{split-master}
& \pr{R_1 R_2}{F}_{a_1 a_2}(x_1,x_2,y, \mu,\mu,\zeta_p)
\nonumber \\
&\quad
= \frac{1}{\pi\ms y^2}\, \sum_{a_0}\,
  \prn{R_1 R_2}{V}_{a_1 a_2, a_0}\bigl( z_1,z_2,
      a_s(\mu), L_y(\mu), L_\zeta(\mu) \bigr) \conv{}
  f_{a_0}(z,\mu)
\end{align}
with
\begin{align}
\label{log-defs}
L_y(\mu) &= \log\frac{\mu^2}{\mu_y^2} \,,
&
\mu_y &= \frac{b_0}{y} \,,
&
L_\zeta(\mu) &= \log \frac{\mu^2}{x_1 x_2\ms \zeta_p}
\end{align}
and the special convolution
\begin{align}
   \label{conv-def}
V(z_1, z_2) \conv{} f(z)
&= \frac{1}{x} \int_{x}^1 d z\;
   V\bigl( u z, (1-u) z \ms\bigr)\, f\biggl( \frac{x}{z} \biggr) \,.
\end{align}
Here we write
\begin{align}
   \label{zu-relations}
z_1 &= u z \,,
&
z_2 &= (1-u)\ms z
\intertext{and recall that}
x_1 &= u x \,,
&
x_2 &= (1-u)\ms x
\end{align}
according to \eqref{xu-defs}.  The analogue of \eqref{split-master} for distributions in the $s$ channel basis is readily obtained using the transformations in \sect{\ref{sec:bases}}.
We limit our attention to the quark-antiquark sector and consider the distributions
\begin{align}
   \label{ud-channels}
F_{u d}^{\overline{3} 3} &= 4 \ms F_{u d}^{6 \overline{6}} \,,
&
F_{u \bar{d}}^{1 1} &= 64 \ms F_{u \bar{d}}^{8 8}
\end{align}
and
\begin{align}
   \label{uu-channels}
F_{u u}^{\overline{3} 3} \,,
\qquad
F_{u u}^{6 \overline{6}} \,,
\qquad
F_{u \bar{u}}^{1 1} \,,
\qquad
F_{u \bar{u}}^{8 8} \,.
\end{align}
Among these, only $F_{u \bar{u}}^{8 8}$ is nonzero at order $a_s$, whilst all others start at order $a_s^2$.

\begin{figure}[ht]
   \begin{center}
   \subfigure[\label{fig:qq-1} $V_{q q'\!,q}$]{\includegraphics[width=0.25\textwidth]{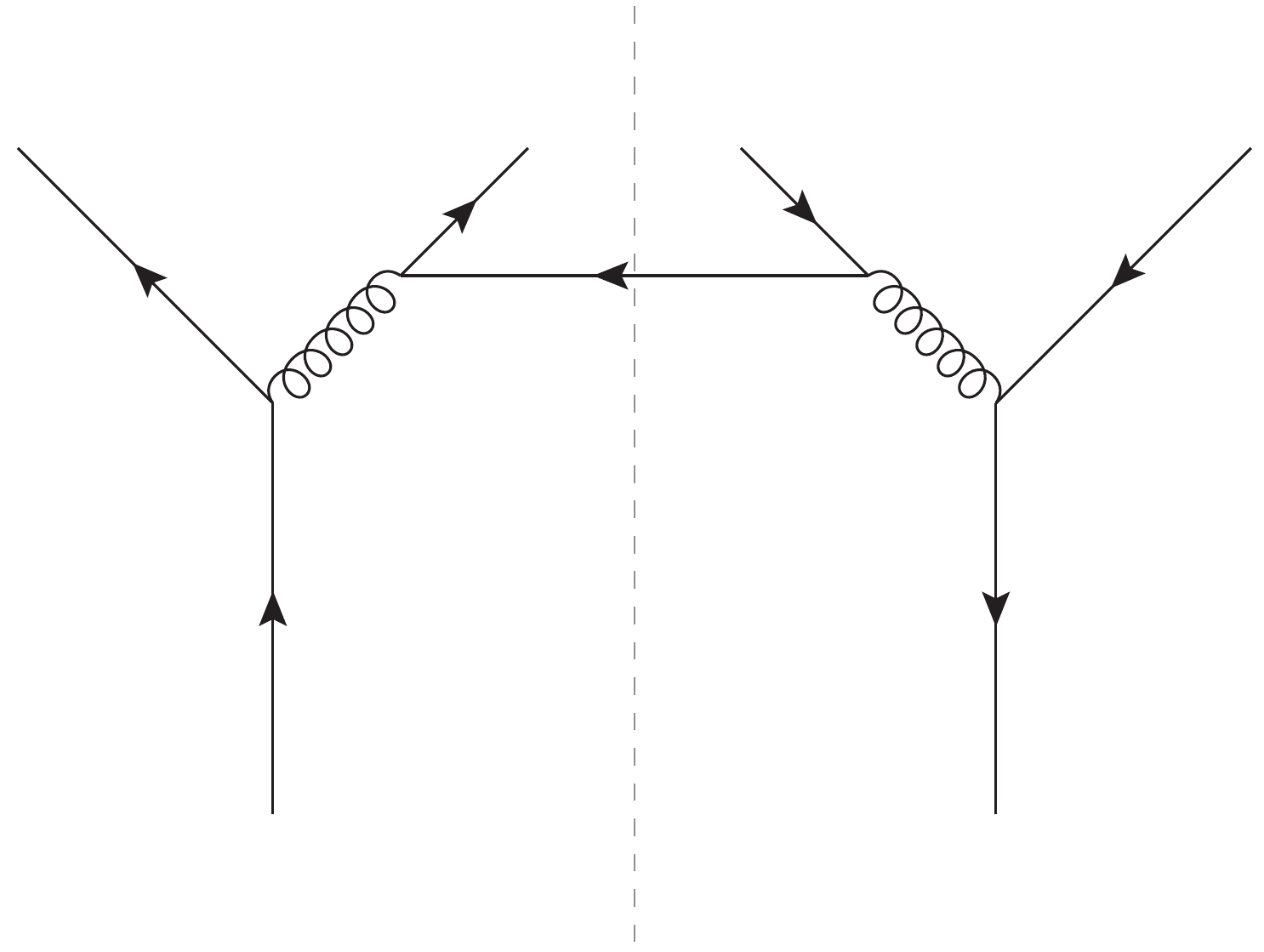}}
   \hspace{1.4em}
   \subfigure[\label{fig:qq-2} $V_{q q,q}^v$]{\includegraphics[width=0.25\textwidth]{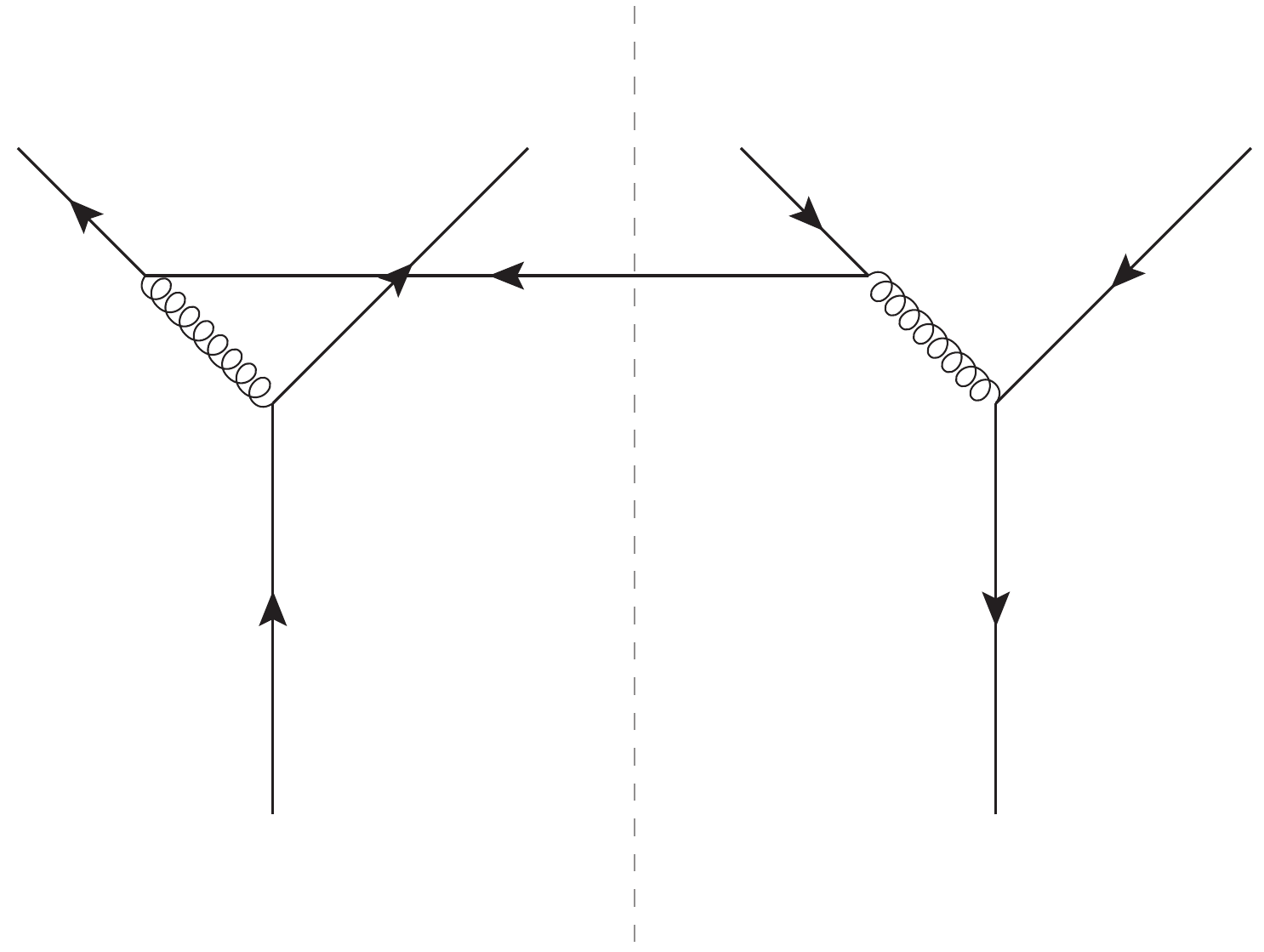}}
\\[1.5em]
   \subfigure[\label{fig:qqbar-1} $V_{q'\! \bar{q}'\!, q}$]{\includegraphics[width=0.25\textwidth]{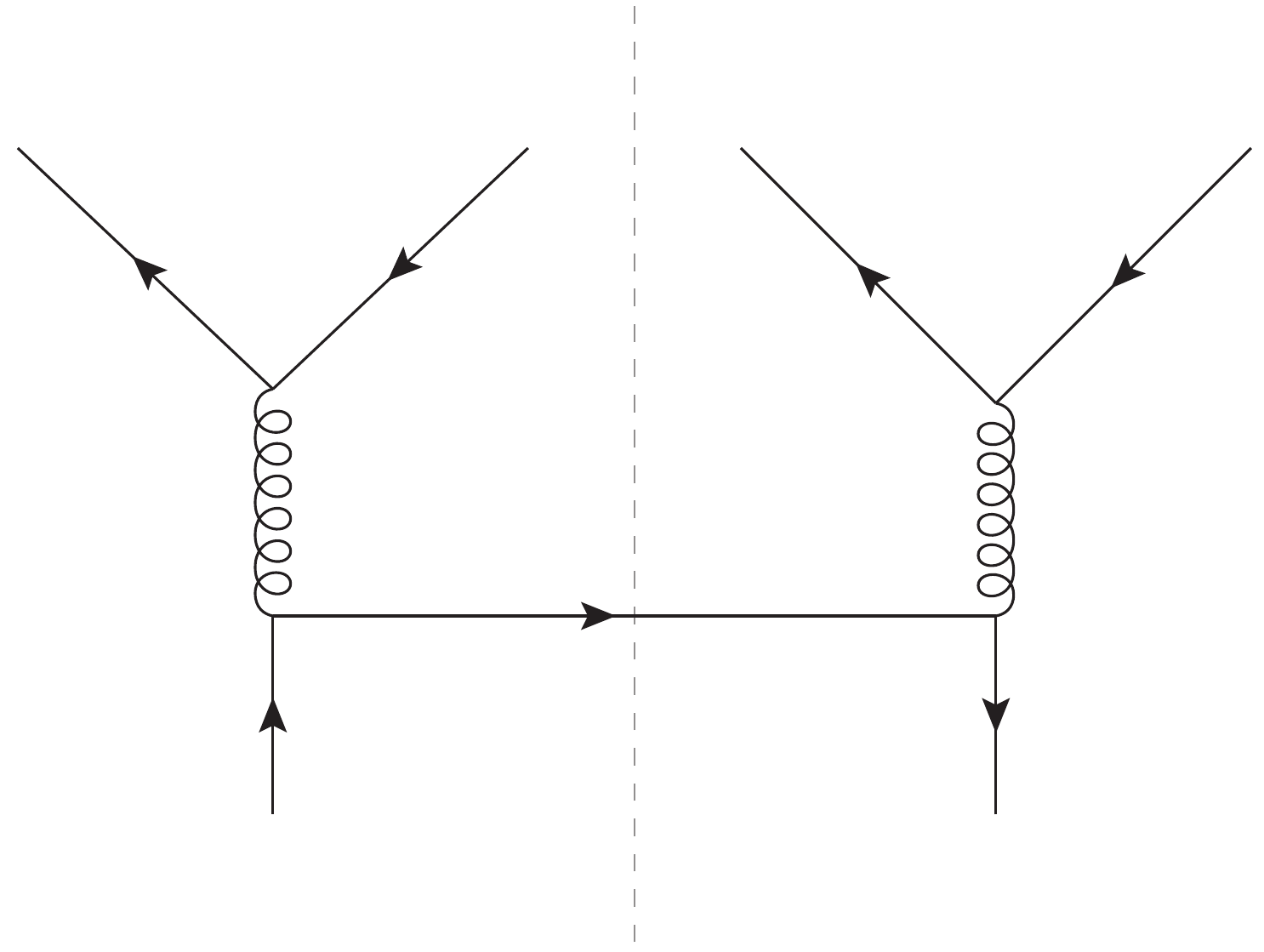}}
   \hspace{1.4em}
   \subfigure[\label{fig:qqbar-2} $V_{q \bar{q}'\!,q}$]{\includegraphics[width=0.25\textwidth]{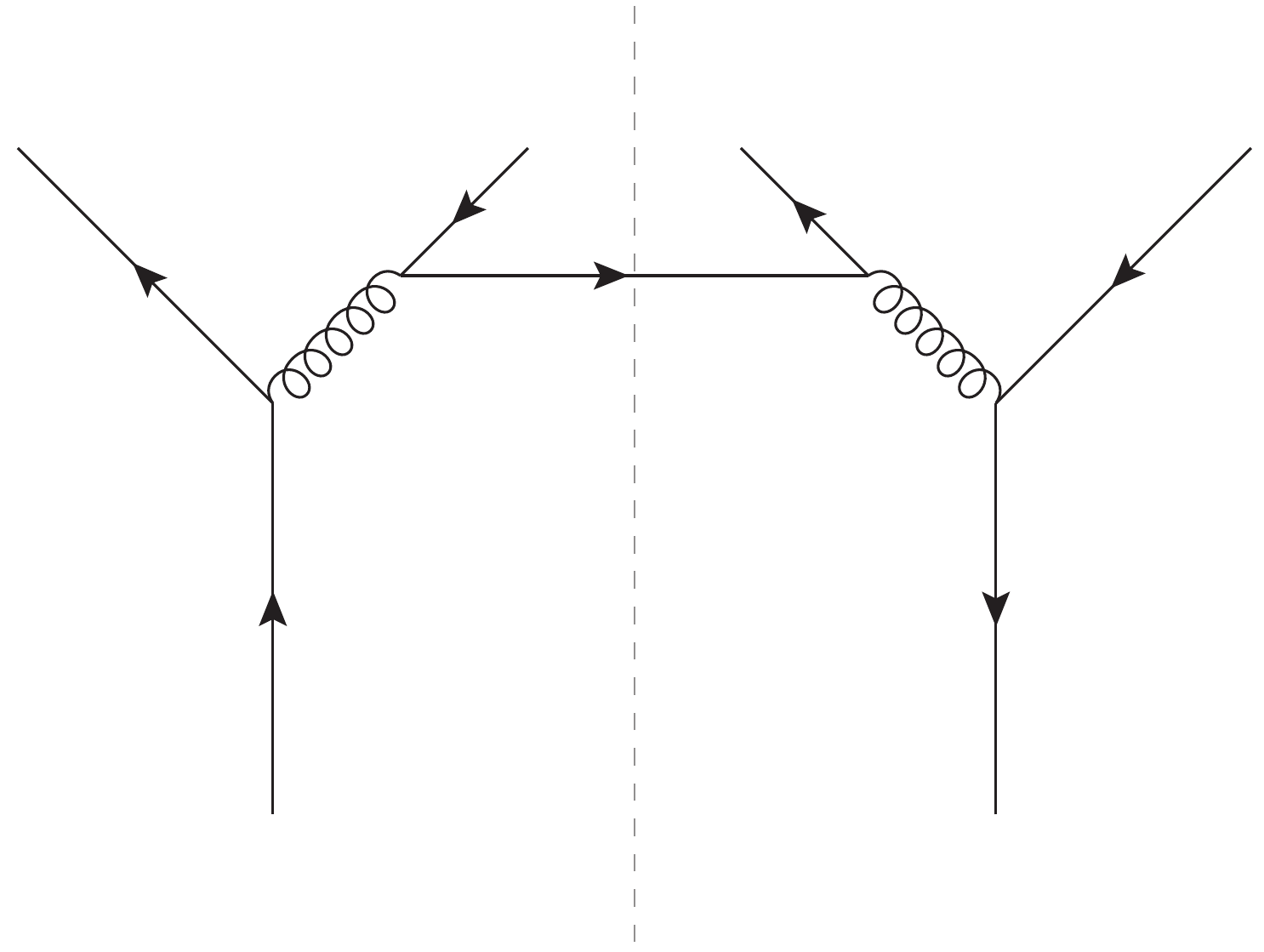}}
   \hspace{1.4em}
   \subfigure[\label{fig:qqbar-3} $V_{q \bar{q},q}^v$]{\includegraphics[width=0.25\textwidth]{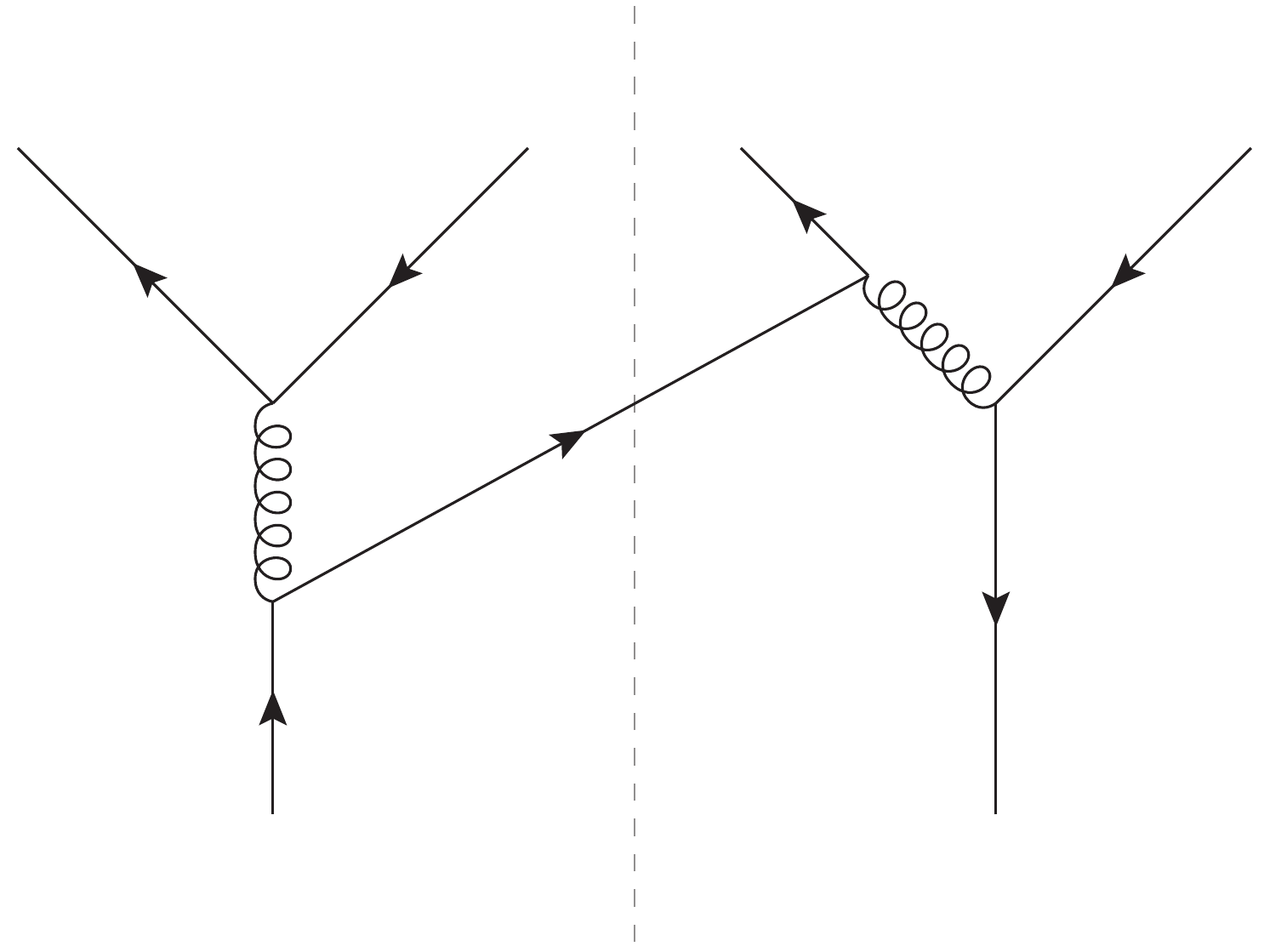}}
   \caption{\label{fig:real-NLO} Real graphs for the splitting $a_0 \to a_1 a_2$ in the pure quark channels.  The parton lines at the bottom of the graph correspond to $a_0$, and those on top to $a_1$, $a_2$, $a_2$ and $a_1$ from left to right.  Additional graphs are obtained by interchanging $a_1 \leftrightarrow a_2$ and by reflection w.r.t.\ the final state cut.}
   \end{center}
\end{figure}

\paragraph{Parton combinations appearing first at two loops.}
At order $a_s^2$ the distributions $F_{u d}$ and $F_{u \smash{\bar{d}}}$ receive a contribution only from the kernels $V_{q q'\!,q}$ or $V_{q \bar{q}'\!,q}$, which respectively correspond to the graphs in \fig{\ref{fig:qq-1}} and \ref{fig:qqbar-2} and further graphs with identical topology.  As a consequence, distributions for different colour representations are proportional to each other, as specified in \eqn{(4.40)} of \cite{Diehl:2021wpp}.  This leads to the relations in \eqref{ud-channels}.

Furthermore, the splitting kernels for $F_{u d}$ and $F_{u \smash{\bar{d}}}$ are proportional to each other, because $\pr{11}{V}_{q q'\!,q} = \pr{11}{V}_{q \bar{q}'\!,q}$ at two-loop accuracy.  Using the basis transform \eqref{quark-trfs} and the proportionality factors between $t$ channel singlet and octet kernels in \eqn{(4.35)} of \cite{Diehl:2021wpp}, one obtains
\begin{align}
   \label{Fud}
F_{u d}^{\overline{3} 3}
&= \frac{1}{\pi\ms y^2}
  \Bigl[\ms V_{q q'\!, q}^{\overline{3} 3}(z_1,z_2) \conv{} f_u(z)
          + V_{q q'\!, q}^{\overline{3} 3}(z_2,z_1) \conv{} f_d(z)
  \ms\Bigr] \,,
\nonumber \\
F_{u \bar{d}}^{1 1}
&= \frac{1}{\pi\ms y^2}
   \Bigl[\ms V_{q \bar{q}'\!, q}^{1 1}(z_1,z_2) \conv{} f_u(z)
           + V_{q \bar{q}'\!, q}^{1 1}(z_2,z_1) \conv{} f_{\bar{d}}(z)
   \ms\Bigr]
\end{align}
with
\begin{align}
   \label{V_qqprime}
V_{q q'\!, q}^{\overline{3} 3}
&= \frac{2}{9} \, \pr{11}{V}_{q q'\!,q} \,,
&
V_{q \bar{q}'\!, q}^{1 1}
&= \frac{8}{9} \, \pr{11}{V}_{q q'\!,q} \,.
\end{align}
Here the convolution $V(z_2,z_1) \otimes f(z)$ is defined as in \eqref{conv-def} with $u$ and $1-u$ interchanged on the r.h.s.  In the last term of \eqref{Fud} we used the relation $V_{\bar{q} q'\!, \bar{q}} = V_{q \bar{q}'\!, q}$ from charge conjugation invariance.  Notice that the kernel $\pr{11}{V}_{q q'\!,q}$ includes an overall factor $a_s^2$.

The distributions $F_{u u}^{\overline{3} 3}$ and $F_{u u}^{6 \overline{6}}$ receive contributions from the kernels $V_{q q'\!,q}$ and $V_{q q,q}^v$ with different weights and are therefore not proportional to each other.  Using \eqn{(3.3)} in \cite{Diehl:2021wpp}, we find that the relevant kernels read
\begin{align}
   \label{V_qq}
V_{q q, q}^{\overline{3} 3}(z_1,z_2)
&= \frac{4}{18} \, \Bigl[ \pr{11}{V}_{q q'\!,q}(z_1,z_2)
                        + \pr{11}{V}_{q q'\!,q}(z_2,z_1) \Bigr]
   + \frac{4}{6} \, \pr{11}{V}_{q q,q}^v(z_1,z_2) \,,
\nonumber \\[0.2em]
V_{q q, q}^{6 \overline{6}}(z_1,z_2)
&= \frac{1}{18} \, \Bigl[ \pr{11}{V}_{q q'\!,q}(z_1,z_2)
                        + \pr{11}{V}_{q q'\!,q}(z_2,z_1) \Bigr]
   - \frac{1}{6} \, \pr{11}{V}_{q q,q}^v(z_1,z_2)
\end{align}
and are to be convolved with $f_u(z)$.

At order $a_s^2$, the kernel $\pr{11}{V}_{q q'\!,q}$ and hence all kernels in \eqref{V_qqprime} and \eqref{V_qq} require ultraviolet renormalisation but no subtraction of rapidity divergences.  As a consequence, they depend linearly on the renormalisation group logarithm $L_y$.
A natural choice of scale in the fixed-order formula \eqref{split-master} is $\mu = \mu_y$, so that $L_y = 0$.  In a numerical study of the two-gluon distributions $\pr{R \Rbar}{F}_{g g}$ in \cite{Diehl:2021wpp}, we found that the size of $a_s^2$ corrections relative to the $a_s$ term is moderate for $\mu = \mu_y$ but grows substantially for $\mu = \mu_y / 2$ or $\mu = 2 \mu_y$.  In the present work, we therefore consider a smaller amount of variation and take $\mu = 1.2 \ms\mu_y$ or $\mu = \mu_y /1.2$ as alternative scales, which corresponds to $L_y \approx \pm 0.36$.  We checked that the size of $a_s^2$ corrections in the two-gluon sector remains moderate for these values.
The DPDs depend on $\mu$ as specified by the relevant DGLAP equations.  When comparing the fixed-order formula \eqref{split-master} for different $\mu$, one thus sees evolution effects truncated at the lowest order in $a_s$.

\begin{figure}[p]
   \begin{center}
   \subfigure[$\pr{11}{V}_{q q'\!, q}$ at $\mu = \mu_y$]{\includegraphics[height=0.28\textwidth,trim=0 0 95 0,clip=true]{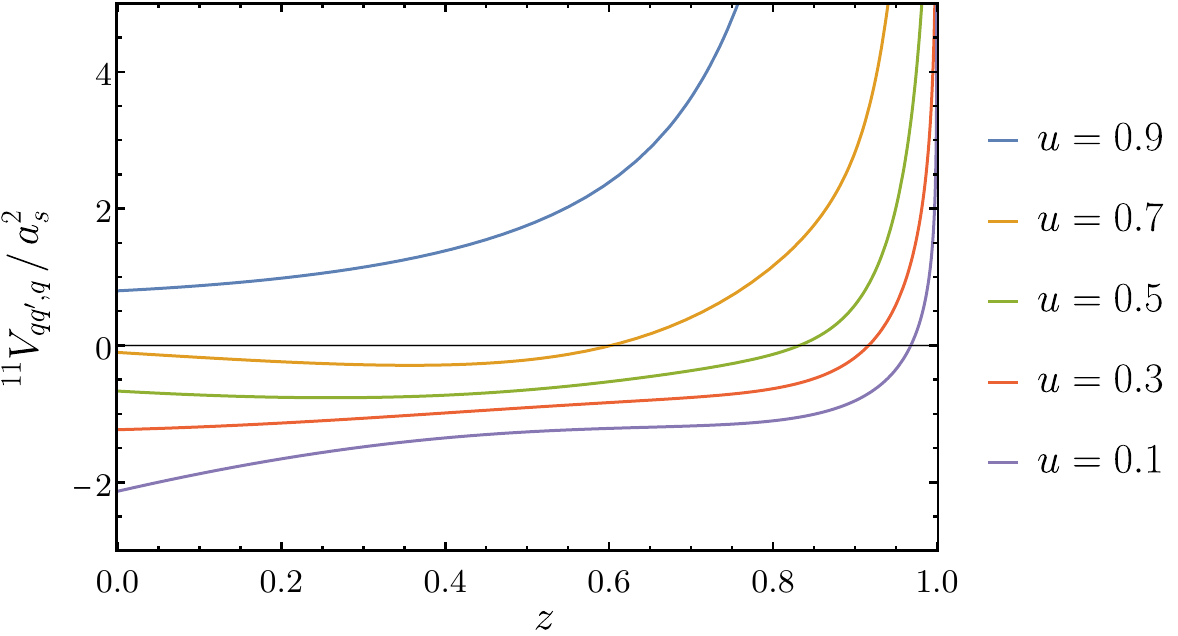}}
   \hspace{1em}
   \subfigure[$\pr{11}{V}_{q q'\!, q}$ at $\mu = 1.2 \ms \mu_y$]{\includegraphics[height=0.28\textwidth,trim=40 0 0 0,clip=true]{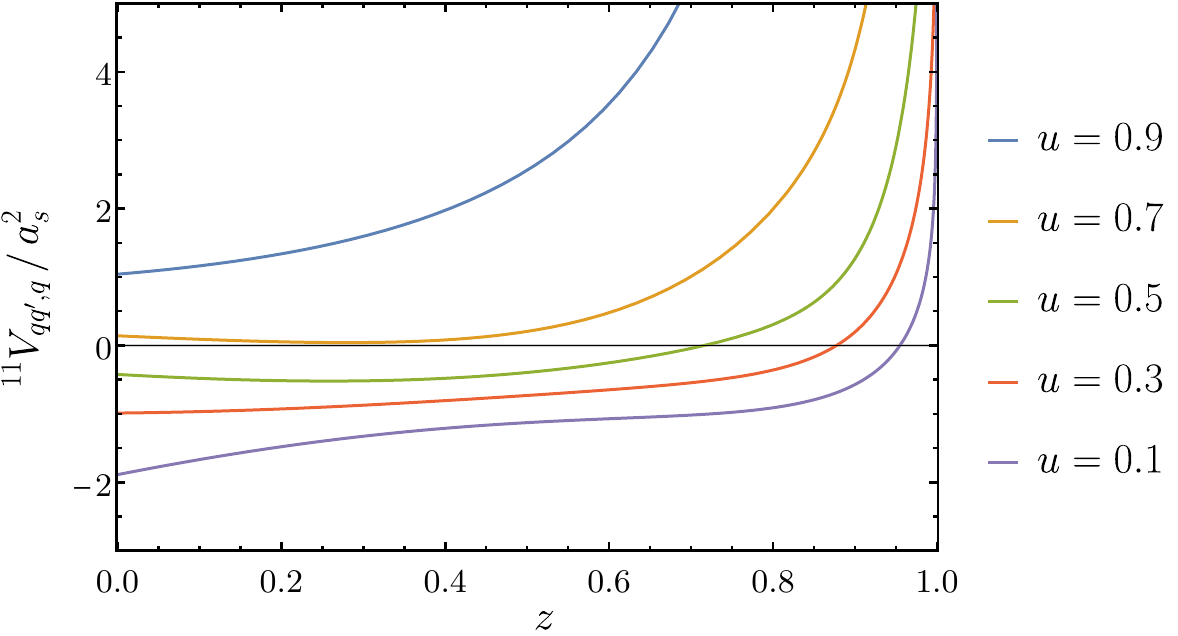}}
   \caption{\label{fig:V_qqprime} The kernel $\pr{11}{V}_{q q'\!, q}(z_1,z_2)$, which appears in the DPDs for the parton combinations $u d$ and $u \bar{d}$ as specified in \protect\eqref{ud-channels}, \protect\eqref{Fud}, and \protect\eqref{V_qqprime}.  A global factor $a_s^2$ has been divided out.  See \protect\eqref{zu-relations} for the relation between $z_1, z_2$ and $u, z$.  The order of curves from top to bottom is identical in the plots and in the legend.}
   \end{center}
%\end{figure}
\vspace{1.5em}
%\begin{figure}[p]
   \begin{center}
   \subfigure[$V_{q q, q}^{\overline{3} 3}$ for $\mu = \mu_y$]{\includegraphics[height=0.28\textwidth,trim=0 0 95 0,clip=true]{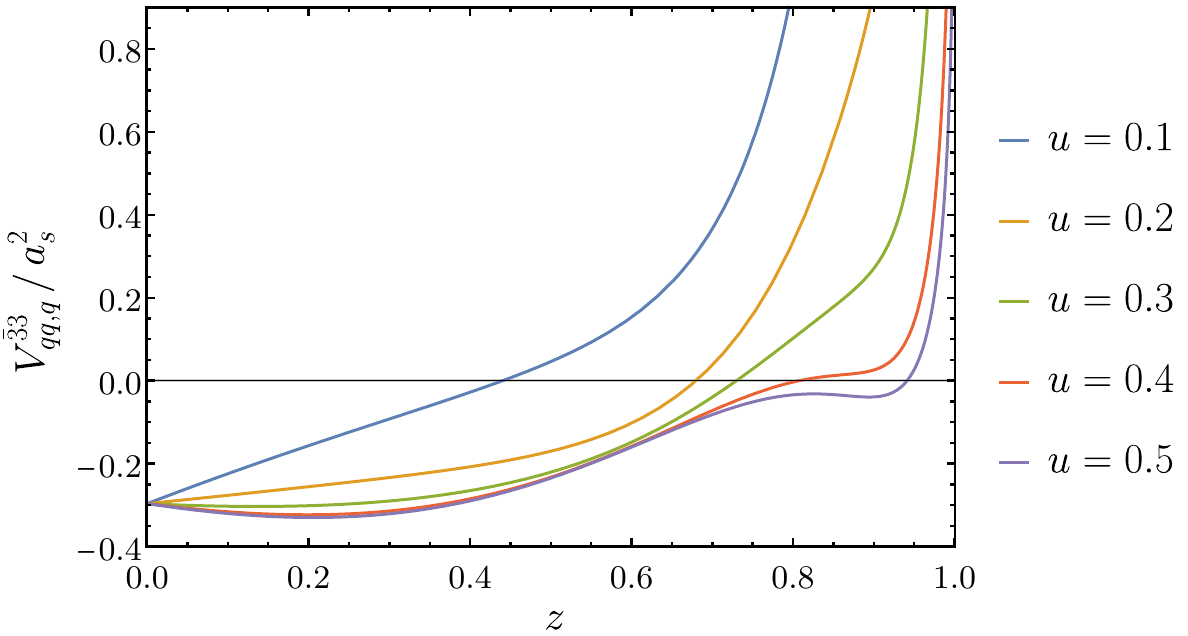}}
   \hspace{1em}
   \subfigure[$V_{q q, q}^{\overline{3} 3}$ for $\mu = 1.2 \ms \mu_y$]{\includegraphics[height=0.28\textwidth,trim=40 0 0 0,clip=true]{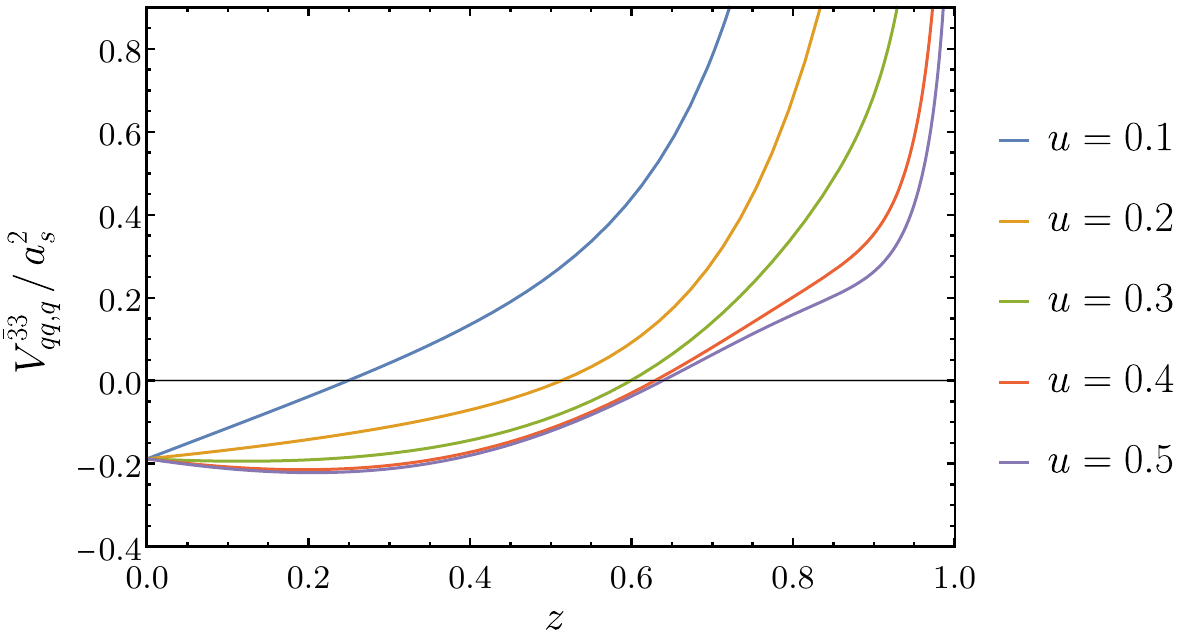}}
\\[0.5em]
   \subfigure[$V_{q q, q}^{6 \overline{6}}$ for $\mu = \mu_y$]{\includegraphics[height=0.28\textwidth,trim=0 0 95 0,clip=true]{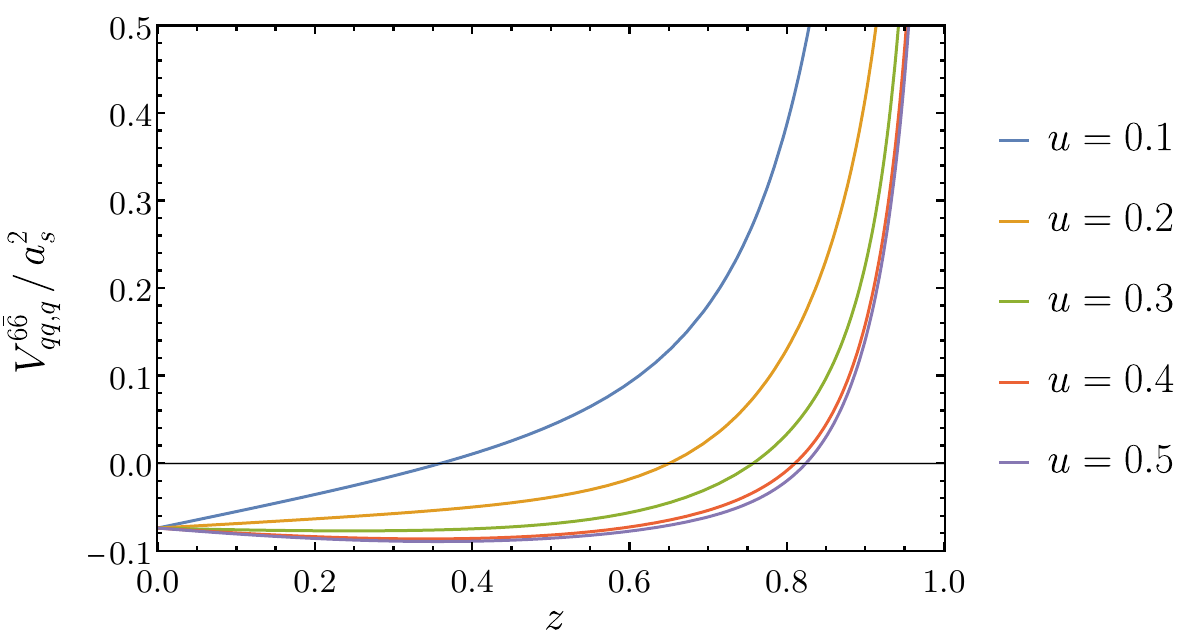}}
   \hspace{1em}
   \subfigure[$V_{q q, q}^{6 \overline{6}}$ for $\mu = 1.2 \ms \mu_y$]{\includegraphics[height=0.28\textwidth,trim=40 0 0 0,clip=true]{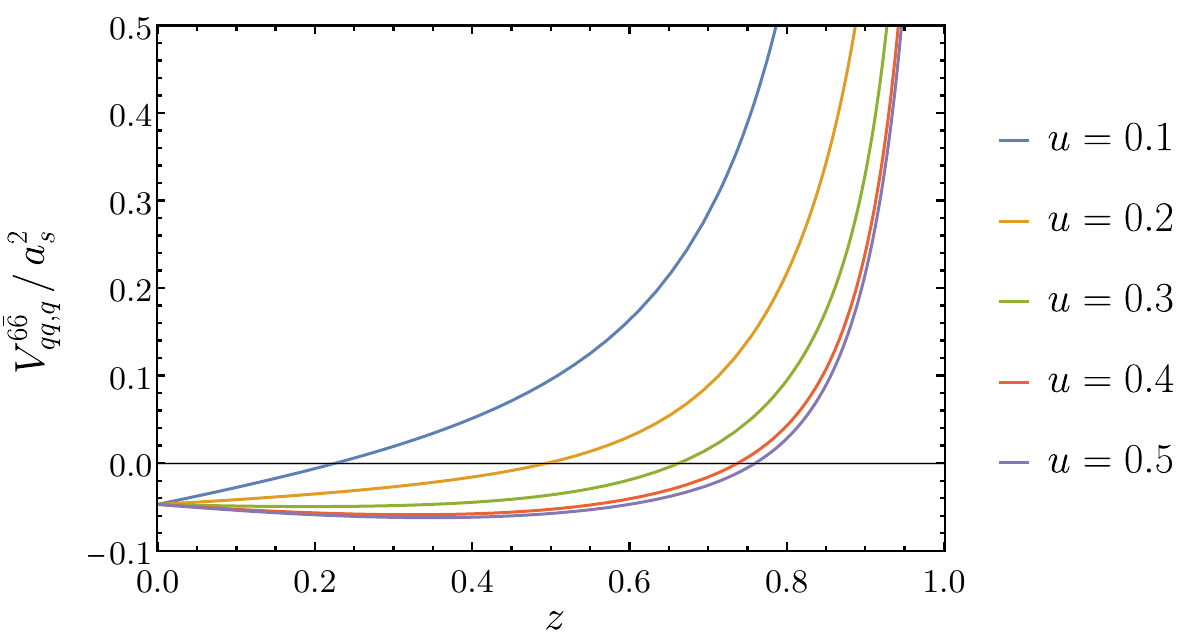}}
   \caption{\label{fig:V_qq} As \fig{\protect\ref{fig:V_qqprime}} but for the kernels in \protect\eqref{V_qq}, which give the DPDs for the parton combination $u u$.  Since these kernels are symmetric under the interchange $z_1 \leftrightarrow z_2$ of the two momentum fractions, we can limit ourselves to the region $u \le 0.5$}
   \end{center}
\end{figure}

In \figs{\ref{fig:V_qqprime}} and \ref{fig:V_qq}, we plot the kernels $\pr{11}{V}_{q q'\!, q}$, $V_{q q, q}^{\overline{3} 3}$, and $V_{q q, q}^{6 \overline{6}}$.  We see that they are negative over wide ranges of $z$ and $u$, although they are obtained from a sum of graphs that correspond to the squared amplitude for $q\to q  q' \bs\bar{q}'$ or $q\to q q \bar{q}$.  This illustrates that the subtraction of ultraviolet divergences can indeed lead to a negative result.  We note that the kernels go to large positive values for $z\to 1$, which can be traced back to terms diverging like $\log 1/(1-z)$ in that limit.  Finally, we observe that the kernels increase with $\mu$ for all $z$ and $u$.

We now investigate to which extent the negative regions in the splitting kernels lead to negative DPDs.  To this end, we evaluate the splitting formula for the distributions in \eqref{ud-channels} and \eqref{uu-channels} with the central PDFs from the \texttt{CT14nlo} set, having checked that the PDFs are positive in the kinematics of interest.  As we did in \sect{\ref{sec:dglap-evol}}, we fix $y$ such that $\mu_y = 10 \gev$.

\begin{figure}
\begin{center}
   \subfigure[$x_2 = x_1$, $\mu = \mu_y / 1.2$]{\includegraphics[height=0.3\textwidth,
      trim=0 0 0 16,clip=true]{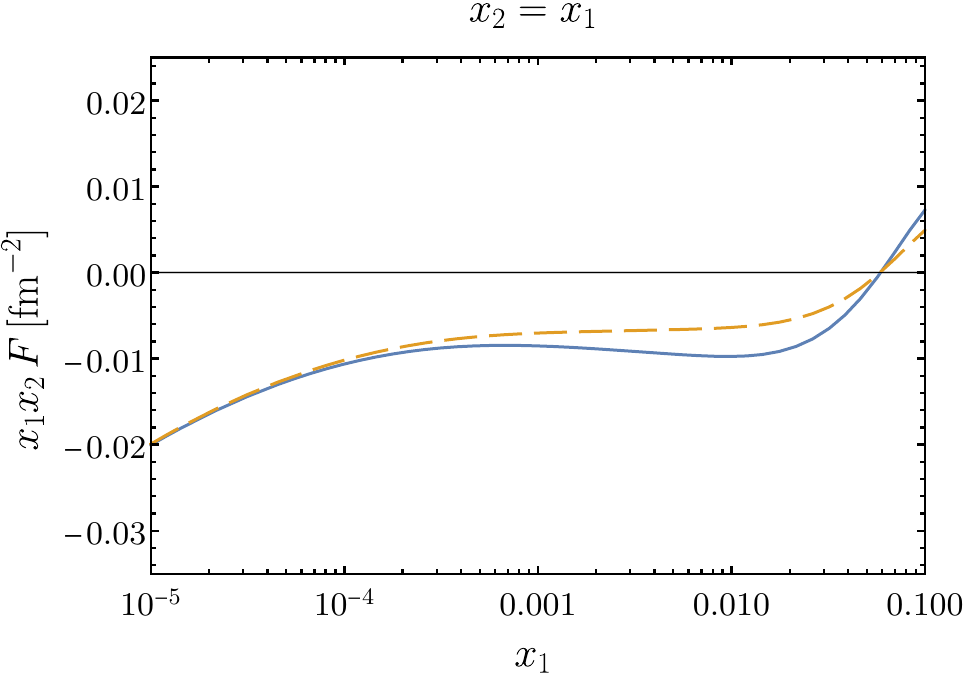}
   \hspace{1em}
   \includegraphics[height=0.3\textwidth,
      trim=30 0 0 16,clip=true]{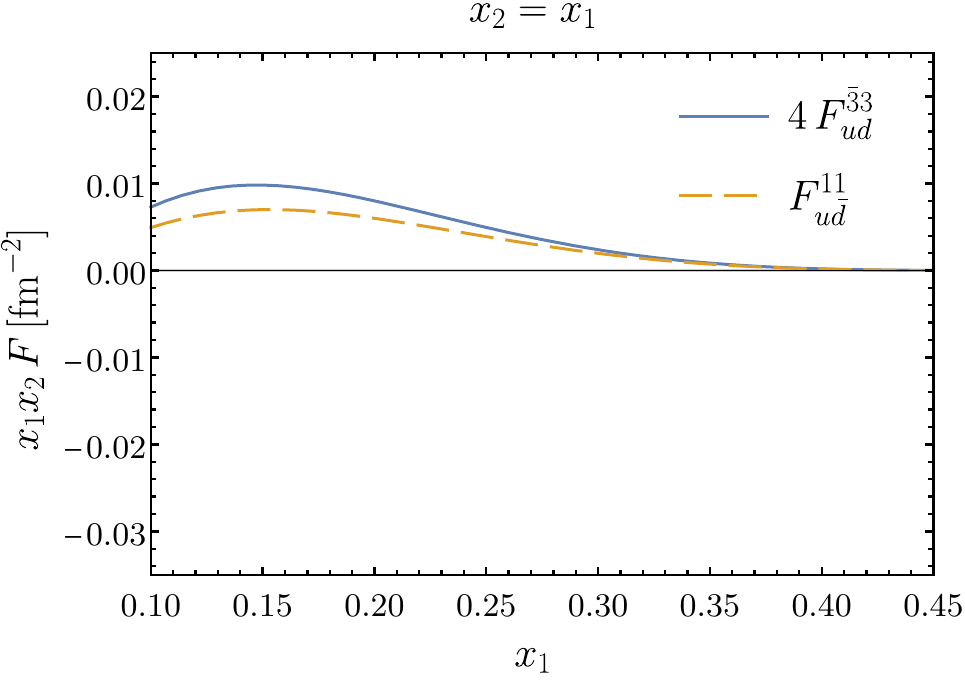}}
\\[0.5em]
   \subfigure[$x_2 = x_1$, $\mu = \mu_y$]{\includegraphics[height=0.3\textwidth,
      trim=0 0 0 16,clip=true]{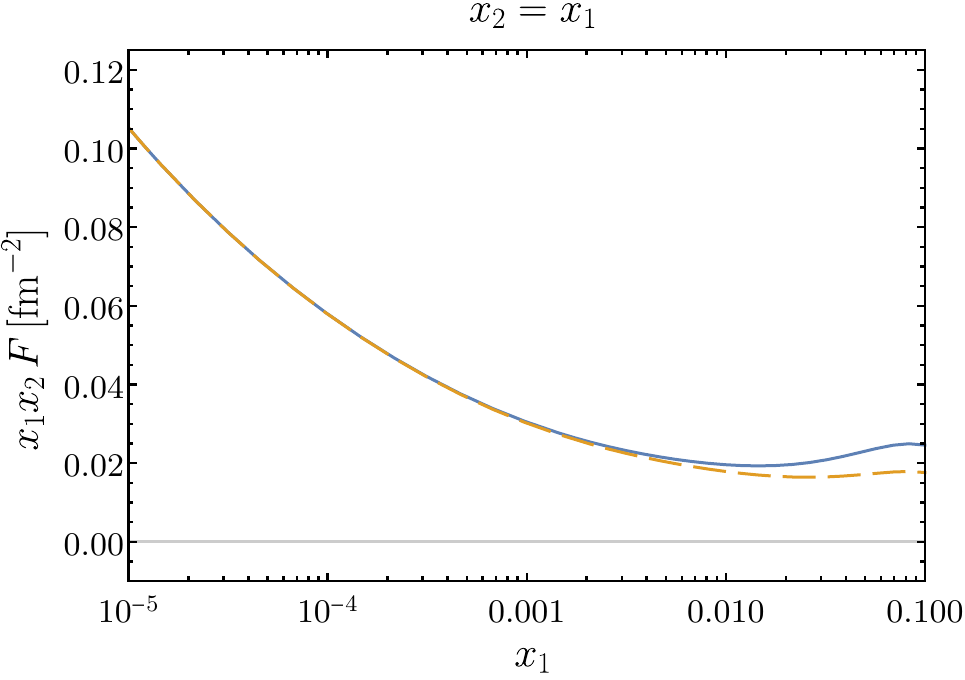}
   \hspace{1em}
   \includegraphics[height=0.3\textwidth,
      trim=30 0 0 16,clip=true]{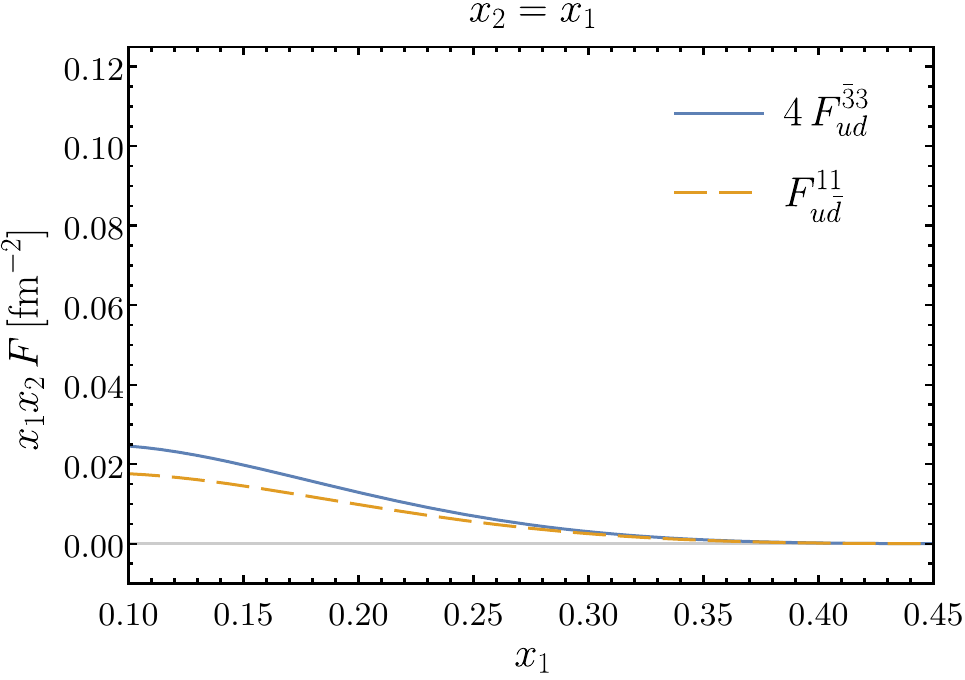}}
\\[0.5em]
   \subfigure[\label{fig:dip-Fud}$x_2 = 0.01$, $\mu = \mu_y / 1.2$]{\includegraphics[height=0.3\textwidth,
      trim=0 0 0 24,clip=true]{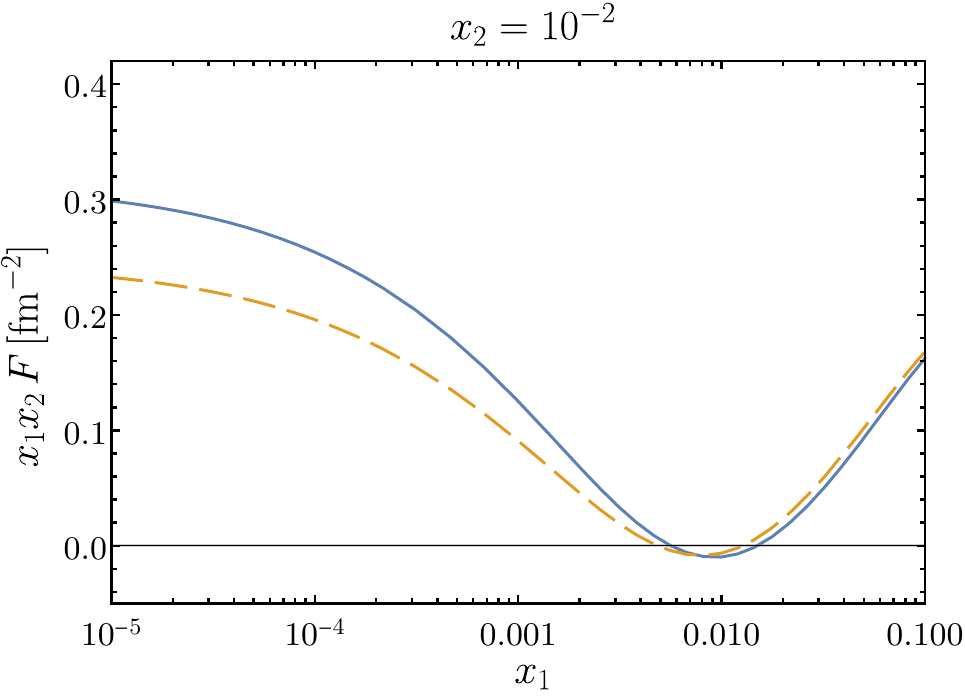}
   \hspace{1em}
   \includegraphics[height=0.3\textwidth,
      trim=30 0 0 24,clip=true]{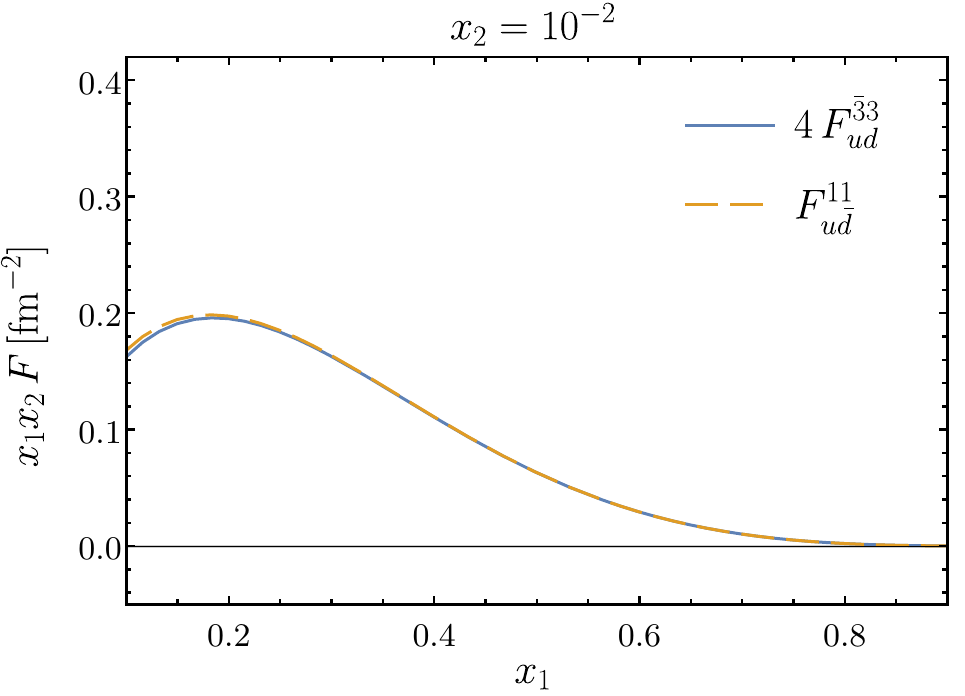}}
\\[0.5em]
   \subfigure[$x_2 = 0.01$, $\mu = \mu_y$]{\includegraphics[height=0.3\textwidth,
      trim=0 0 0 24,clip=true]{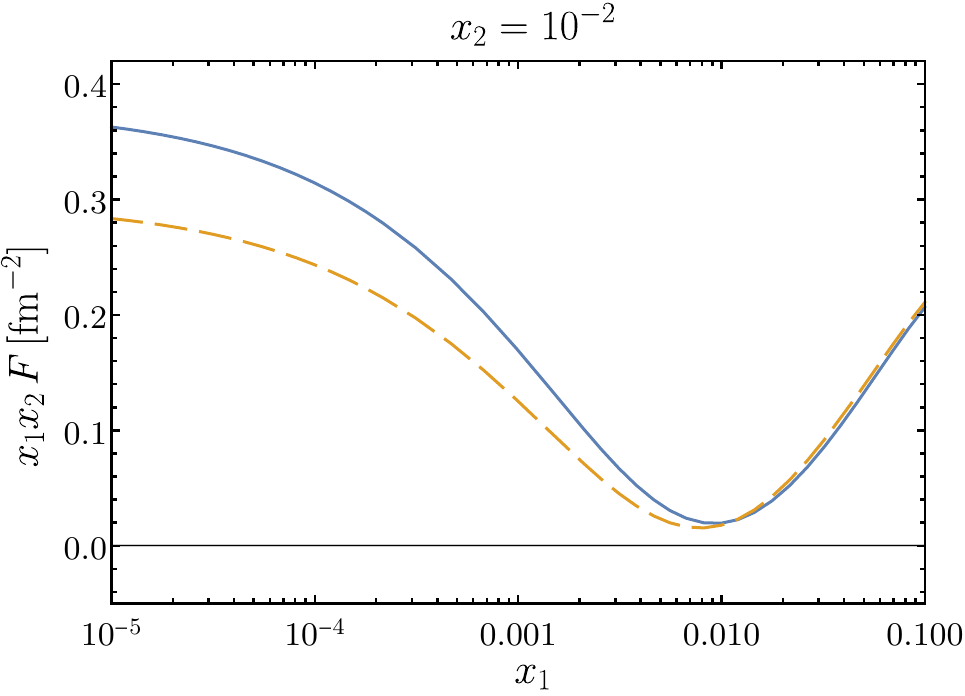}
   \hspace{1em}
   \includegraphics[height=0.3\textwidth,
      trim=30 0 0 24,clip=true]{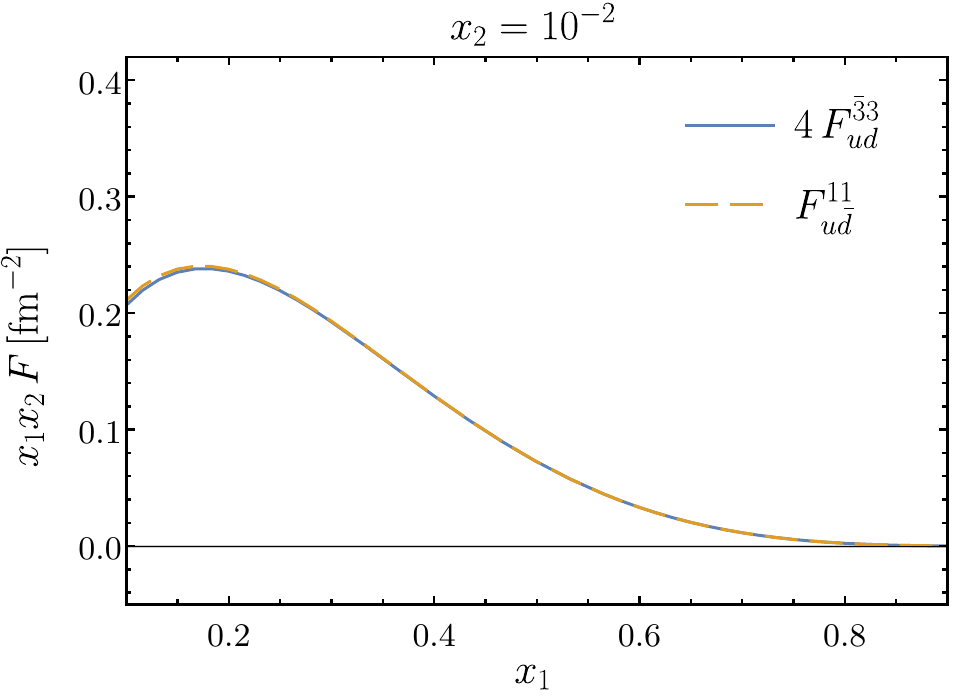}}
\caption{\label{fig:Fud} The distributions $F_{u d}^{\overline{3} 3}$ and $F_{u \bar{d}}^{1 1}$ computed with the two-loop splitting formula for $\mu_y = 10 \gev$ and two choices of $\mu$.  The other two colour channels are given in \eqref{ud-channels}.}
\end{center}
\end{figure}

\begin{figure}
\begin{center}
   \subfigure[$x_2 = x_1$, $\mu = \mu_y$]{\includegraphics[height=0.3\textwidth,
      trim=0 0 0 16,clip=true]{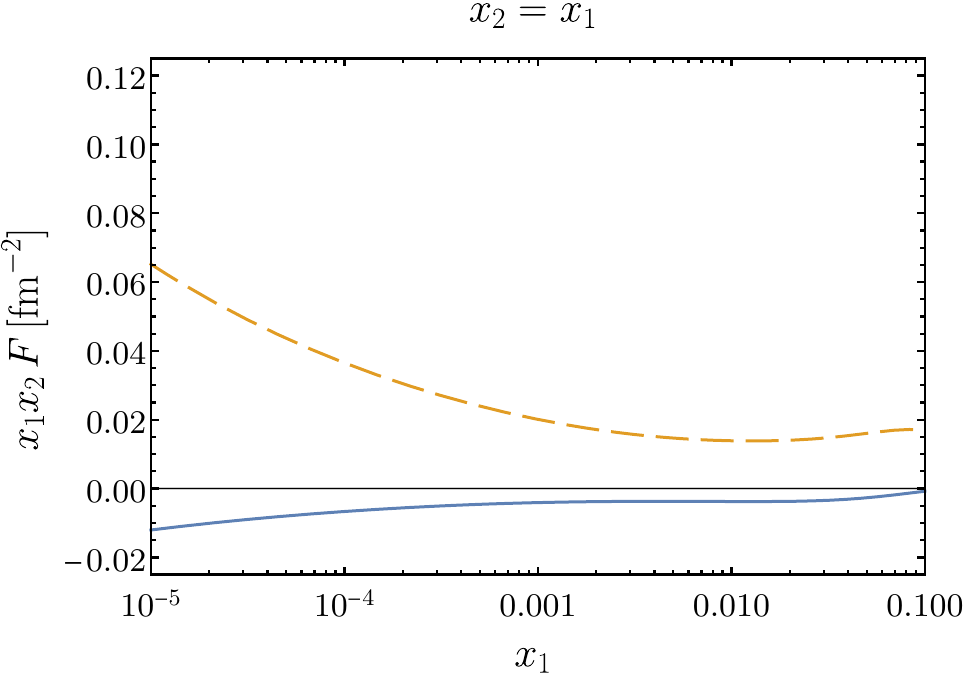}
   \hspace{1em}
   \includegraphics[height=0.3\textwidth,
      trim=30 0 0 16,clip=true]{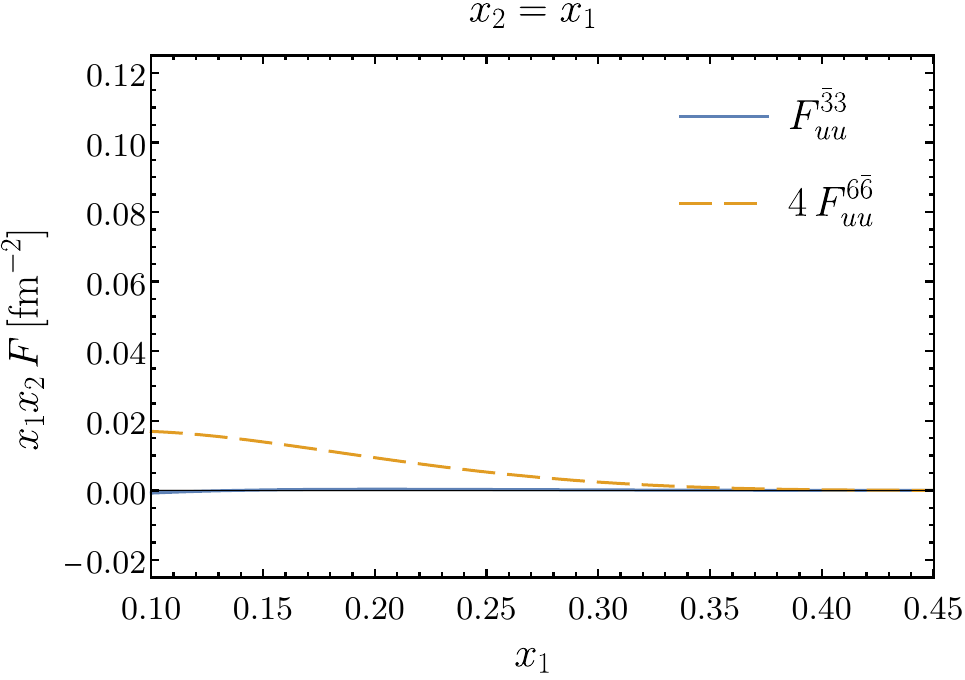}}
\\[0.5em]
   \subfigure[$x_2 = x_1$, $\mu = 1.2 \ms\mu_y$]{\includegraphics[height=0.3\textwidth,
      trim=0 0 0 16,clip=true]{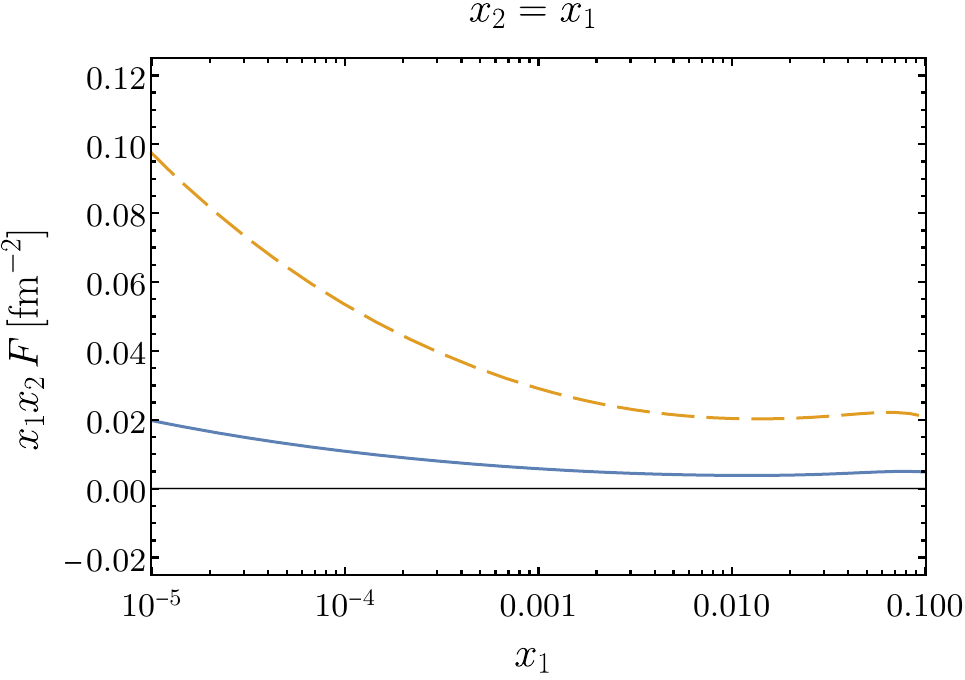}
   \hspace{1em}
   \includegraphics[height=0.3\textwidth,
      trim=30 0 0 16,clip=true]{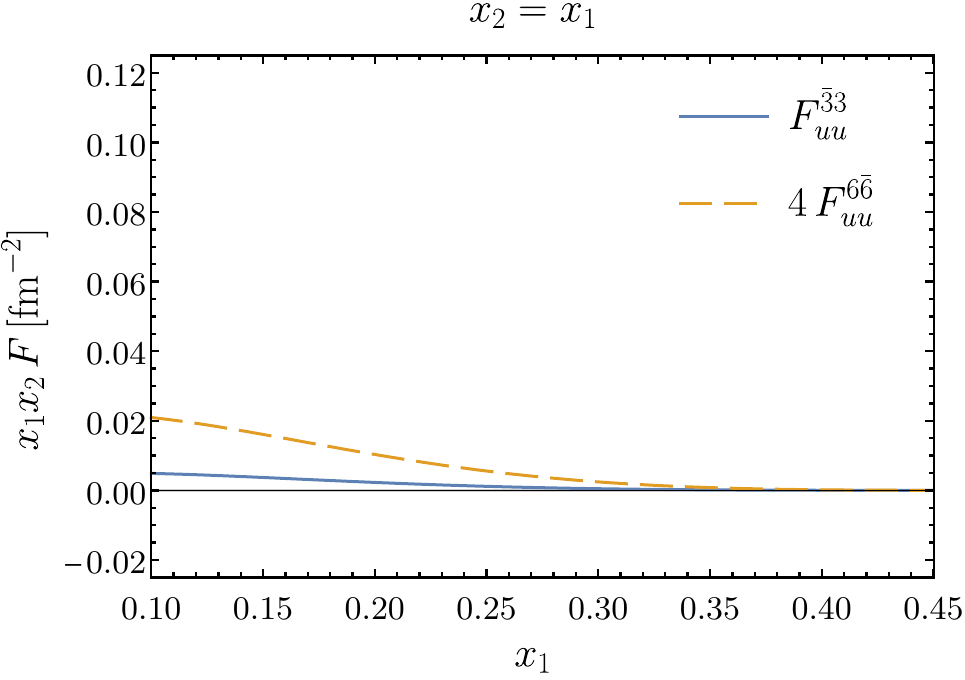}}
\\[0.5em]
   \subfigure[\label{fig:dip-Fuu}$x_2 = 0.01$, $\mu = \mu_y$]{\includegraphics[height=0.3\textwidth,
      trim=0 0 0 24,clip=true]{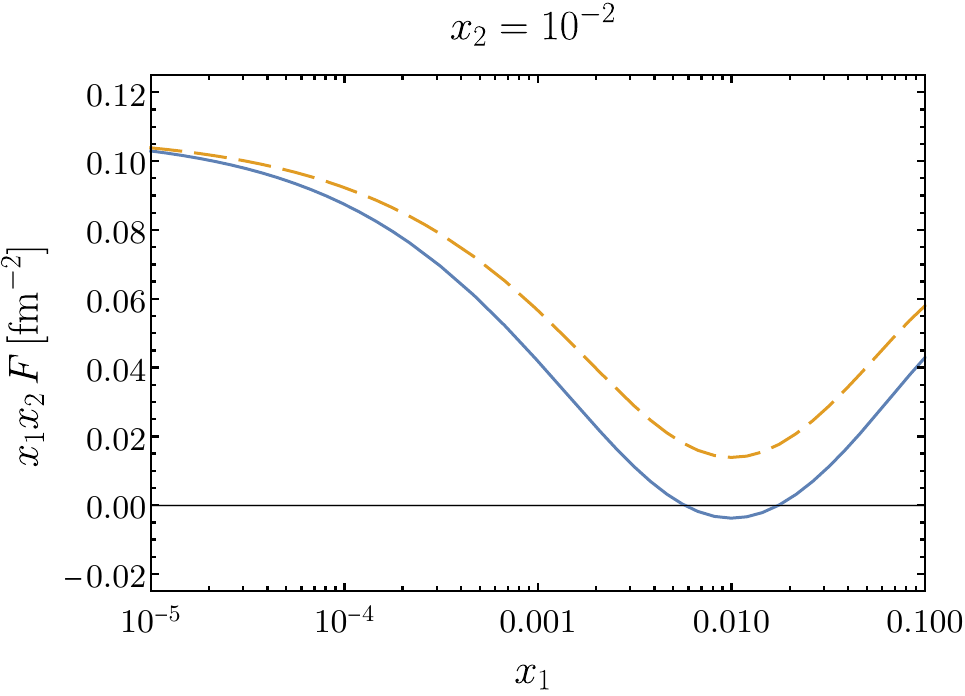}
   \hspace{1em}
   \includegraphics[height=0.295\textwidth,
      trim=30 0 0 24,clip=true]{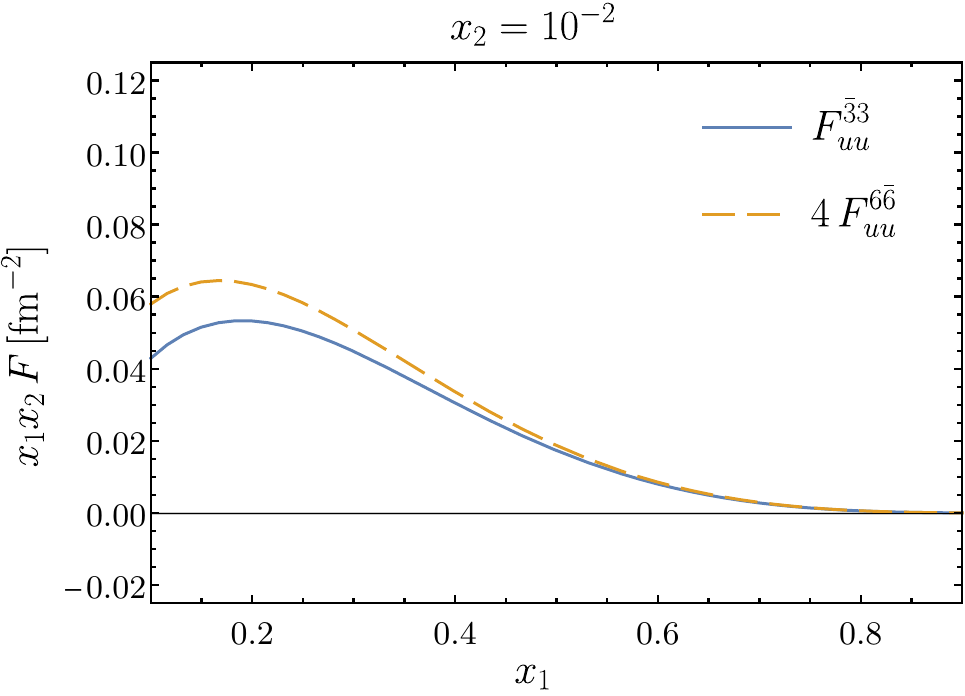}}
\\[0.5em]
   \subfigure[$x_2 = 0.01$, $\mu = 1.2 \ms\mu_y$]{\includegraphics[height=0.3\textwidth,
      trim=0 0 0 24,clip=true]{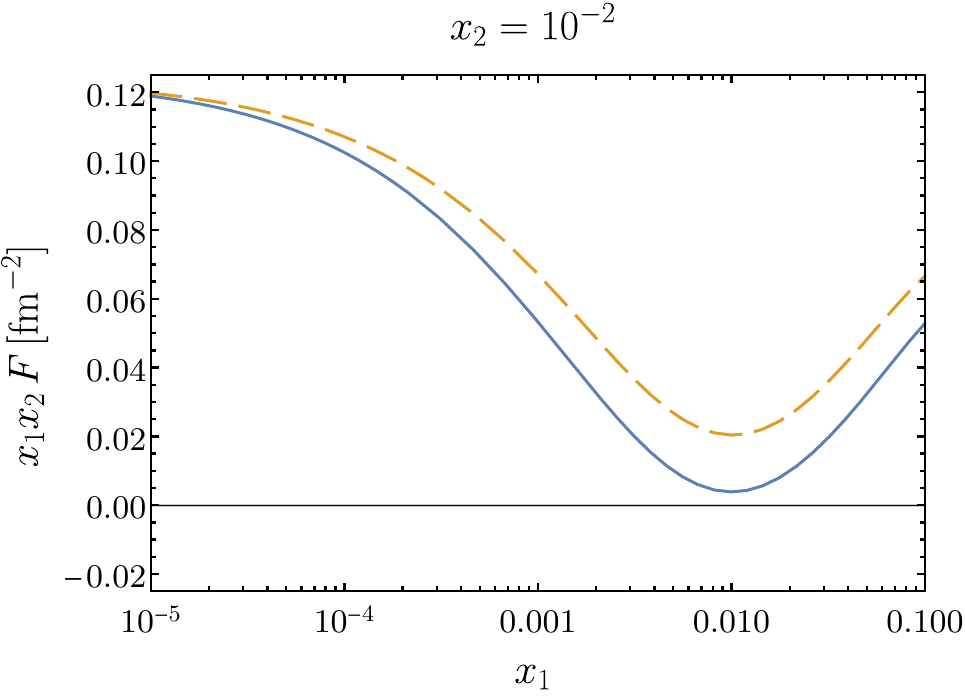}
   \hspace{1em}
   \includegraphics[height=0.295\textwidth,
      trim=30 0 0 24,clip=true]{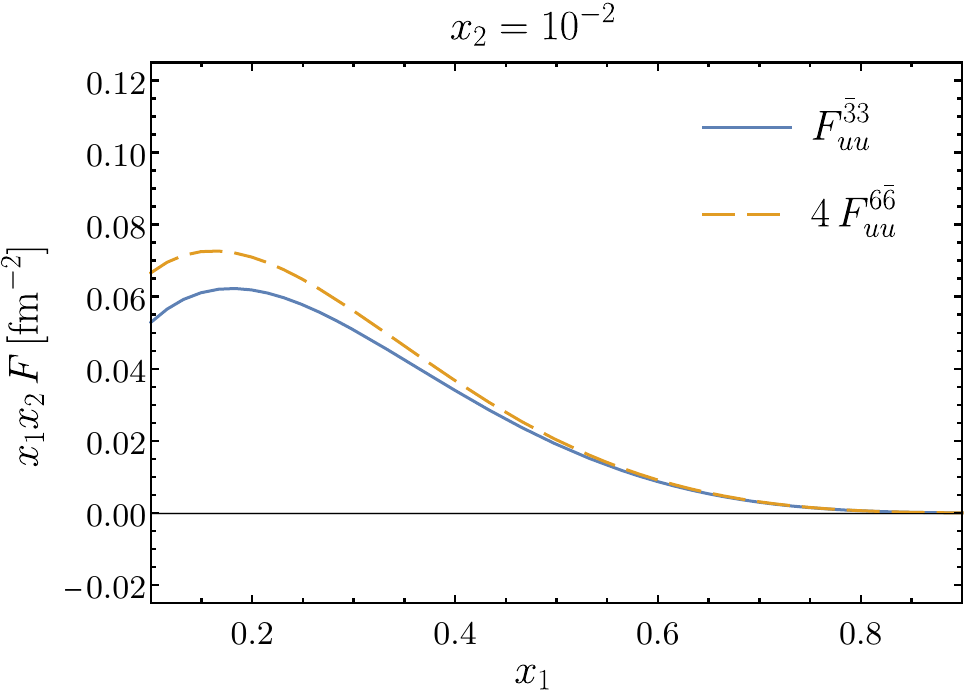}}
\caption{\label{fig:Fuu} The distributions $F_{u u}^{\overline{3} 3}$ and $F_{u u}^{6 \overline{6}}$ computed with the two-loop splitting formula for $\mu_y = 10 \gev$.  Note that the two values of $\mu$ are not the same as in \fig{\protect\ref{fig:Fud}}.}
\end{center}
\end{figure}

In \fig{\ref{fig:Fud}} we show $F_{u d}^{\overline{3} 3}$ and $F_{u \bar{d}}^{1 1}$ for $\mu = \mu_y /1.2$ and $\mu = \mu_y$.  The curves are scaled such that the difference between them originates from the different PDFs in \eqref{Fud} and not from the different normalisation of the kernels in \eqref{V_qqprime}.  We observe a strong effect of scale evolution: for $x_1 \sim x_2$ the distributions are negative at the smaller scale but positive at $\mu = \mu_y$ (and also at $\mu = 1.2 \ms\mu_y$, which is not shown in the figure).  We also note that the negative values in \fig{\ref{fig:dip-Fud}} are tiny compared with the size of the same distributions at other values of $x_1$ for the same $x_2$.

The distributions $F_{u u}^{\overline{3} 3}$ and $F_{u u}^{6 \overline{6}}$ are shown in \fig{\ref{fig:Fuu}}, with a scaling factor such that the difference between the curves is due to the contribution of $\pr{11}{V}_{q q, q}^{v}$ to the kernels in \eqref{V_qq}.  The situation for $\mu = \mu_y /1.2$ (not shown in the figure) is qualitatively similar to the one at $\mu = \mu_y$, where we find negative values at $x_1 \sim x_2$ for $F_{u u}^{\overline{3} 3}$ but not for $F_{u u}^{6 \overline{6}}$.  At the higher scale $\mu = 1.2 \ms\mu_y$, all values are positive.  As in \fig{\ref{fig:dip-Fud}}, the negative values in \fig\ref{fig:dip-Fuu} are tiny compared with the size of the distribution at other momentum fractions.  In this sense, the violations of positivity we have shown so far may be regarded as minor.

\paragraph{Quark-antiquark distributions.}
The last two distributions in \eqref{uu-channels} are for a quark and an antiquark of equal flavour.  $F_{u \bar{u}}^{1 1}$ receives contributions from all three kernels in the bottom row of \fig{\ref{fig:real-NLO}} and from the real two-loop graphs for the splitting $g \to q \bar{q}$, an example of which is shown in \fig{\ref{fig:qqbar-NLO-re}}.  $F_{u \bar{u}}^{8 8}$ receives contributions from the same graphs, from the LO graph in \fig{\ref{fig:qqbar-LO}}, and from virtual two-loop graphs such as the one in \fig{\ref{fig:qqbar-NLO-vi}}.  The virtual graphs depend on the colour of the observed partons in the same way as the LO graph and hence do not contribute to $F_{u \bar{u}}^{1 1}$.

The distributions $F_{u \bar{u}}^{1 1}$ and $F_{u \bar{u}}^{8 8}$ require both ultraviolet renormalisation and the subtraction of rapidity divergences.  The latter appears in the splitting process $g \to q \bar{q}$, whose kernels have a more complicated structure than the ones considered so far.  Up to order $a_s^2$, they can be written as
\begin{align}
   \label{V_uubar}
V_{q \bar{q}, g}^{1 1}
&= a_s^2 \, D\; \pr{8}{\gamma}_J^{(0)} \,
   \delta(1-z)\, \pr{11}{V}_{q \bar{q}, g}^{(1)}(u)
   + a_s^2 \, \Bigl(
      V_{q \bar{q}, g}^{1 1 \, [2,0]}(z, u)
      + L_y \, V_{q \bar{q}, g}^{1 1 \, [2,1]}(z, u) \Bigr) \,,
\nonumber \\[0.3em]
V_{q \bar{q}, g}^{8 8}
&= \frac{a_s}{8} \, \Bigl( 1 - a_s \ms D \; \pr{8}{\gamma}_J^{(0)} \Bigr) \,
   \delta(1-z)\, \pr{11}{V}_{q \bar{q}, g}^{(1)}(u)
   + a_s^2 \, \Bigl(
      V_{q \bar{q}, g}^{8 8 \, [2,0]}(z, u)
      + L_y \, V_{q \bar{q}, g}^{8 8 \, [2,1]}(z, u) \Bigr) \,,
\nonumber \\
\end{align}
where $\pr{11}{V}_{q \bar{q}, g}^{(1)}(u) = \bigl[ u^2 + (1-u)^2 \ms\bigr] \big/ 2$ is the LO splitting kernel appearing in \eqref{split-LO} and
\begin{align}
   \label{double-log}
D &= \frac{1}{36} \ms
     \biggl( L_y^2 - 2 L_y \ms L_\zeta + \frac{\pi^2}{6} \biggr)
\end{align}
contains the double logarithms associated with rapidity divergences.  We note that \eqref{double-log} is obtained with the standard definition of the \msbar scheme, and that the term $\pi^2 /6$ is absent if one instead uses the definition proposed by Collins in \sect{3.2.6} of \cite{Collins:2011zzd}.
The coefficients $V^{[2,0]}$ and $V^{[2,1]}$ in \eqref{V_uubar} are smooth functions of $u$ but  distributions in $z$.  They consist of a regular part, a part proportional to the plus-distribution $1 / [1-z]_+$, and a term proportional to $\delta(1-z)$.  The regular part is a smooth function of $z$ but may have a $\log(1-z)$ singularity for $z \to 1$, similar to what we saw for the pure quark kernels in \figs{\ref{fig:V_qqprime}} and \ref{fig:V_qq}.

\begin{figure}
\begin{center}
   \subfigure[$\mu = \mu_y$]{\includegraphics[height=0.28\textwidth,
      trim=0 0 160 16,clip=true]{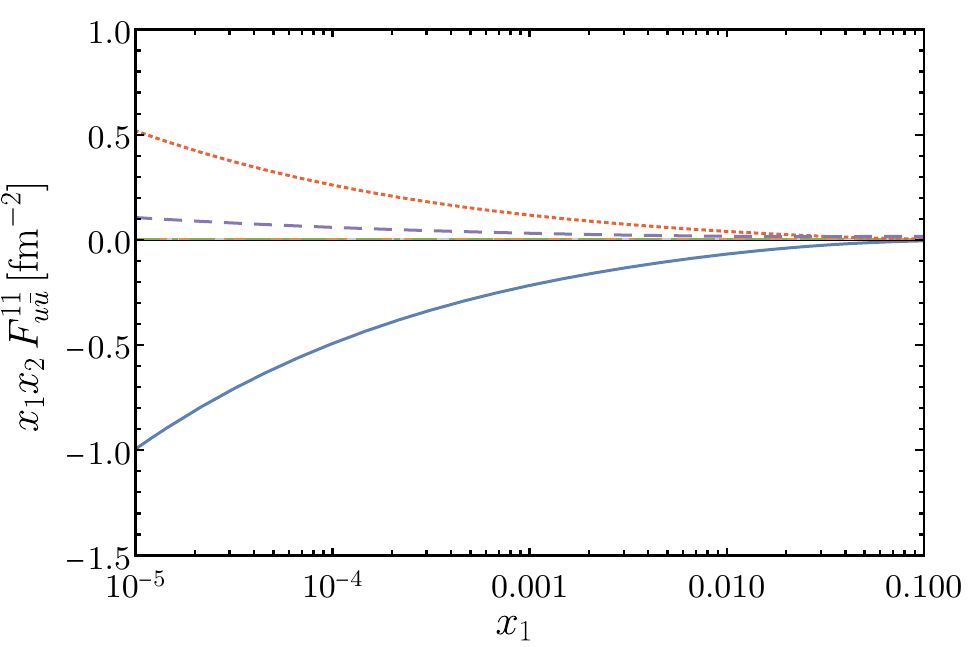}}
   \hspace{0.1em}
   \subfigure[$\mu = 1.2 \ms\mu_y$]{\includegraphics[height=0.28\textwidth,
      trim=30 0 0 16,clip=true]{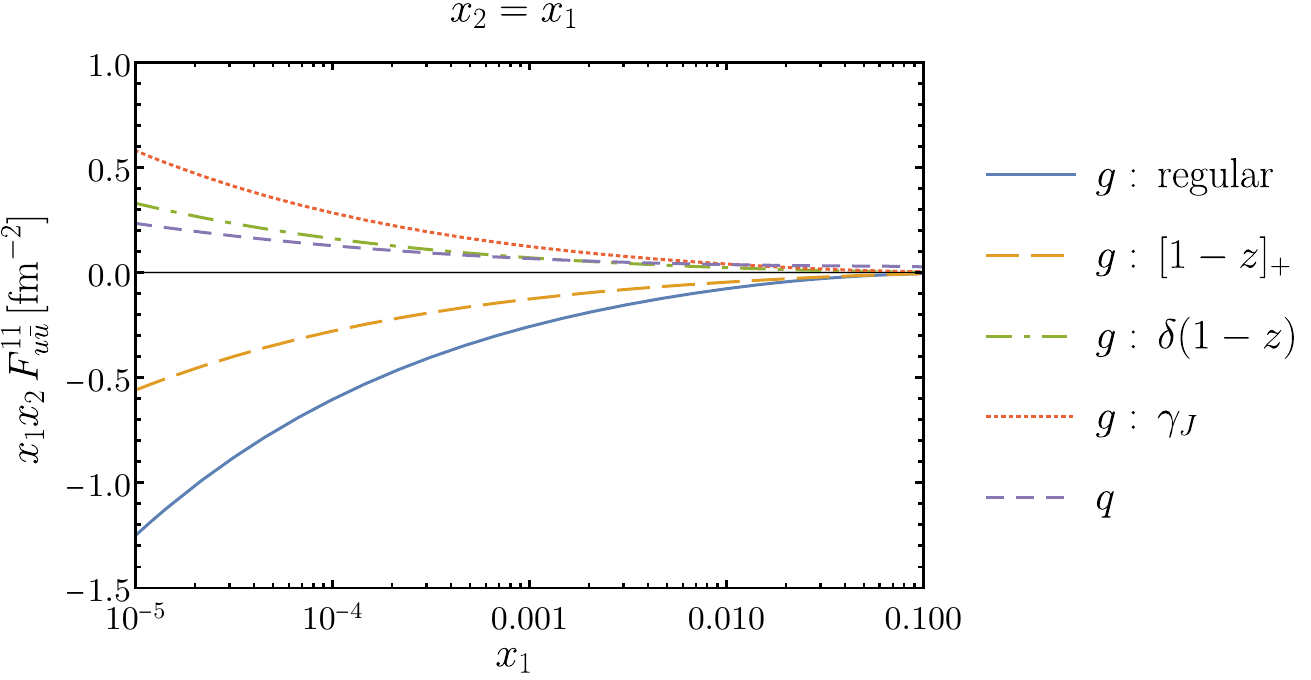}}
\caption{\label{fig:Fuubar-11-parts} Individual contributions to $F_{u \bar{u}}^{1 1}$ at $x_1 = x_2$ at two-loop level, evaluated with $x_1 x_2\ms \zeta_p = \mu^2$ for two values of $\mu$.  The curves labelled ``$g$'' are for the splitting $g\to q\bar{q}$, and the curve labelled ``$q$'' is for the sum of all splitting contributions initiated by a quark or an antiquark.
The labels ``regular'', ``$[1-z]_+$'', and ``$\delta(1-x)$'' refer to the different parts of the kernels $V^{1 1 \, [2,0]} + L_y \, V^{1 1 \, [2,1]}$ discussed below \eqn{\protect\eqref{double-log}}.}
\end{center}
\end{figure}

In \fig{\ref{fig:Fuubar-11-parts}} we show the different contributions to $F_{u \bar{u}}^{1 1}$ for $x_1 = x_2$.  The regular part of the $g \to q\bar{q}$ kernel turns out to be negative for all values of $z$ and $u$ and results in a large negative contribution to the DPD.  At $\mu = 1.2 \ms\mu_y$, another negative contribution comes from the part of the kernel that goes with the plus-distribution $1 /[1-z]_+$.

\begin{figure}
\begin{center}
   \subfigure[$x_2 = x_1$, $\mu = \mu_y$]{\includegraphics[height=0.3\textwidth,
      trim=0 0 0 16,clip=true]{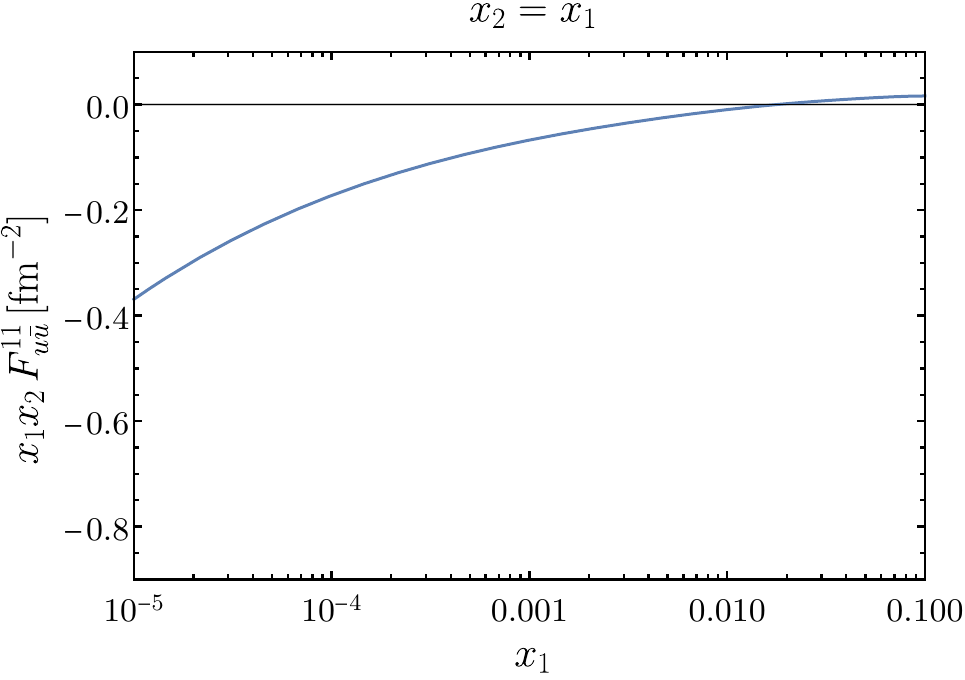}
   \hspace{1em}
   \includegraphics[height=0.3\textwidth,
      trim=25 0 0 16,clip=true]{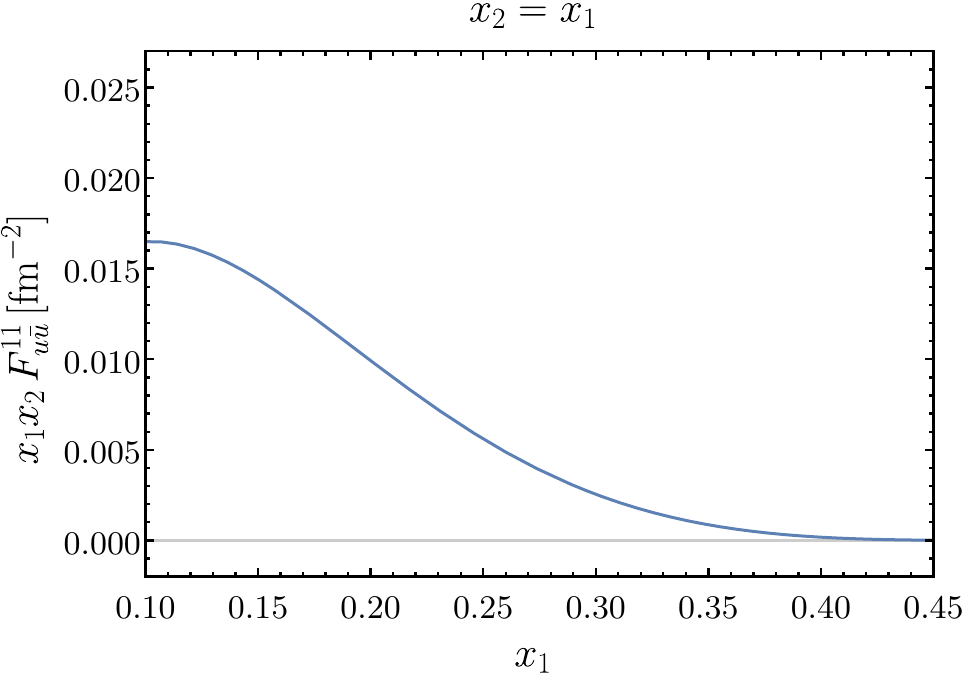}}
\\[0.5em]
   \subfigure[$x_2 = x_1$, $\mu = 1.2 \ms\mu_y$]{\includegraphics[height=0.3\textwidth,
      trim=0 0 0 16,clip=true]{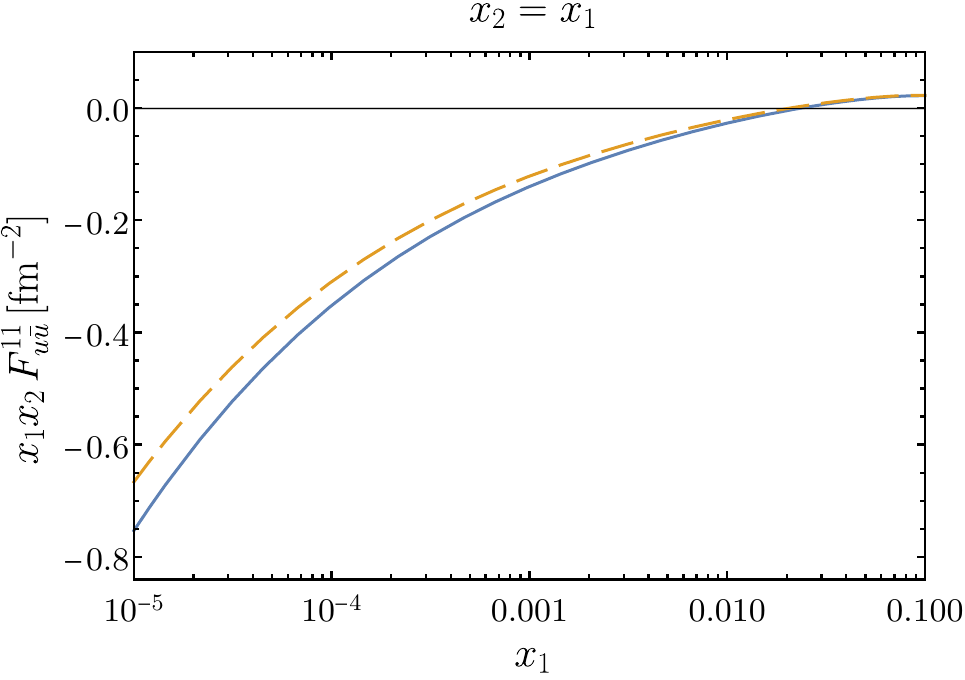}
   \hspace{1em}
   \includegraphics[height=0.3\textwidth,
      trim=25 0 0 16,clip=true]{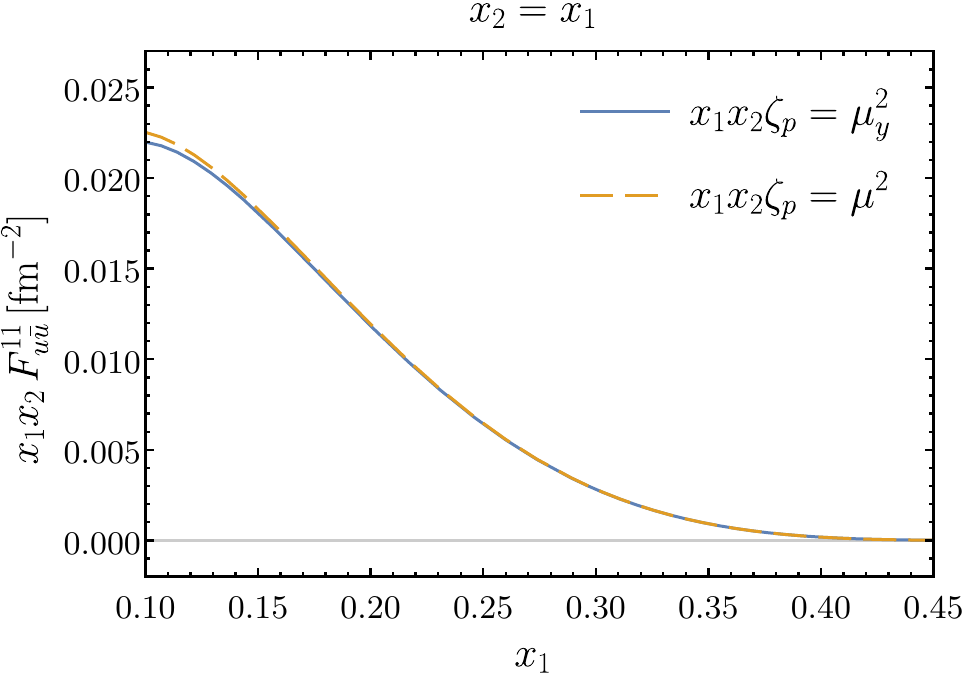}}
\\[0.5em]
   \subfigure[$x_2 = 0.01$, $\mu = \mu_y$]{\includegraphics[height=0.3\textwidth,
      trim=0 0 0 24,clip=true]{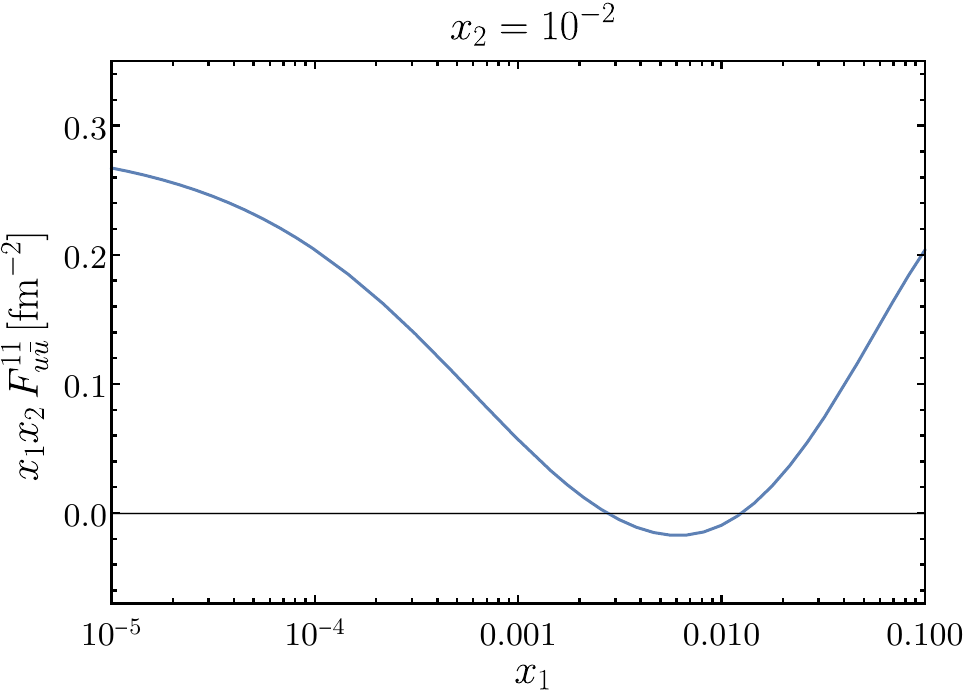}
   \hspace{1em}
   \includegraphics[height=0.3\textwidth,
      trim=30 0 0 24,clip=true]{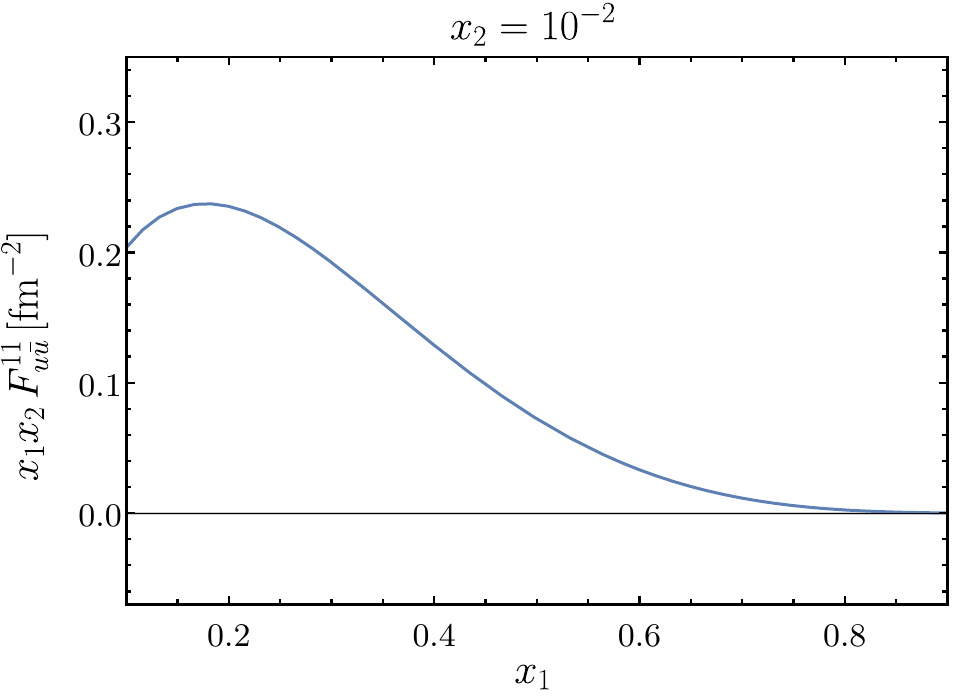}}
\\[0.5em]
   \subfigure[$x_2 = 0.01$, $\mu = 1.2 \ms\mu_y$]{\includegraphics[height=0.3\textwidth,
      trim=0 0 0 24,clip=true]{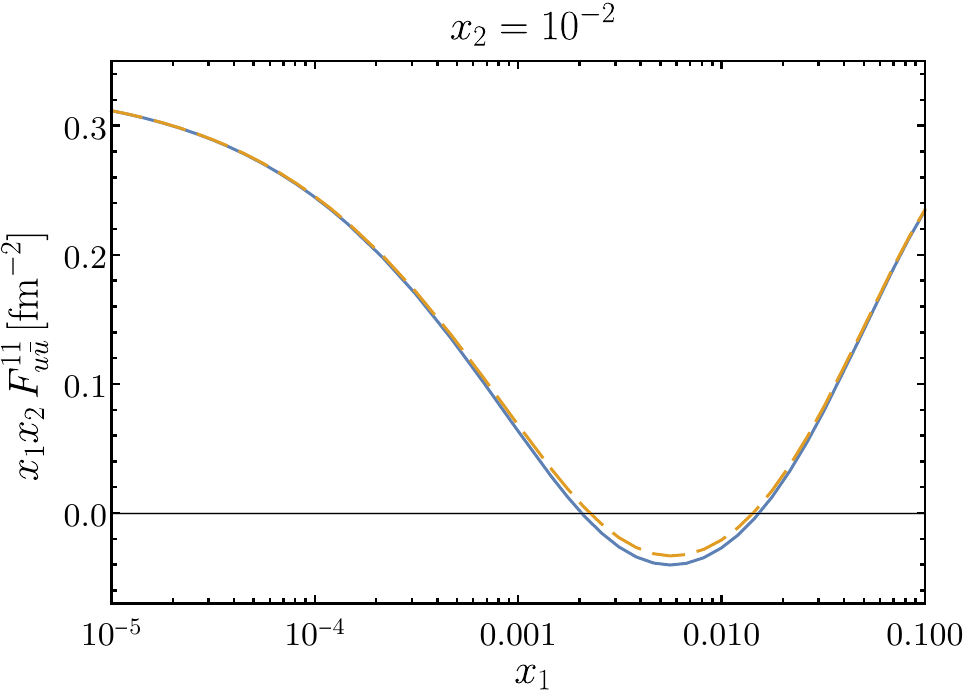}
   \hspace{1em}
   \includegraphics[height=0.3\textwidth,
      trim=30 0 0 24,clip=true]{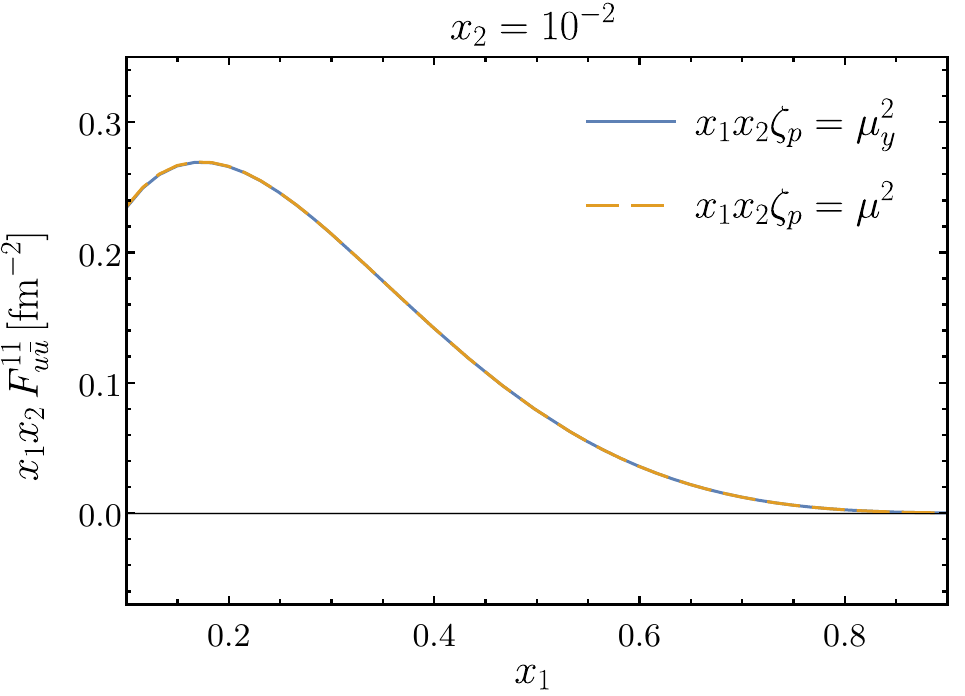}}
\caption{\label{fig:Fuubar-11} The distribution $F_{u \bar{u}}^{1 1}$ computed with the two-loop splitting formula for $\mu_y = 10 \gev$ and two choices of $\mu$.}
\end{center}
\end{figure}

Adding up all contributions, one obtains the distribution shown in the two upper rows of \fig{\ref{fig:Fuubar-11}}.  The negative contributions dominate for $x_1 = x_2$ up to a few $0.01$, whereas for larger momentum fractions the positive contributions from quark or antiquark splitting gradually takes over.  In the two lower rows of \fig{\ref{fig:Fuubar-11}}, we see that the regions of negative $F_{u \bar{u}}^{1 1}$ are centred around $x_1 \sim x_2$, with values that are small compared with the size of the distribution at other values of $x_1$ at the same $x_2$.  This is the same phenomenon that we observed earlier for $F_{u d}^{}$, $F_{u \smash{\bar{d}}}^{}$, and $F_{u u}^{\overline{3} 3}$.  However, in the present case the negative values around $x_1 \sim x_2$ decrease with $\mu$, so that the violation of positivity in $F_{u \bar{u}}^{1 1}$ becomes more pronounced as $\mu$ becomes larger.

In \fig{\ref{fig:Fuubar-11}} we also see that a mild variation of the rapidity parameter $\zeta_p$ has a rather small effect in the kinematics considered here.  The choice of $\zeta_p$ is only relevant if $\mu \neq \mu_y$, because $L_{\zeta}$ is multiplied by $L_y$ in \eqref{double-log}.

Let us finally turn to the distribution $F_{u \bar{u}}^{8 8}$.  Evaluated at one-loop accuracy, this distribution is positive, so that negative values at two- loop level can only appear when the contribution of order $a_s^2$ is larger in size than the one of order $a_s$.  In such a case, one may of course worry whether the unknown contributions of yet higher orders will change the sign of the distribution again.  This situation is qualitatively different from the one for the distributions discussed so far, where the terms of order $a_s^2$ give the first non-vanishing contribution.

\begin{figure}
\begin{center}
\subfigure[\label{fig:Fuubar-1a}$x_2 = x_1$, $\mu = \mu_y$]{\includegraphics[height=0.3\textwidth,
      trim=0 0 0 16,clip=true]{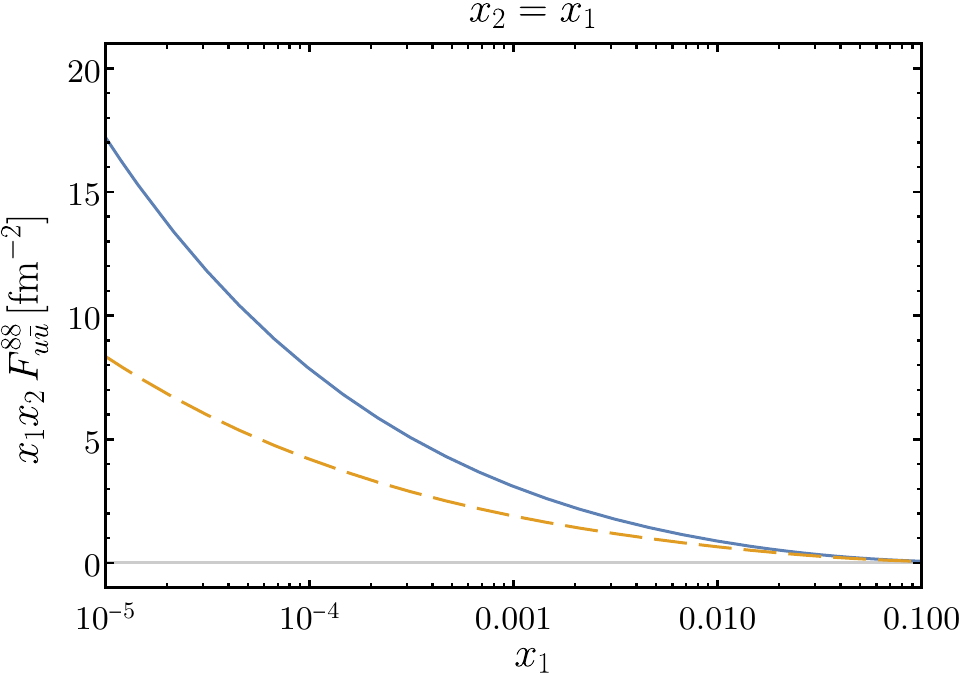}
   \hspace{1em}
   \includegraphics[height=0.3\textwidth,
      trim=30 0 0 16,clip=true]{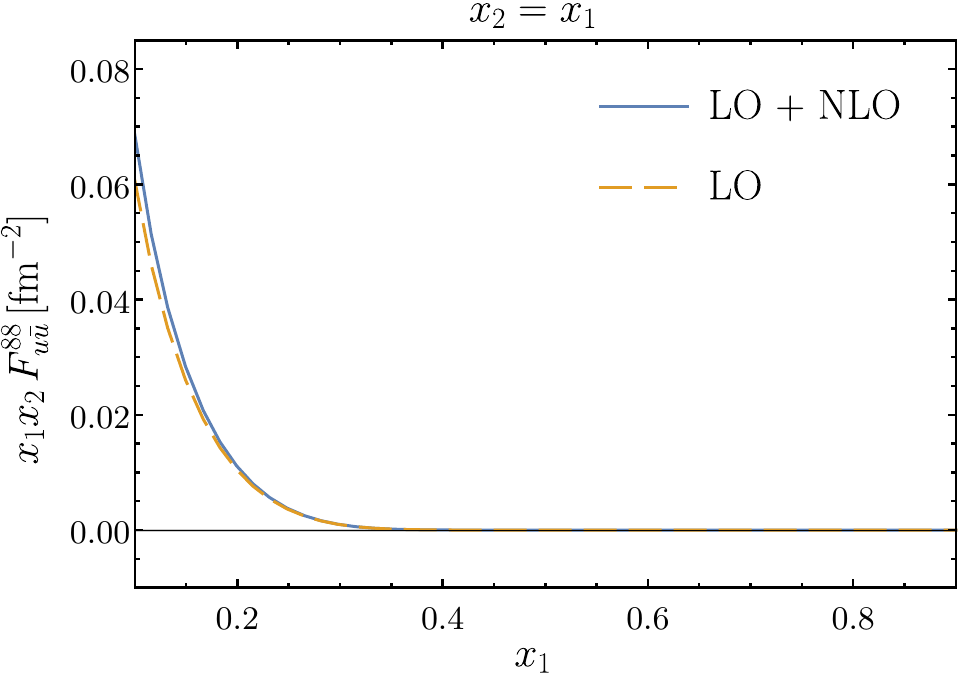}}
\\[0.5em]
\subfigure[\label{fig:Fuubar-1b}$x_2 = x_1$, $\mu = 1.2 \ms\mu_y$]{\includegraphics[height=0.3\textwidth,
      trim=0 0 0 16,clip=true]{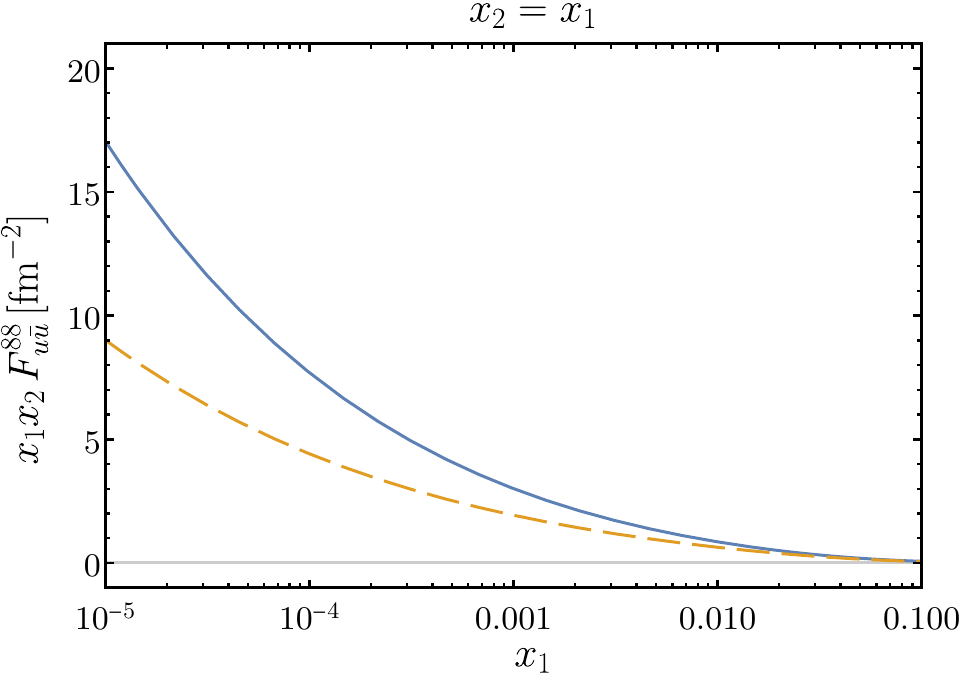}
   \hspace{1em}
   \includegraphics[height=0.3\textwidth,
      trim=30 0 0 16,clip=true]{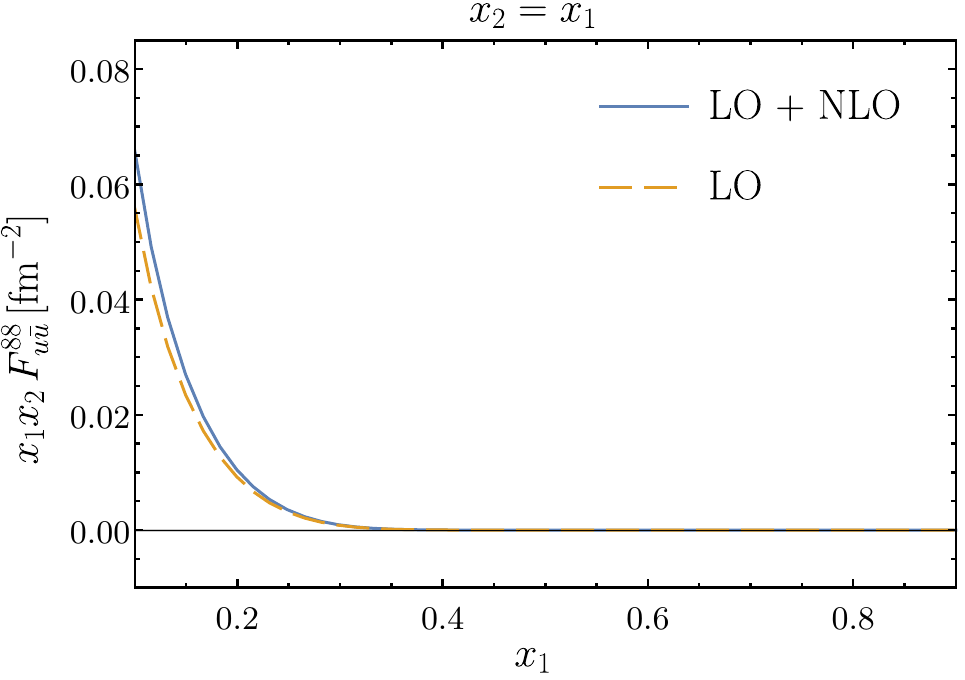}}
\\[0.5em]
\subfigure[$x_2 = 0.01$, $\mu = \mu_y$]{\includegraphics[height=0.3\textwidth,
      trim=0 0 0 24,clip=true]{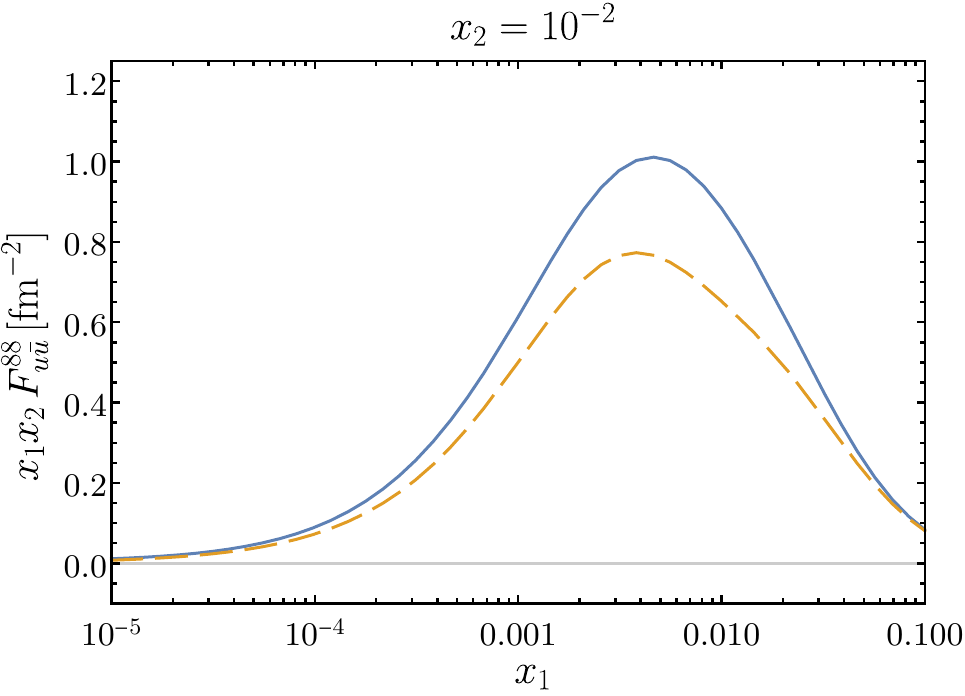}
   \hspace{1em}
   \includegraphics[height=0.3\textwidth,
      trim=30 0 0 24,clip=true]{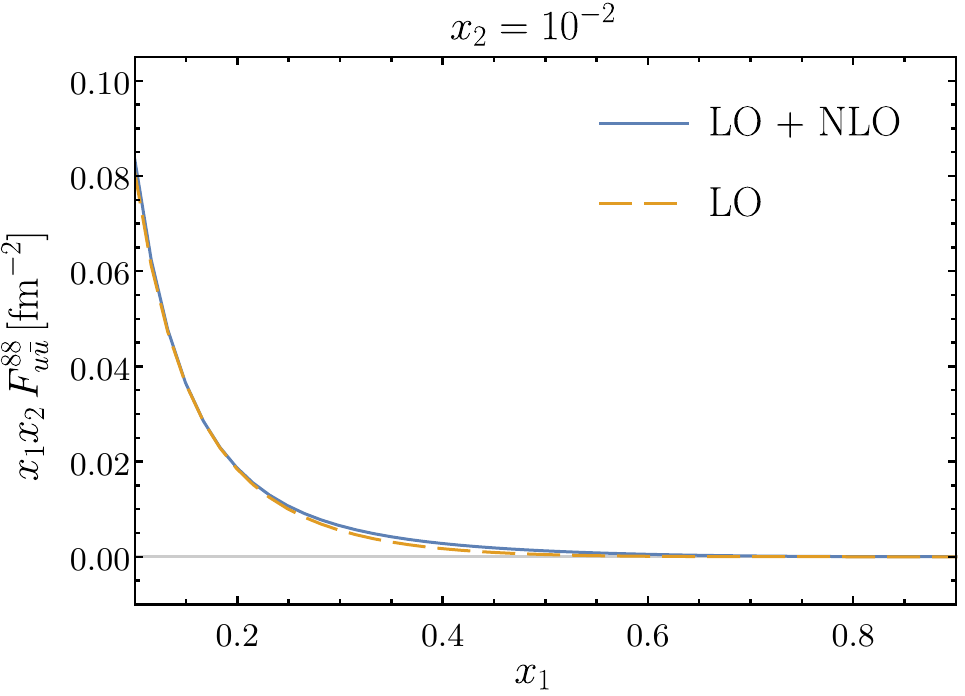}}
\\[0.5em]
\subfigure[$x_2 = 0.01$, $\mu = 1.2 \ms\mu_y$]{\includegraphics[height=0.3\textwidth,
      trim=0 0 0 24,clip=true]{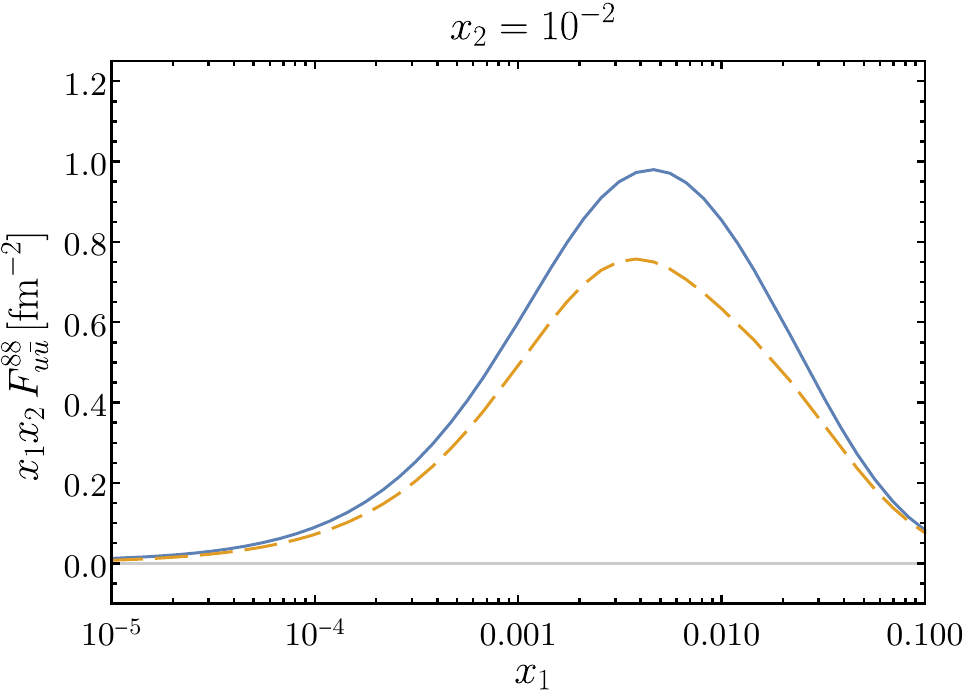}
   \hspace{1em}
   \includegraphics[height=0.3\textwidth,
      trim=30 0 0 24,clip=true]{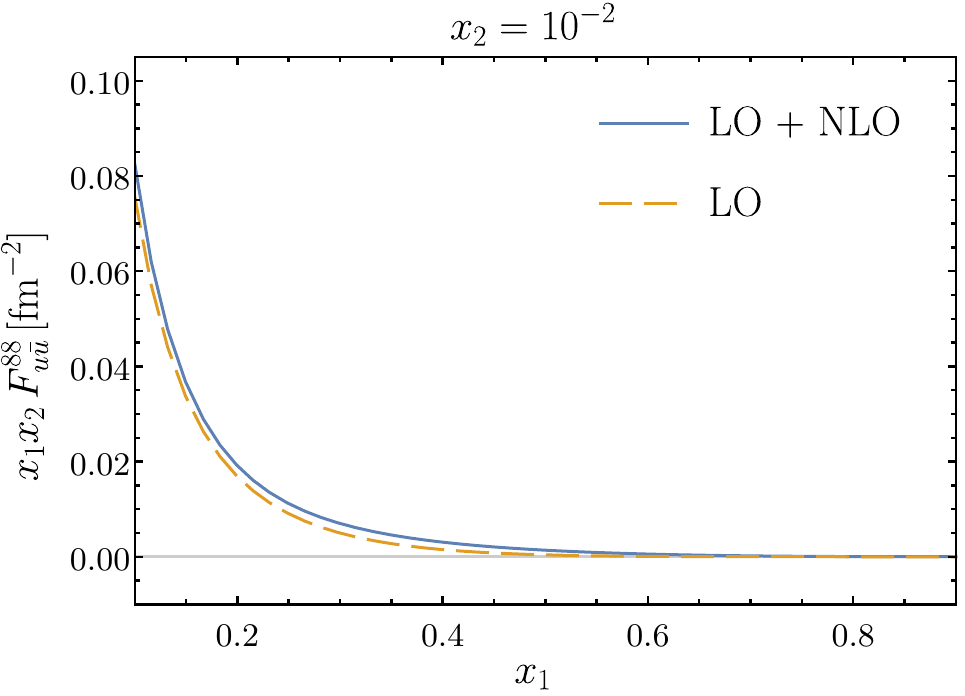}}
\caption{\label{fig:Fuubar-88-part1} The distribution $F_{u \bar{u}}^{8 8}$ computed for $\mu_y = 10 \gev$, either with one-loop (LO) or with the sum of one- and two-loop splitting kernels (LO+NLO).  All plots are for $x_1 x_2\ms \zeta_p = \mu^2$.}
\end{center}
\end{figure}

\begin{figure}
\begin{center}
\subfigure[\label{fig:Fuubar-2a}$x_1 = 0.1$, $\mu = 1.2 \ms\mu_y$]{\includegraphics[height=0.3\textwidth,
      trim=0 0 0 24,clip=true]{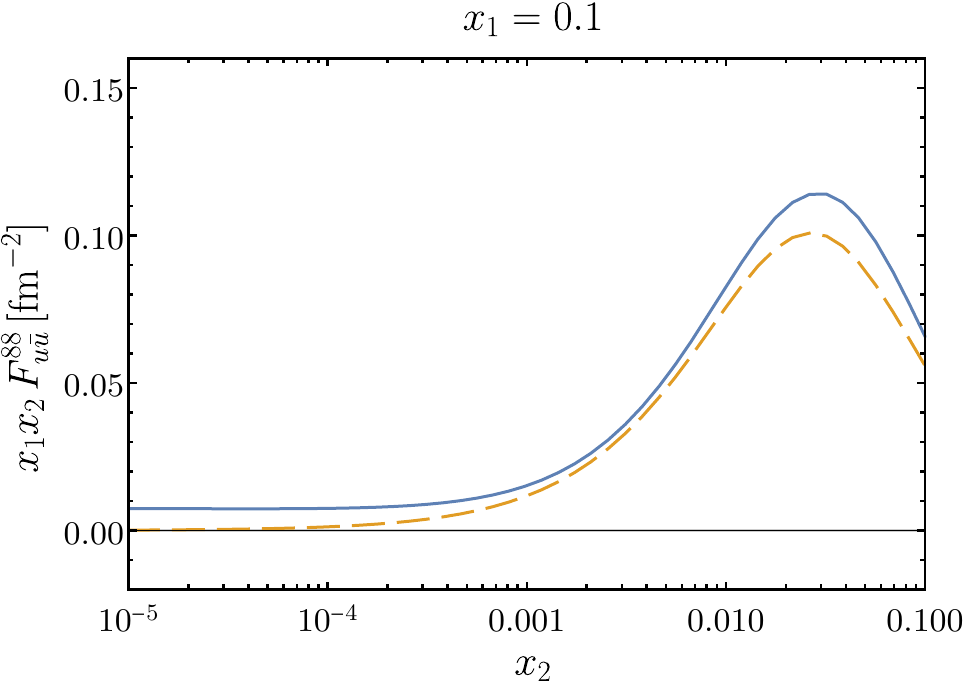}
   \hspace{1em}
   \includegraphics[height=0.298\textwidth,
      trim=30 0 0 22,clip=true]{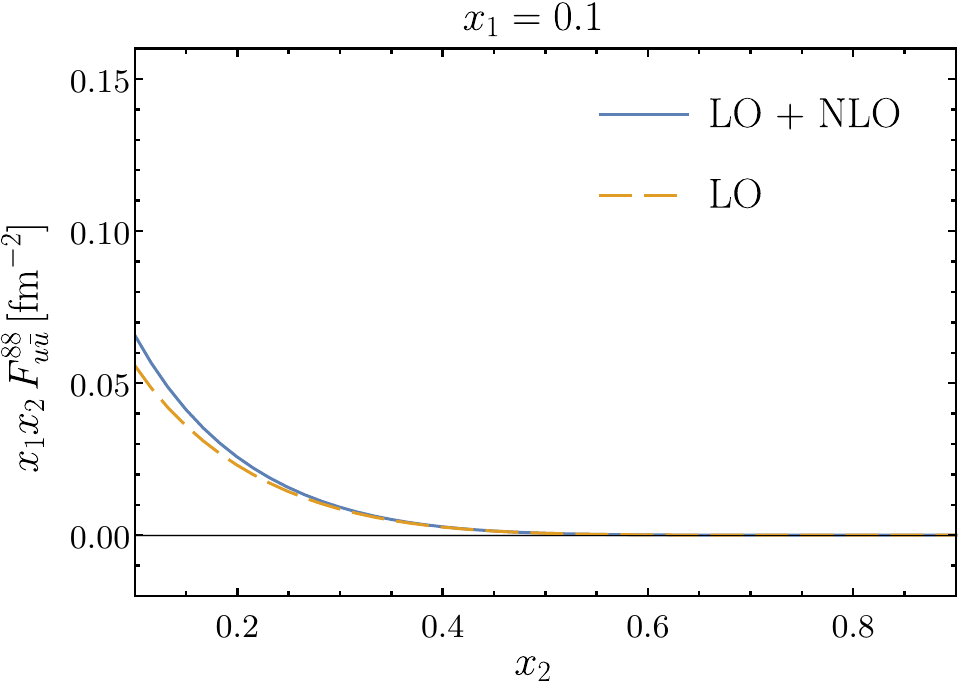}}
\\[0.5em]
\subfigure[\label{fig:Fuubar-2b}$x_2 = 10^{-3}$, $\mu = 1.2 \ms\mu_y$]{\includegraphics[height=0.3\textwidth,
      trim=0 0 0 24,clip=true]{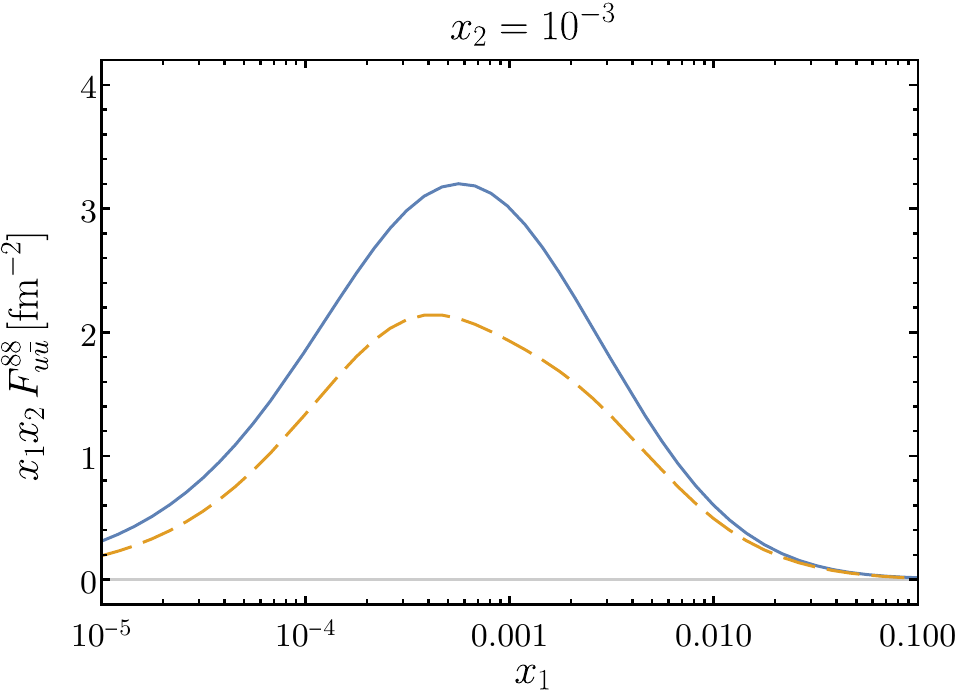}
   \hspace{1em}
   \includegraphics[height=0.3\textwidth,
      trim=30 0 0 24,clip=true]{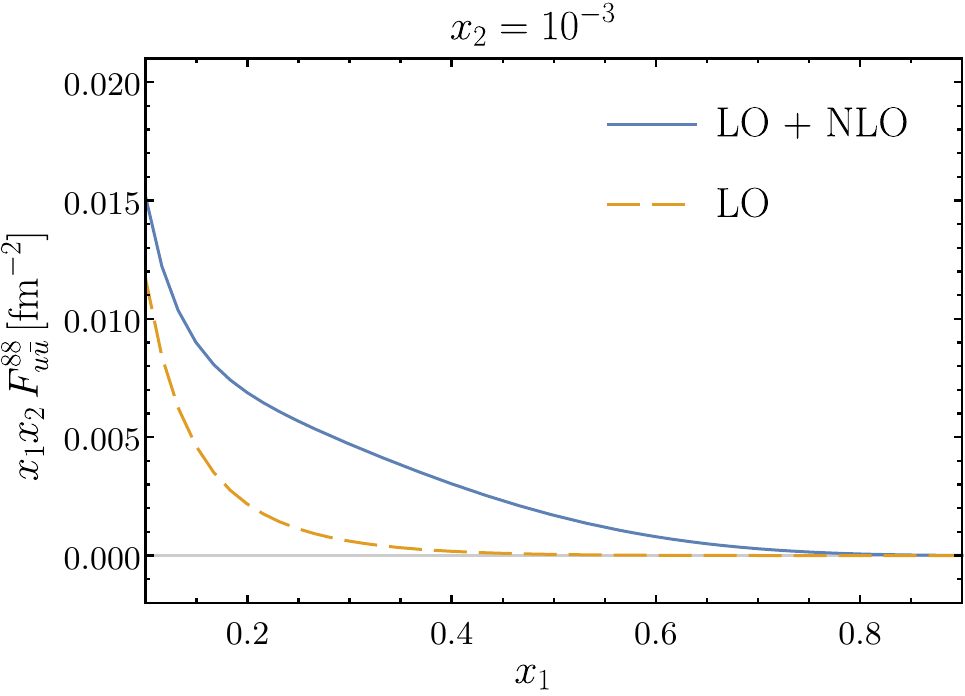}}
\\[0.5em]
\subfigure[\label{fig:Fuubar-2c}$x_1 = 10^{-3}$, $\mu = \mu_y / 1.2$]{\includegraphics[height=0.3\textwidth,
      trim=0 0 0 24,clip=true]{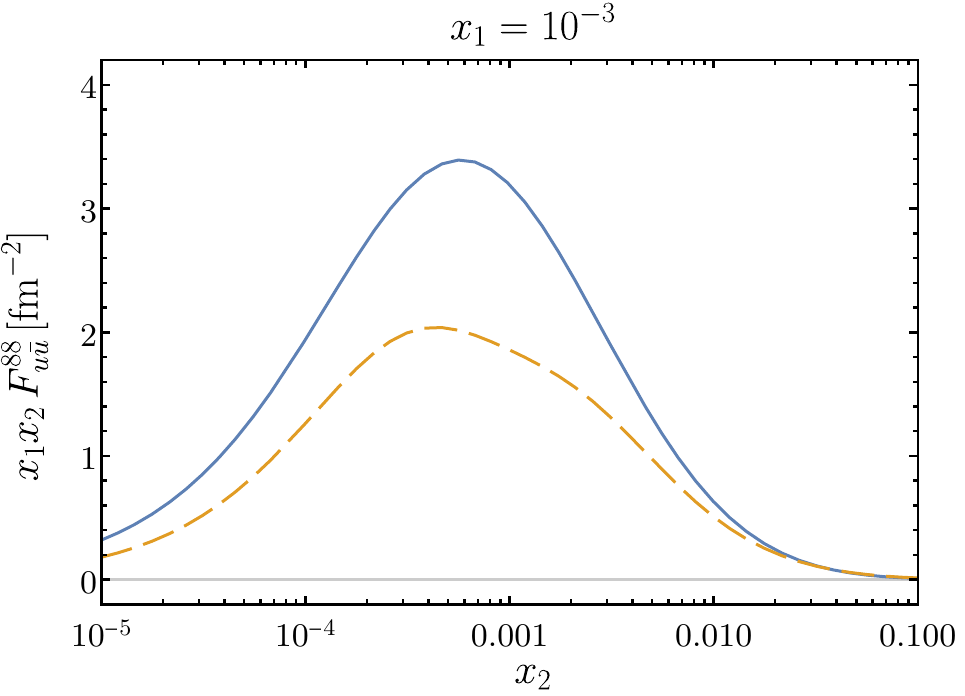}
   \hspace{1em}
   \includegraphics[height=0.3\textwidth,
      trim=30 0 0 24,clip=true]{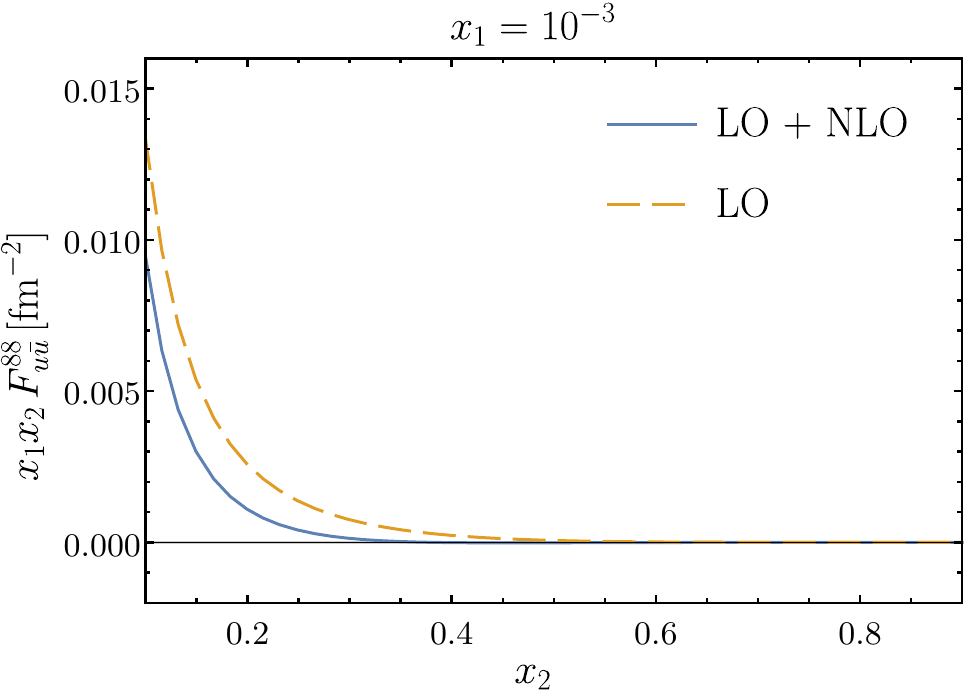}}
\\[0.5em]
\subfigure[\label{fig:Fuubar-2d}detail of (c)]{\includegraphics[height=0.3\textwidth,
      trim=0 0 0 24,clip=true]{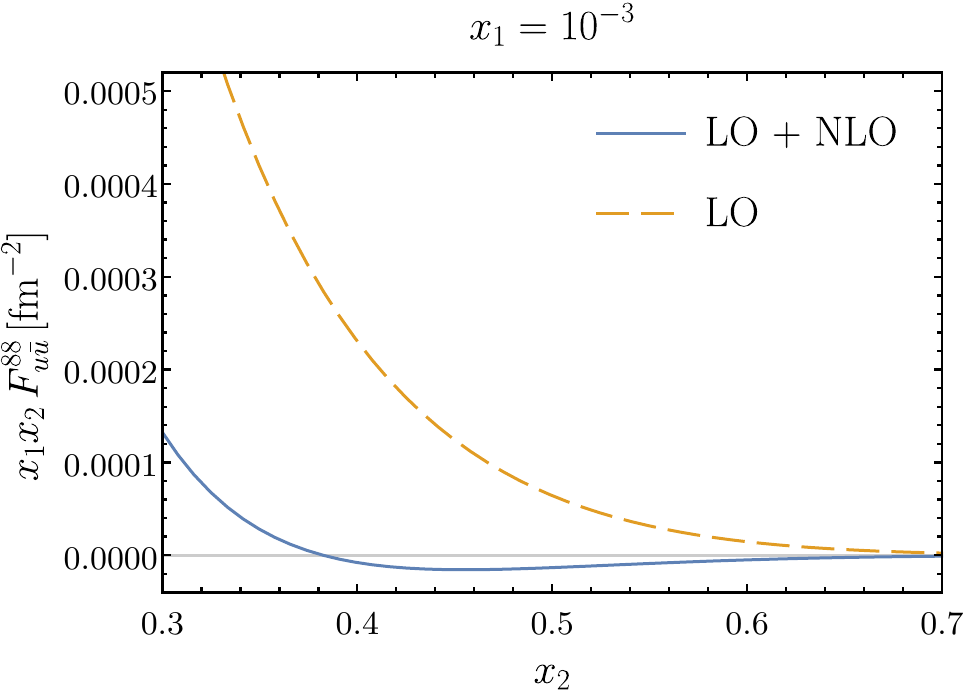}}
\caption{\label{fig:Fuubar-88-part2} Continuation of \fig{\protect\ref{fig:Fuubar-88-part1}}.  Notice that $x_1$ is fixed in panels (a), (c), and (d), whilst $x_2$ is fixed in panel (b).}
\end{center}
\end{figure}

In \figs{\ref{fig:Fuubar-88-part1}} and \ref{fig:Fuubar-88-part2} we show $F_{u \bar{u}}^{8 8}$ for different values of the momentum fractions.  We always take $x_1 x_2\ms \zeta_p = \mu^2$, bearing in mind that according to \eqref{V_uubar} the effect of varying $\zeta_p$ is eight times smaller for $F_{u \bar{u}}^{8 8}$ than it is for $F_{u \bar{u}}^{1 1}$.  As can be seen in \fig{\ref{fig:Fuubar-88-part1}}, the difference between the distributions at $\mu = \mu_y$ to $\mu = 1.2\ms \mu_y$ is rather small, in contrast to what we found for the other distributions discussed so far.  This is not surprising, because for $F_{u \bar{u}}^{8 8}$ the relative effect of changing the scale from $\mu_a$ to $\mu_b$ is of order $a_s \log(\mu_a/\mu_b)$, whereas it is of order $\log(\mu_a/\mu_b)$ for distributions that receive their first nonzero contribution at two loops.

In all panels of \figs{\ref{fig:Fuubar-88-part1}} and \ref{fig:Fuubar-88-part2}, we show the result obtained with either the one-loop kernel or with the sum of one- and two-loop kernels.  In both cases we take the same NLO PDFs, so that the difference between the LO and the LO+NLO curves directly shows the impact of the two-loop kernels.  We find that the size of the $a_s^2$ corrections is often moderate but becomes large in several kinematic situations.
\begin{enumerate}
\item As discussed in \sect{4.3} of \cite{Diehl:2021wpp}, the two-loop corrections for $g \to q \bar{q}$ and $q \to q'\bs \bar{q}'$ are enhanced at small $x_1 + x_2$.  This is seen in the left panels of \figs{\ref{fig:Fuubar-1a}}, \ref{fig:Fuubar-1b}, \ref{fig:Fuubar-2b}, and \ref{fig:Fuubar-2c}.  As follows from \eqs{(4.46)} and (4.48) in \cite{Diehl:2021wpp}, the enhanced corrections  provide a positive contribution to $F_{u \bar{u}}^{8 8}$.

An all-order resummation of the enhanced corrections using techniques from small-$x$ factorisation may be possible, but details of this have not been worked out.
\item As explained in \sect{4.4} of \cite{Diehl:2021wpp}, the splitting graph for $u\to u \bar{u}$ in \fig{\ref{fig:qqbar-2}} leads to a behaviour of the DPD like $1/(1-u) \approx x_1/x_2$ for $x_2 \ll x_1$, which is absent at order~$a_s$.  This explains why in \fig{\ref{fig:Fuubar-2a}} the LO+NLO result for the scaled DPD $x_1 x_2\ms F_{u \bar{u}}^{8 8}$ goes to a finite value when $x_2 \ll x_1$, whereas the LO result goes to zero.  It also explains the huge relative NLO corrections seen in the right panel of \fig{\ref{fig:Fuubar-2b}}.  In the latter case, the splitting process $u\to u \bar{u}$ is further enhanced by the fact that the $u$ quark distribution becomes the dominant PDF with increasing $x$.

The appearance of an additional power $1/(1-u)$ is unique for the step from LO to NLO in this channel and will not repeat itself at yet higher orders.
\item For $x_2 \gg x_1$ we find a large negative two-loop contribution to $F_{u \bar{u}}^{8 8}$, which is seen in the right panel of \fig{\ref{fig:Fuubar-2c}} and more clearly in \fig{\ref{fig:Fuubar-2d}}, where the LO+NLO result becomes just slightly negative.  This can be traced back to a negative term
\begin{align}
- \frac{5 a_s^2}{24}\, \delta(1-z)\,
   \pr{11}{V}_{q \bar{q}, g}^{(1)}(u)\, \log^2\bs u
\end{align}
in the kernel $V_{q \bar{q}, g}^{8 8}$, which is enhanced by two powers of $\log u$ compared with the LO expression.

It would require further analysis to understand whether this type of enhancement repeats itself at yet higher orders and, if so, whether it can be resummed to all orders.  We therefore cannot say whether the negative values seen in \fig{\ref{fig:Fuubar-2d}} will disappear when higher order terms are included.

Notice that \figs{\ref{fig:Fuubar-2c}} and \fig{\ref{fig:Fuubar-2d}} refer to the lowest scale $\mu = \mu_y /1.2$ considered in this study.  The corresponding plots for $\mu = \mu_y$ or $1.2\ms \mu_y$ show large negative NLO corrections as well, but the LO+NLO curves do not become negative any more.
\end{enumerate}
We remark in passing that the enhanced two-loop contributions described in points 1 and 3 scale like the LO kernel for $g \to q \bar{q}$ and hence cancel out in $F_{u \bar{u}}^{1 1}$.  By contrast, the $1/(1-u)$ behaviour discussed in point 2 appears both in $F_{u \bar{u}}^{8 8}$ and $F_{u \bar{u}}^{1 1}$.

\paragraph{Colour summed distributions}
Let us finally consider the colour summed distributions
$\pr{11}{F}_{u d}$,
$\pr{11}{F}_{u \smash{\bar{d}}}$,
$\pr{11}{F}_{u u}$, and
$\pr{11}{F}_{u \bar{u}}$.
The relations \eqref{singlet-sum} and \eqref{ud-channels} imply that $\pr{11}{F}_{u d}$ and $\pr{11}{F}_{u \smash{\bar{d}}}$ are negative whenever their $s$ channel counterparts are.  By contrast, we find that $\pr{11}{F}_{u u}$ and $\pr{11}{F}_{u \bar{u}}$ remain positive for the kinematic settings shown in \figs{\ref{fig:Fuu}} to \ref{fig:Fuubar-88-part2} .

\FloatBarrier

\section{Summary}
\label{sec:sum}

In the context of the parton model, DPDs in the $s$ channel colour basis are probability densities for finding two partons in a definite colour state (with specified longitudinal momenta and specified transverse distance from each other).  This leads to the expectation that $s$ channel DPDs for unpolarised partons should be non-negative.  If it is satisfied, this positivity property provides valuable constraints on the colour dependence of DPDs, along with a strategy to model them by saturating the positivity bounds at a certain scale.

In the present work, we investigate whether evolution of DPDs to higher scales preserves the positivity property under the assumption that it holds at the starting scale.  We limit ourselves to unpolarised partons, and we exclude two-gluon DPDs from our consideration because their colour structure is much more involved than the one of pure quark or quark-gluon distributions.

DPDs in the $s$ channel colour basis are subject to Collins-Soper evolution in the rapidity parameter $\zeta_p$, and this evolution includes mixing between different colour channels.  Provided that the renormalisation scales $\mu_1$ and $\mu_2$ associated with the two partons are large enough compared with $1/y$, the Collins-Soper kernel $\pr{8}{J}(y, \mu_1, \mu_2)$ for DPDs is negative.  Under this condition, evolution to higher $\zeta_p$ preserves positivity.  Conversely, backward evolution to sufficiently small $\zeta_p$ eventually leads to negative distributions (except for the special case in which all $t$ channel distributions other than $\pr{11}{F}_{a_1 a_2}$ are zero and hence all $s$ channel distributions are independent of $\zeta_p$).

We next consider the DGLAP equations for evolution in one of the scales $\mu_1$ or $\mu_2$, with the evolution kernels taken at LO.  We find that evolution to higher scales is not guaranteed to preserve positivity: there are initial conditions that satisfy positivity but lead to negative $s$~channel distributions at higher scales.  This is due to the convolution of a plus distribution in the evolution kernel with a DPD different from the one being evolved.  There is no  contribution of this type in the LO evolution equations for polarised colour summed DPDs $\pr{11}{F}_{a_1 a_2}$, which conserve positivity in the same way as the LO evolution of polarised PDFs \cite{Diehl:2013mla}.  In a numerical illustration, we choose initial conditions where certain $s$ channel DPDs are zero and see that they turn negative at slightly higher scales.  We also study joint DGLAP and Collins-Soper evolution in the common scale $\mu = \mu_1 = \mu_2 = \sqrt{x_1 x_2\ms \zeta_p}$ and find that positivity is not preserved, for the same reasons as above.

At small inter-parton distance $y$, the initial conditions for DPD evolution can be computed using the perturbative splitting mechanism, setting $\mu = \mu_1 = \mu_2 \approx 1/y$ and using a fixed-order truncation of the DPD splitting kernels.  It is easy to see that at order $a_s$ one obtains DPDs that satisfy positivity in colour space.  At order $a_s^2$ this no longer holds: in a numerical study we obtain negative values for colour channels in which the distributions are zero at order $a_s$, and also for the distribution $F_{u \bar{u}}^{8 8}$, which is nonzero at order $a_s$.  Negative values are also found for the colour summed distributions $\pr{11}{F}_{u d}$ and $\pr{11}{F}_{u \smash{\bar{d}}}$.  In several cases, the considered distributions have no $\zeta_p$ dependence at order $a_s^2$, and negative values of them can be uniquely traced back to the subtraction of ultraviolet divergences implied in the definition of twist-two operators.  The explicit form of this subtraction is given in \sect{2.6} of \cite{Diehl:2021wpp}.  The negative values we find for the distributions are small compared with the size of the same distributions at other values of the momentum fractions $x_1$ and $x_2$.  In this sense, the violations of positivity we have seen may be regarded as ``relatively small''.  In view of this, we should also caution that negative values obtained with the splitting formula at order $a_s^2$ may turn into positive ones when yet higher orders are included.

Note that the violations of positivity just discussed refer to DPDs defined with \msbar renor\-malisation of twist-two operators.  By contrast, the violation of positivity by forward DGLAP evolution described earlier occurs at LO and is hence not specific to the \msbar scheme.

We conclude that the positivity of DPDs in full colour space cannot be taken for granted and can be violated in physically realistic settings.  Using positivity as a guide for modelling DPDs at large $y$ may still be an option when there is a lack of better information.  It should however be done with due caution, and one should check whether the chosen initial conditions give positive distributions when evolved to higher scales.

\appendix
\section{Colour space projectors}

In this appendix, we list the colour space projectors that appear in the definitions \eqref{s-channel-F} and \eqref{t-channel-F} of DPDs in the $s$ and $t$ channel bases.  In the pure quark and the mixed quark-gluon sector, we have
\begin{align}
   \label{qq-pro}
\Pro{11}{i i' \ms j j'}
   &= \frac{1}{3} \, \delta_{i' i} \, \delta_{j'\bs j} \,,
&
\Pro{88}{i i' \ms j j'}
   &= 2 \ms t^a_{i' i} \, t^a_{j'\bs j} \,,
\\
   \label{qg-pro}
\Pro{11}{a a' \ms i i'}
   &= \frac{1}{\sqrt{24}}\, \delta_{i'\bs i} \, \delta^{a a'} \,,
&
\Pro{S8}{a a' \ms i i'}
   &= \sqrt{\frac{6}{5}}\, d^{a a' c}\, t^c_{i'\bs i} \,,
&
\Pro{A8}{a a' \ms i i'}
   &= \sqrt{\frac{2}{3}}\, f^{a a' c} \, t^c_{i'\bs i} \,,
\end{align}
and
\begin{align}
\label{qq-pro-s}
\Pro{3\ms \overline{3}}{i j\, i' j'} &= \frac{1}{2}\,
   \bigl( \delta_{i' i} \delta_{j' j} - \delta_{i' j} \delta_{j' i} \bigr) \,,
&
\Pro{\overline{6}\ms 6}{i j\, i' j'} &= \frac{1}{2}\,
   \bigl( \delta_{i' i} \delta_{j' j} + \delta_{i' j} \delta_{j' i} \bigr) \,,
\\
\Pro{\overline{3}\ms 3}{i a \, i'\bs a'}
   &= \frac{3}{4}\, \bigl( t^{a'} t^a \bigr)_{i'\bs i} \,,
&
\label{qg-pro-s}
\Pro{6\ms \overline{6}}{i a \, i'\bs a'}
   &= \frac{1}{2}\, \delta_{i'\bs i} \, \delta^{a a'}
      - \bigl( t^{a} t^{a'} \bigr)_{i'\bs i}
      - \frac{1}{2}\, \bigl( t^{a'} t^a \bigr)_{i'\bs i} \,,
\nonumber \\
& &
\Pro{\overline{15}\, 15}{i a \, i'\bs a'}
   &= \frac{1}{2}\, \delta_{i'\bs i} \, \delta^{a a'}
      + \bigl( t^{a} t^{a'} \bigr)_{i'\bs i}
      - \frac{1}{4}\, \bigl( t^{a'} t^a \bigr)_{i'\bs i} \,.
\end{align}
For completeness, we also give the projectors for the pure gluon sector, although they are not used in the present work.
\begin{align}
\label{gg-pro}
\Pro{11}{a a'\, b b'} &= \frac{1}{8}\, \delta^{a a'} \delta^{b b'} \,,
\nonumber \\
\Pro{SS}{a a'\, b b'} &= \frac{3}{5}\, d^{a a' c} d^{b b' c} \,,
\qquad\qquad
\Pro{AA}{a a'\, b b'} = \frac{1}{3}\, f^{a a' c} f^{b b' c} \,,
\qquad\qquad
\Pro{S\bs A}{a a'\, b b'} = \frac{1}{\sqrt{5}}\, d^{a a' c} f^{b b' c} \,,
\nonumber \\
P_{10\ms \overline{10}}^{a a'\, b b'}
   &= \frac{1}{4}\, \bigl( \delta^{a b}
      \delta^{a' b'} - \delta^{a b'} \delta^{a' b} \bigr)
      - \frac{1}{2}\, P_{AA}^{a a'\, b b'}
      - \frac{i}{4}\, \bigl( d^{a b c} f^{a' b' c}
         + f^{a b c} d^{a' b' c} \bigr) \,,
\nonumber \\
\Pro{27\ms 27}{a a'\, b b'}
&= \frac{1}{2}\, \bigl( \delta^{ab} \delta^{a' b'}
   + \delta^{a b'} \delta^{a'b} \bigr)
   - P_{SS}^{a a'\, b b'} - P_{11}^{a a'\, b b'} \,.
\end{align}
In all cases, we have set the number of colours to $N=3$.
Further projectors are obtained by exchanging the representation labels and the corresponding indices:
\begin{align}
\Pro{R \Rp}{r_1^{} r_2^{}\, r_1' r_2'}
   &= \Pro{\Rp\bs R}{r_1' r_2'\, r_1^{} r_2^{}} \,.
\end{align}

One readily verifies the completeness relations
\begin{align}
   \label{complete-rel}
\sum_{R} \Pro{R \Rbar}{r_1^{} r_2^{}\, r_2' r_1'}
   &= \delta_{r_1^{} r_1'}\, \delta_{r_2^{} r_2'}
&& \text{for $P_{R \Rbar}$ in \protect\eqref{qq-pro}} \,,
\nonumber \\
\sum_{R} \Pro{R \Rbar}{r_1^{} r_2^{}\, r_1' r_2'}
   &= \delta_{r_1^{} r_1'}\, \delta_{r_2^{} r_2'}
&& \text{for $P_{R \Rbar}$ in \protect\eqref{qq-pro-s},
   \protect\eqref{qg-pro-s}, \protect\eqref{gg-pro}} \,,
\end{align}
where in each case the sum runs over all available representations.
The multiplicity of a representation $R$ can be computed from the trace
\begin{align}
  \label{mR-def}
m(R) &=
\begin{cases}
P_{R \Rbar}^{r_1 r_2\, r_2 r_1}
   & \text{for $P_{R \Rbar}$ in \protect\eqref{qq-pro},} \\[0.6em]
P_{R \Rbar}^{r_1 r_2\, r_1 r_2}
   & \text{for $P_{R \Rbar}$ in \protect\eqref{qq-pro-s},
   \protect\eqref{qg-pro-s}, \protect\eqref{gg-pro}.}
\end{cases}
\end{align}

%%%%%%%%%%%%%%%%%%%%%%%%%%%%%%%%%%%%%%%%%%%%%%%%%%%%%%%%%%%%%%%%%%%%%%%%%%

\section*{Acknowledgements}

This work was in part supported by the Deutsche Forschungsgemeinschaft (DFG, German Research Foundation) -- Research Unit FOR 2926, grant number 409651613.  The work of JRG is supported by the Royal Society through Grant URF\textbackslash{}R1\textbackslash{}201500.

%%%%%%u%%%%%%%%%%%%%%%%%%%%%%%%%%%%%%%%%%%%%%%%%%%%%%%%%%%%%%%%%%%%%%%%%%%%

% the following lines create an entry in the table of contents
\phantomsection
\addcontentsline{toc}{section}{References}

\bibliographystyle{JHEP}
\bibliography{bounds.bib}

\end{document}